\documentclass[11pt, a4paper]{article}
\usepackage{jheppub}
\input{epsf.tex}

\usepackage{colortbl}
\usepackage{booktabs}
\usepackage{multirow}
\usepackage{anysize}
\usepackage{array}
\usepackage{latexsym, amscd, amsthm}
\usepackage{pdfsync}
\usepackage{epsf}
\usepackage[all]{xy}
\usepackage[usenames,dvipsnames]{xcolor}
\usepackage{float}
\restylefloat{table}

\makeatletter
\def\@fpheader{\relax}
\makeatother

\newcommand{\Hom}{{\rm Hom}}
\newcommand{\A}{{\Bbb{A}}}

\newcommand{\im}{\mathrm{im}}

\newcommand{\bp}{\begin{Proposition}}
\newcommand{\ep}{\end{Proposition}}
\newcommand{\bl}{\begin{Lemma}}
\newcommand{\el}{\end{Lemma}}
\newcommand{\bt}{\begin{Theorem}}
\newcommand{\et}{\end{Theorem}}
\newcommand{\bd}{\begin{Definition}}
\newcommand{\ed}{\end{Definition}}
\newcommand{\End}{\mathrm{End}}
\newcommand{\Aut}{\mathrm{Aut}}

\newcommand{\Mat}{\mathrm{Mat}}
\newcommand{\ev}{\mathrm{ev}}
\newcommand{\eqdef}{\stackrel{{\rm def.}}{=}}
\newcommand{\Aord}{{A=\mathrm{ordered}}}
\newcommand{\KA}{K\"{a}hler-Atiyah~}
\newcommand{\be}{\begin{equation*}}
\newcommand{\ee}{\end{equation*}}
\newcommand{\ben}{\begin{equation}}
\newcommand{\een}{\end{equation}}
\newcommand{\beqa}{\begin{eqnarray*}}
\newcommand{\eeqa}{\end{eqnarray*}}
\newcommand{\beqan}{\begin{eqnarray}}
\newcommand{\eeqan}{\end{eqnarray}}
\newcommand{\nn}{\nonumber}

\newcommand{\id}{\mathrm{id}}

\newcommand{\tr}{\mathrm{tr}}

\newcommand{\e}{\mathbf{e}}
\newcommand{\cinf}{{{\cal \cC}^\infty(M,\R)}}

\renewcommand{\Re}{\mathrm{Re}}
\renewcommand{\Im}{\mathrm{Im}}
\newcommand{\pqarray}{\begin{array}{c} p-q\\ {\rm mod}~8\end{array}}
\newcommand{\twopartdef}[4]
{
	\left\{
		\begin{array}{ll}
			#1 & \mbox{if } #2 \\
			#3 & \mbox{if } #4
		\end{array}
	\right.
}

\DeclareFontFamily{U}{rsf}{}
\DeclareFontShape{U}{rsf}{m}{n}{<5> <6> rsfs5 <7> <8> <9> rsfs7 <10-> rsfs10}{}
\DeclareMathAlphabet\Scr{U}{rsf}{m}{n}

\def\Z{{\Bbb Z}}
\def\C{{\Bbb C}}
\def\R{{\Bbb R}}

\def\H{{\Bbb H}}

\def\S{{\Bbb S}}

\def\rk{\mathrm{rk}}

\def\vol{\mathrm{vol}}

\def\cR{{\mathcal R}}
\def\cC{{\mathcal C}}
\def\cB{\Scr B}

\def\Cl{\mathrm{Cl}}

\def\cK{\mathrm{\cal K}}
\def\odd{\mathrm{odd}}
\def\even{\mathrm{even}}
\def\Spin{\mathrm{Spin}}

\def\Pin{\mathrm{Pin}}

\def\O{\mathrm{O}}

\def\cI{\mathcal{I}}
\def\cJ{\mathcal{J}}
\def\cP{\mathcal{P}}
\def\cR{\mathcal{R}}

\def\cC{\mathcal{C}}

\def\G_2{\mathrm{G_2}}
\def\cO{\mathcal{O}}

\def\cU{\mathcal{U}}
\def\fJ{\mathfrak{J}}
\def\fI{\mathfrak{I}}
\def\fR{\mathfrak{R}}
\def\fP{\mathfrak{P}}

\def\fZ{\mathfrak{Z}}

\def\f{\mathfrak{f}}
\def\j{\mathbf{j}}

\DeclareSymbolFont{extraup}{U}{zavm}{m}{n}
\DeclareMathSymbol{\varheart}{\mathalpha}{extraup}{86}
\DeclareMathSymbol{\vardiamond}{\mathalpha}{extraup}{87}
\newcommand{\bdiamond}{{\scriptstyle \vardiamond}}

\newcommand{\bcE}{\boldsymbol{\check{E}}}

\title{The geometric algebra of Fierz identities in arbitrary dimensions and signatures}

\author{C. I. Lazaroiu$^{1,2}$, E. M. Babalic$^1$, I. A. Coman$^1$}
\affiliation{$^1$Horia Hulubei National Institute for Physics and Nuclear Engineering (IFIN-HH), Str. Reactorului No. 30, 
Magurele 077125 , Romania\\$^2$Center for Geometry and Physics, Institute for Basic Science and 
Department of Mathematics, POSTECH, Pohang, Gyeongbuk 790-784, Korea
}
\emailAdd{lcalin@theory.nipne.ro, mbabalic@theory.nipne.ro, icoman@theory.nipne.ro} 

\abstract{We use geometric algebra techniques to give a synthetic and
computationally efficient approach to Fierz identities in arbitrary
dimensions and signatures, thus generalizing previous work. Our
approach leads to a formulation which displays the underlying real,
complex or quaternionic structure in an explicit and conceptually
clear manner and is amenable to implementation in various symbolic
computation systems.  We illustrate our methods and results with a few
examples which display the basic features of the three classes of pin
representations governing the structure of such identities in various
dimensions and signatures.}

\preprint{}

\begin{document}

\maketitle

\pagebreak

\vskip .6in

\section{Introduction}
\label{sec:intro}

Computations involving Fierz identities in curved backgrounds for
various dimensions and signatures are a cumbersome ingredient of
supergravity and string theories and their applications. As any
student of the subject knows all too well, the very construction of
such theories relies in crucial ways on such identities, whose
expert manipulation is often essential for answering various questions.

So far, little progress appears to have been made in giving a
conceptually unified and computationally efficient treatment of Fierz
identities in various dimensions and signatures, though partial
steps in this direction were taken from various perspectives. In the
present paper, we initiate such a unified treatment by using concepts 
and techniques borrowed from a certain
approach to spinors known as ``geometric algebra''. Employing such
methods, we give a systematic and unified treatment of Fierz
identities for form-valued pinor bilinears, which can be applied in
curved backgrounds (including flux backgrounds) of any dimension and
signature. We also show how various results which were obtained
previously can be recovered quite efficiently through our approach.

As typical in the geometric algebra approach to pinors (see
\cite{ga1} for our formulation), we start by viewing a bundle $S$ of
pinors over a pseudo-Riemannian manifold $(M,g)$ as a bundle of
modules over the (real) \KA bundle $(\wedge T^\ast M,\diamond)$ of
$(M,g)$, with module structure specified by a morphism $\gamma:(\wedge
T^\ast M,\diamond)\rightarrow (\End(S),\circ)$ of bundles of
algebras. We give a uniform description of the image and kernel of
this morphism using the {\em Schur bundle} $\Sigma_\gamma$ of
$\gamma$, which we define as the commutant sub-bundle of the image of
$\gamma$ inside the bundle of algebras $(\End(S),\circ)$.  When
$\gamma$ is irreducible (which is the case of interest in many
applications), Schur's lemma implies that the Schur bundle is a bundle of
simple associative algebras, thus --- by the Frobenius
theorem on the classification of such algebras --- having fiber
isomorphic with either of the algebras $\R$, $\C$ or $\H$. It turns
out that the image of $\gamma$ equals the commutant sub-bundle
$\End_{\Sigma_\gamma}(S)$ of $\Sigma_\gamma$ inside $(\End(S),\circ)$,
while the kernel of $\gamma$ equals a (possibly vanishing) sub-bundle
$\wedge^{-\gamma} T^\ast M$ of the \KA bundle, whose complement in the
latter is denoted by $\wedge^\gamma T^\ast M$ and whose construction
we explain in the main text.  The bundle $\wedge^\gamma T^\ast M$
plays the role of {\em effective domain} of definition of $\gamma$,
allowing us to define a partial inverse
$\gamma^{-1}:\End_{\Sigma_\gamma}(S)\rightarrow \wedge ^\gamma T^\ast
M$ of $\gamma$ which can be used to transport sections of
$\End_{\Sigma_\gamma}(S)$ to inhomogeneous forms belonging to the
subalgebra $\Omega^\gamma(M)\eqdef \Gamma(M,\wedge ^\gamma T^\ast M)$
of the \KA algebra $(\Omega(M),\diamond)$ of $(M,g)$. This generalizes
the `dequantization' procedure which was used in \cite{ga1} in a
particular case. Using the partial inverse $\gamma^{-1}$ and the
results of \cite{Okubo1, Okubo2, Okubo3, AC1, AC2}, we show how basic Fierz
identities constraining differential forms constructed as bilinears in
sections of $S$ admit a systematic formulation in terms of so-called
{\em Fierz isomorphisms}, thereby providing algebraic constraints on
certain systems of differential forms defined on $(M,g)$. We also show how the 
algebra of constraints on differential forms extracted in this manner can be formulated 
in a concise form which allows for easy analysis of their structural properties. 

The paper is organized as follows. In Section \ref{sec:pin}, we
systematize the basic properties of pin bundles within the geometric
algebra approach, taking the theory of representations of {\em real}
Clifford algebras as a starting point. The discussion is organized
into the normal, almost complex and quaternionic cases, according to
the type of the corresponding Schur algebra.  Section
\ref{sec:pairings} discusses admissible bilinear forms on such bundles
using the language and results of \cite{AC1, AC2}. Section
\ref{sec:fierz} gives our treatment of basic Fierz identities for
form-valued (s)pinor bilinears in the normal, almost complex and
quaternionic case. Section \ref{sec:examples} illustrates our
treatment by considering three examples, one for each of the three cases
mentioned above --- while explaining how the more traditional
treatment of the examples used to illustrate the almost complex
and quaternionic cases can be recovered within our approach. Section
\ref{sec:conclusions} contains our conclusions while Appendix
\ref{sec:appsystematics} summarizes some properties of real (s)pinors in
those dimensions of signatures which are of most direct physical
interest.

\paragraph{Notations.} We work within the smooth differential category, so all
manifolds, vector bundles, maps, morphisms of bundles, differential forms
etc. are taken to be smooth. We further assume that our smooth manifolds $M$
are connected, paracompact and Hausdorff \footnote{As is well-known, this
implies that $M$ is second countable, $\sigma$-compact and admits countable
atlases. Furthermore, it implies that $M$ has finite Lebesgue
covering dimension, equal to its usual dimension as a manifold. In particular,
we have partitions of unity of finite multiplicity subordinate to any open
cover and the smooth version of the Serre-Swan correspondence applies for
finite rank vector bundles over $M$.}. If $V$ is an $\R$-vector bundle over $M$, we let
$\Gamma(M,V)$ denote the space of smooth ($\cC^\infty$) sections of
$V$. We also let $\End(V)=\Hom(V,V)=V\otimes V^\ast $ denote the 
bundle of endomorphisms of $V$, where
$V^\ast=\Hom(V,\cO_\R)$ is the dual vector bundle to $V$ while
$\cO_\R$ denotes the trivial $\R$-line bundle on $M$. The unital ring
of smooth $\R$-valued functions defined on $M$ is denoted by
$\cinf=\Gamma(M,\cO_\R)$. The tensor product of $\R$-vector bundles is denoted 
by $\otimes$, while the tensor product of $\cinf$-modules is denoted by
$\otimes_{\cinf}$; hence $\Gamma(M,V_1\otimes
V_2)=\Gamma(M,V_1)\otimes_{\cinf} \Gamma(M,V_2)$.  The space of $\R$-valued smooth inhomogeneous
globally-defined differential forms on $M$ is denoted by
$\Omega(M)\eqdef \Gamma(M,\wedge T^\ast M)$ and is a $\Z$-graded
module over the commutative ring $\cinf$.  The fixed rank
components of this graded module are denoted by
$\Omega^k (M)=\Gamma(M,\wedge^k T^\ast M)$ ($k=0\ldots d$, where
$d$ is the dimension of $M$).

The kernel and image of any $\R$-linear map
$T:\Gamma(M,V_1)\rightarrow \Gamma(M,V_2)$ will be
denoted by $\cK(T)$ and $\cI(T)$; these are $\R$-linear subspaces of
$\Gamma(M,V_1)$ and $\Gamma(M,V_2)$, respectively.  In the particular
case when $T$ is $\cinf$-linear (i.e. when it is a morphism of
$\cinf$-modules), the subspaces $\cK(T)$ and $\cI(T)$ are 
$\cinf$-submodules of $\Gamma(M,V_1)$ and $\Gamma(M,V_2)$,
respectively --- even in those cases when $T$ is not induced by any
bundle morphism from $V_1$ to $V_2$. We always denote a morphism
$f:V_1\rightarrow V_2$ of $\R$-vector bundles and the 
$\cinf$-linear map $\Gamma(M,V_1)\rightarrow \Gamma(M,V_2)$ induced by
it between the modules of sections through the same
symbol. Because of this convention, we clarify that the notations
$\cK(f)\subset \Gamma(M,V_1)$ and $\cI(f)\subset \Gamma(M,V_2)$ denote
the kernel and the image of the corresponding map on sections
$\Gamma(M,V_1)\stackrel{f}{\rightarrow} \Gamma(M,V_2)$, which in this
case are $\cinf$-submodules of $\Gamma(M,V_1)$ and $\Gamma(M,V_2)$,
respectively.  In general, there does {\em not} exist any sub-bundle
$\ker f$ of $V_1$ such that $\cK(f)=\Gamma(M,\ker f)$ nor any
sub-bundle $\im f$ of $V_2$ such that $\cI(f)=\Gamma(M,\im f)$ ---
though there exist sheaves $\ker f$ and $\im f$ with the corresponding
properties.

Given a pseudo-Riemannian metric $g$ on $M$ of signature $(p,q)$,
we let $(e_a)_{a=1\ldots d}$ (where $d=\dim M$) denote a local frame
of $T M$, defined on some open subset $U$ of $M$. We let $(e^a)_{a=1\ldots d}$ be the
dual local coframe ($=$ local frame of $T^\ast M$), which satisfies
$e^a(e_b)=\delta^a_b$ and ${\hat g}(e^a,e^b)=g^{ab}$, where ${\hat g}$ is the
metric induced on the cotangent bundle and $(g^{ab})$
is the inverse of the matrix $(g_{ab})$. The contragradient frame
$(e^a)^{\#}$ and contragradient coframe $(e_a)_{\#}$ are given by:
\be
(e^a)^{\#}=g^{a b}e_b~~,~~(e_a)_{\#}=g_{ab}e^b~~,
\ee
where the ${\#}$ subscript and superscript denote the (mutually
inverse) musical isomorphisms between $T M$ and $T^\ast M$ given
respectively by lowering and raising indices with the metric $g$.  We
set $e^{a_1\ldots a_k}\eqdef e^{a_1}\wedge \ldots \wedge e^{a_k}$ and
$e_{a_1\ldots a_k}\eqdef e_{a_1}\wedge \ldots \wedge e_{a_k}$ for any
$k=0\ldots d$. A general $\R$-valued inhomogeneous form $\omega\in
\Omega (M)$ expands as:
\ben
\label{FormExpansion}
\omega=\sum_{k=0}^d\omega^{(k)}=_{_U}\sum_{k=0}^{d}\frac{1}{k!}\omega^{(k)}_{a_1\ldots
a_k} e^{a_1\ldots a_k}~~,
\een
where the symbol $=_{_U}$ means that the equality holds only after
restriction of $\omega$ to $U$ and where we used the expansion:
\ben
\label{HomFormExpansion}
\omega^{(k)}=_{_U}\frac{1}{k!}\omega^{(k)}_{a_1\ldots a_k} e^{a_1\ldots a_k}~~.
\een
The locally-defined smooth functions $\omega^{(k)}_{a_1\ldots a_k}\in
\cC^\infty(U,\R)$ (the `strict coefficient functions' of $\omega$) are
completely antisymmetric in $a_1\ldots a_k$. Given a pinor bundle on
$M$ with underlying fiberwise representation $\gamma$ of the Clifford
bundle of $T^\ast M$, the corresponding gamma `matrices' in the
coframe $e^a$ are denoted by $\gamma^a\eqdef \gamma(e^a)$, while the
gamma matrices in the contragradient coframe $(e_a)_{\#}$ are denoted
by $\gamma_a\eqdef \gamma((e_a)_{\#})=g_{ab}\gamma^b$. We will
occasionally assume that the frame $(e_a)$ is {\em pseudo-orthonormal}
in the sense that $e_a$ satisfy:
\be
g(e_a,e_b)~(=g_{ab})~=\eta_{ab}~~,
\ee
where $(\eta_{ab})$ is a diagonal matrix with $p$ diagonal
entries equal to $+1$ and $q$ diagonal entries equal to $-1$.

\section{Real (s)pin bundles over a pseudo-Riemannian manifold}
\label{sec:pin} 

Let $(M,g)$ be an {\em oriented} pseudo-Riemannian manifold of dimension $d=p+q$,
where $p$ and $q$ are the numbers of positive and negative eigenvalues
of the metric tensor $g$. Let $\nu$ be the (real) volume form of $(M,g)$. 

\subsection{Basics}
\label{sec:pinbasics}
We start by recalling the basics of our  approach to ``geometric algebra'' and
spin geometry, which is based on the theory of \KA bundles. We refer the reader to
\cite{ga1} for a detailed discussion of this approach and for the
derivation of some results which are used in the present paper. 

\paragraph{The \KA algebra and \KA bundle of $(M,g)$.}

Recall that the \KA bundle of $(M,g)$ is a bundle of $\Z_2$-graded
associative and unital $\R$-algebras $(\wedge T^\ast M,\diamond)$
whose underlying vector bundle coincides with the exterior bundle of
$M$ (endowed with its natural $\Z_2$-grading induced by rank, namely
$\wedge T^\ast M=\wedge^\ev T^\ast M\oplus \wedge ^\odd T^\ast M$) and
whose fiberwise $\R$-bilinear, associative and unital multiplication
$\diamond$ is the so-called {\em geometric product} of $(M,g)$ (see \cite{ga1}
for a detailed discussion).  The
fibers of the \KA bundle are unital and associative algebras which are
isomorphic with the real Clifford algebra $\Cl(p,q)$\footnote{Our convention
  is that the canonical generators $e_1,\ldots, e_{p+q}$ of $\Cl(p,q)$ satisfy
  $e_i^2=+1$ for $i=1,\ldots, p$ and $e_j^2=-1$ for $j=p+1,\ldots, p+q$. This is
  the same convention used in \cite{ga1}, to which we refer the reader for
  further details.}, which we view as
a $\Z_2$-graded algebra in the usual manner. Note that $(\wedge^\ev
T^\ast M,\diamond)$ is a bundle of unital subalgebras of the \KA
bundle, which we shall call the {\em even \KA bundle of $(M,g)$}.

Let $\pi$ be the {\em main automorphism} (a.k.a. the {\em signature}, or {\em grading} automorphism), 
i.e. that involutive automorphism of the \KA bundle which is
uniquely determined by the property that it acts as minus the identity
on all one-forms:
\ben
\label{piDef}
\pi(\omega)\eqdef \sum_{k=0}^{d}(-1)^k\omega^{(k)} ~~,
~~\forall \omega=\sum_{k=0}^d\omega^{(k)}\in \Omega(M)~~,~~{\rm where}~~\omega^{(k)}\in \Omega^k(M)~~
\een
and $\tau$ be the {\em main anti-automorphism} (a.k.a. {\em reversion}), i.e. the involutive anti-automorphism of the \KA bundle 
given by:
\ben
\label{taudef}
\tau(\omega)\eqdef (-1)^{\frac{k(k-1)}{2}}\omega~~,~~\forall \omega\in \Omega^k (M)~~.
\een
It is the unique anti-automorphism of $(\wedge T^\ast M,\diamond)$
which acts trivially on all one-forms. The fact that the exterior product is recovered from
the geometric product in the limit of infinite metric (through a trivial direct
computation\footnote{Namely, take $1\leq a_1<\ldots <a_k \leq d$. Then, on
 a sufficiently small open subset $U$ around any point, we have
  $e^{a_1}\diamond \ldots \diamond e^{a_k}= e^{a_1}\wedge \ldots \wedge
  e^{a_k}$ since $e^a$ is an orthonormal coframe. Since $\pi$ is an
automorphism of the \KA bundle, we have $\pi(e^{a_1}\diamond \ldots \diamond
e^{a_k})=\pi(e^{a_1})\diamond \ldots \diamond
\pi(e^{a_k})$. Using the fact that $\pi(e^a)=-e^a$ and the previous
observation, this gives $\pi(e^{a_1}\wedge \ldots \wedge
  e^{a_k})=\pi(e^{a_1}\diamond \ldots \diamond
e^{a_k})=(-1)^k e^{a_1}\diamond \ldots \diamond
e^{a_k}= (-1)^k  e^{a_1}\wedge \ldots \wedge
  e^{a_k}=\pi(e^{a_1})\wedge \ldots \wedge
  \pi(e^{a_k})$. This implies that $\pi$ acts as an automorphism of the
  restricted exterior algebra $(\Omega(U), \wedge)$. Using a partition of
  unity, we find that $\pi$ is an automorphism of the full exterior algebra
  $(\Omega(M),\wedge)$. A similar argument shows that $\tau$ is an anti-automorphism
of the exterior algebra.}) implies that $\tau$
and $\pi$ are also (anti-)automorphisms of the exterior bundle $(\wedge T^\ast
M,\wedge)$ --- see Sections 3.2. and 3.3. of \cite{ga1}. We have the relations:
\be\pi \circ \tau=\tau \circ \pi~~,~~\pi\circ \pi=\tau\circ\tau=\id_{\Omega (M)}~~.
\ee
The (real) volume form $\nu=\vol_M\in \Omega^d(M)$ of $(M,g)$ satisfies the following properties (see Table \ref{table:AlgClassif}):
\ben
\label{NuSquared}
\boxed{\nu\diamond
\nu=(-1)^{q+\left[\frac{d}{2}\right]}1_M=\twopartdef{(-1)^{\frac{p-q}{2}}
  1_M~,~}{d={\rm even}~}{(-1)^{\frac{p-q-1}{2}}1_M~,~}{d=\odd~}}~~
\een
and: 
\ben
\label{nucomm} 
\boxed{\nu \diamond \omega=\pi^{d-1}(\omega)\diamond \nu ~~,~~\forall \omega\in \Omega(M)}~~.
\een
Hence $\nu$ is central in the \KA algebra when $d$ is odd and twisted central (i.e., we have $\nu \diamond \omega=\pi(\omega)\diamond \nu$) in the \KA algebra 
when $d$ is even. 
\begin{table}
\centering
\begin{tabular}{|c|c|c|}
\hline
~~ & $\nu\diamond\nu=+1$ & $\nu\diamond\nu=-1$ \\
\hline\hline
$\nu$ is central & \cellcolor{cyan}$\mathbf{1(\R),5(\H)}$ & $\mathbf{3(\C)}, \mathbin{\textcolor{ForestGreen}{\mathbf{7(\C)}}}$ \\
\hline
$\nu$ is not central & $\mathbin{\textcolor{ForestGreen}{\mathbf{0(\R),4(\H)}}}$ & $\mathbf{2(\R)}, \mathbin{\textcolor{ForestGreen}{\mathbf{6(\H)}}}$ \\
\hline
\end{tabular}
\caption{Properties of the (real) volume form $\nu$ according to the mod 8 reduction of
$p-q$.  At the intersection of each row and column, we indicate the
values of $p-q~({\rm mod}~8)$ for which the volume form $\nu$ has the
corresponding properties. In parentheses, we also indicate the Schur
algebra (see Subsection \ref{sec:pinschur}) $\S$ for that value
of $p-q~({\rm mod}~8)$. The real Clifford algebra $\Cl(p,q)$ is
non-simple iff. $p-q\equiv_8 1,5$, which corresponds to the upper left
cell of the table and is also indicated through the blue shading of that
table cell.  In the non-simple cases, there are two choices for
$\gamma$, which are distinguished by the signature
$\epsilon_\gamma=\pm 1$; these are also the only cases when $\gamma$
fails to be injective. Notice that $\nu$ is central iff. $d$ is
odd. The green color indicates those values of $p-q~({\rm mod}~8)$
for which a spin endomorphism can be defined (see next page).}
\label{table:AlgClassif}
\end{table}

\paragraph{Real pin bundles on $(M,g)$.} A {\em bundle of real pinors} 
on $(M,g)$ is a real vector bundle $S$ over $M$ endowed with a unital morphism of
bundles of algebras $\gamma:(\wedge T^\ast M,\diamond)\rightarrow
(\End(S),\circ)$ from the \KA bundle of $(M,g)$ to the bundle of
endomorphisms of $S$, i.e. a bundle of modules over the \KA bundle of
$(M,g)$. The map induced on global sections (which we denote by the
same letter):
\be
\gamma:(\Omega(M),\diamond)\rightarrow (\Gamma(M, \End(S)),\circ)
\ee
is a unital morphism of $\cinf$-algebras from the \KA algebra of
$(M,g)$ to the algebra of globally-defined endomorphisms of $S$.  For
each point $x\in M$, the fiber $\gamma_x$ is a representation of the
Clifford algebra $(\wedge T^\ast_x M,\diamond_x)\approx \Cl(p,q)$ in
the $\R$-vector space $S_x$ (the fiber of $S$ at $x$). We say that $S$
is a real {\em pin} bundle if this representation is irreducible for
each $x\in M$, i.e. when the fibers of $S$ are simple
modules. Similarly, a {\em bundle of real spinors} of $(M,g)$ is a
bundle of modules over the {\em even} \KA bundle $(\wedge^\ev T^\ast M,\diamond)$ of $(M,g)$; it is
called a {\em spin bundle} when its fibers are simple modules. Notice
that any bundle of pinors is a bundle of spinors in a natural
way\footnote{Indeed, the restriction $\gamma_\ev$ of $\gamma$ to the
sub-bundle $\wedge^\ev T^\ast M\subset \wedge T^\ast M$ makes any
pinor bundle $(S,\gamma)$ into a spinor bundle
$(S,\gamma_\ev)$.}. From now on, we let $S$ be a pin bundle of
$(M,g)$, so we assume that $\gamma$ is fiberwise irreducible.

\paragraph{Spin projectors and spin bundles.} 
Giving a direct sum bundle decomposition $S=S_+\oplus S_-$ amounts to
giving a {\em product structure} on $S$, i.e. a bundle endomorphism
$\cR\in \Gamma(M,\End(S))$ such that $\cR\circ \cR=\id_S$. Indeed, $S_\pm$
determine $\cR$ as that product structure whose eigenbundles associated
with the eigenvalues $+1$ and $-1$ of $\cR$ equal $S_+$ and $S_-$, while a
product structure $\cR$ determines $S_\pm$ as the sub-bundles associated
with its two eigenvalues.  We say that $\cR$ is non-trivial if $S_+$ and $S_-$
are both non-zero, i.e. if $\cR$ differs from $+\id_S$ as well as from
$-\id_S$.  It is easy to see that the restriction:
\ben
\label{gammaev}
\gamma_\ev\eqdef \gamma|_{\wedge^\ev T^\ast M}: \wedge^\ev T^\ast M\rightarrow \End(S)
\een
is reducible on the fibers as a morphism of bundles of algebras
iff. there exists a nontrivial product structure on $S$ which lies in
the commutant of $\gamma(\Omega^\ev(M))$, i.e. a globally-defined
endomorphism $\cR\in \Gamma(M,\End(S))\setminus \{-\id_S,\id_S\}$ which
satisfies:
\ben
\label{Rprops}
\cR^2=\id_S~~{\rm and}~~[\cR,\gamma(\omega)]_{-,\circ}=0~~,~~\forall \omega\in \Omega^\ev(M)~~. 
\een
Such an endomorphism (when defined) is called a {\em spin endomorphism}. 
When $\gamma_\ev$ is fiberwise reducible,  we define the {\em  spin projectors} determined by $\cR$ to be the globally-defined endomorphisms: 
\be
\cP^\cR_\pm\eqdef \frac{1}{2}(\id_S \pm \cR)~~,
\ee
which are complementary idempotents in $\Gamma(M,\End(S))$. Then the eigen sub-bundles:
\be
S^\pm\eqdef \cP^\cR_\pm (S)\subset S~~
\ee
corresponding to the eigenvalues $+1$ and $-1$ of $\cR$ are complementary in $S$:
\be
S=S^+\oplus S^-~~
\ee
and we have: 
\be
\gamma(\omega)(S^\pm)\subset S^\pm~~,~~\forall \omega\in \Omega^\ev(M)~~.
\ee
This gives a nontrivial direct sum decomposition $\gamma_\ev=\gamma^+\oplus \gamma^-$ of $\gamma_\ev$ into morphisms of bundles of algebras: 
\be
\gamma^\pm\eqdef \gamma|_{\wedge^\ev T^\ast M}^{\End(S^\pm)}:(\wedge^\ev T^\ast M,\diamond) \rightarrow (\End(S^\pm),\circ)~~.
\ee
Of course, the vector bundles $S^\pm$ are spin bundles with underlying
morphisms given by $\gamma^\pm$. Such a nontrivial decomposition of $\gamma$ (and
hence a spin endomorphism $\cR$) exists iff. $p-q\equiv_8 0,4,6,7$. In the
Physics literature, the sections of $S^\pm$ are addressed by historically-motivated names. This terminology is summarized below:
\begin{itemize}
\item When $p-q\equiv_8 0$, the sections of $S^\pm$ are called {\em Majorana-Weyl spinors}  (of positive and negative chirality).
\item When $p-q\equiv_8 4$, the sections of $S^\pm$ are called {\em symplectic Majorana-Weyl spinors} (of positive and negative chirality).
\item When $p-q\equiv_8 6$, the sections of $S^+$ are called {\em symplectic Majorana spinors}. 
\item When $p-q\equiv_8 7$, the sections of $S^+$ are called {\em Majorana spinors}.
\end{itemize}

\paragraph{Local expressions.} 

Let $e_m$ be an oriented local pseudo-orthonormal frame of $(M,g)$. We set: 
\ben
\label{gammadef}
\gamma(e^m)\eqdef \gamma^m~,~~\gamma_m=\eta_{mn}\gamma^n
\een
For $A=(m_1,...,m_k)$ with $1\leq m_1<...<m_k\leq d$, we let $|A|\eqdef k$ denote the {\em length} of $A$ and: 
\beqan
\label{eAgammaAdef}
e^A\eqdef e^{m_1}\wedge...\wedge e^{m_k}~~,~~\gamma^A\eqdef  \gamma^{m_1}\circ\ldots \circ \gamma^{m_k}~~,~~
\gamma_A\eqdef \gamma_{m_1}\circ \ldots \circ \gamma_{m_k}~~.
\eeqan
Since $\gamma^{-1}_m=\gamma^m$ and $[\gamma_m,\gamma_n]_{+,\circ}=2\eta_{m n}$, we have $\gamma^{-1}_{m_k}\circ ...\circ \gamma^{-1}_{m_1}=\gamma^{m_k}\circ ...\circ \gamma^{m_1}$, which 
gives: 
\ben
\label{gammaAinv}
\boxed{\gamma^{-1}_A=(-1)^{\frac{|A|(|A|-1)}{2}}\gamma^A}~~.
\een
Also note the relation: 
\ben
\label{gamma_top}
\boxed{\gamma(\nu)=\gamma^{(d+1)}\eqdef \gamma^1\circ \ldots \circ \gamma^d}~~.
\een

\subsection{The Schur bundle and algebra}
\label{sec:pinschur}

As before, let $S$ be a real pin bundle with underlying morphism $\gamma$. We let: 
\ben
\label{N}
N\eqdef \rk_\R S~~
\een
denote the rank of $S$. 

\paragraph{Definition.} Let $x$ be any point in $M$. The {\em Schur
algebra} of $\gamma_x$ is the unital subalgebra $\Sigma_{\gamma,x}$
defined as the commutant of the image $\gamma_x(\wedge T^\ast_x M )$
inside the algebra $(\End(S_x),\circ_x)$:
\be
\Sigma_{\gamma, x}\eqdef \{T_x\in \End (S_x)~|~ [T_x,\gamma_x(\omega_x)]_{-,\circ}=0~,~\forall \omega_x\in \wedge T^\ast_x M \}~~.
\ee
It is easy to see that the subset: 
\be
\Sigma_\gamma=\{(x,T_x)~|~x\in M~,~ T_x\in \Sigma_{\gamma,x}\}=\sqcup_{x\in M}\Sigma_{\gamma,x}
\ee
is a sub-bundle of unital algebras of the bundle of algebras
$(\End(S),\circ)$, which we shall call the {\em Schur bundle} of
$\gamma$.  In particular, the isomorphism type (as a unital
associative algebra) of the fiber $(\Sigma_{\gamma, x},\circ_x)$ is
independent of $x$ and is denoted by\footnote{Since $\gamma$ is fiberwise irreducible, it turns out that $\S$ depends 
only on $p-q~{\rm mod}~8$.} $\S$, being called the
{\em Schur algebra} of $\gamma$. Notice that the space
$\Gamma(M,\Sigma_\gamma)$ of globally-defined smooth sections
of the Schur bundle is a unital subalgebra of the $\cinf$-algebra
$(\Gamma(M,\End(S)),\circ)$, which coincides with the commutant of 
$\gamma(\Omega(M))$ inside $\Gamma(M,\End(S))$:
\be
\Gamma(M,\Sigma_\gamma)=\{T\in \Gamma(M,\End(S))~|~[T,\gamma(\omega)]_{-,\circ}=0~,~\forall \omega\in \Omega(M)\}~~.
\ee

Since $\gamma$ is irreducible on the fibers, Schur's lemma implies
that $\S$ is a division algebra over $\R$ and hence (by the Frobenius
theorem on classification of such algebras) the abstract fiber of $\Sigma_\gamma$ 
is isomorphic with $\R, \C$ or $\H$. Notice that $S$ becomes a bundle of left $\Sigma_\gamma$-modules if
one defines left fiberwise multiplication with elements of
$\Sigma_\gamma$ as evaluation of the corresponding $\R$-linear
operator. On global sections, this gives:
\be
C \xi \eqdef C(\xi)\in \Gamma(M,S)~~,~~\forall \xi \in \Gamma(M,S)~~,~~\forall C\in \Gamma(M,\Sigma_\gamma)~~.
\ee
As a consequence, the dual vector bundle $S^\ast=\Hom(S,\cO_\R)$ becomes a bundle of right $\Sigma_\gamma$-modules with external multiplication given by postcomposition, which on global sections gives: 
\be
\eta C\eqdef \eta\circ C\in \Gamma(M,S^\ast)~~,~~\forall \eta\in \Gamma(M,S^\ast)~~,~~\forall C\in \Gamma(M,\Sigma_\gamma)~~.
\ee
Finally, $\End(S)\approx S\otimes S^\ast$ becomes a bundle of
$\Sigma_\gamma$-bimodules if one uses the module structures on $S$ and
$S^\ast$, i.e. if one defines left and right multiplication with
elements of $\Sigma_\gamma$ through composition from the left and
right with the corresponding $\R$-linear operators. On global
sections, this gives:
\be
C_1 T C_2\eqdef C_1\circ T\circ C_2~~,~~\forall T\in \Gamma(M,\End(S))~~,~~\forall C_1,C_2\in \Gamma(M,\Sigma_\gamma)~~.
\ee
The fibers of $S$ are in fact {\em free} as left modules over the fibers of $\Sigma_\gamma$ (since the latter is a field or a skew-field), so
we have an isomorphism of $\R$-vector bundles:
\be
S\approx \Sigma_\gamma\otimes S_0~~,
\ee 
where $S_0$ is an $\R$-vector bundle over $M$ whose rank we denote through: 
\ben
\label{Delta}
\Delta \eqdef \rk_\R S_0=\rk_{\Sigma_\gamma} S~~
\een
and call the {\em Schur rank} of $S$. We have the relation: 
\be
\rk_\R S=\rk_\R \Sigma_\gamma~\rk_{\Sigma_\gamma}S~\Longleftrightarrow~ N=\Delta \dim_\R \S~~.
\ee
The bundle of bimodule endomorphisms of $S$ can be identified with the sub-bundle
$\End_{\Sigma_\gamma}(S)$ of those endomorphisms of $S$ which commute
with all elements of $\Sigma_\gamma$. On global sections, this gives:
\be
\Gamma(M,\End_{\Sigma_\gamma}(S))=\{T\in \Gamma(M,\End(S))~|~ [T,M]_{-,\circ}=0~,~\forall M\in \Gamma(M,\Sigma_\gamma)\}\subset \Gamma(M,\End(S))~~,
\ee
which is a unital subalgebra of the $\cinf$-algebra $(\Gamma(M,\End(S)),\circ)$. Notice the bundle isomorphism:
\be
\End_{\Sigma_\gamma}(S)\approx \Sigma_\gamma\otimes \End (S_0)~~.
\ee
In the following (and especially in Sections \ref{sec:fierz} and
\ref{sec:examples}) we will sometimes denote $\End_{\Sigma_\gamma}(S)$
through $\End_{\S}(S)$; this is justified by the fact that left
multiplication with elements of $\Sigma_\gamma$ induces an $\S$-module
structure on each of the fibers of $S$.

\subsection{The image and kernel of $\gamma$}
\label{sec:pinkerimage}

\paragraph{Twisted (anti-)selfdual forms.} When $p-q\equiv_8 0,1, 4, 5$,
we have $\nu\diamond \nu=+1$ and we can consider the sub-bundles
$\wedge^\pm T^\ast M\subset \wedge T^\ast M$, whose spaces of smooth
global sections:
\be
\Omega^\pm(M)\eqdef \{\omega\in \Omega(M)~|~\omega\diamond \nu=\pm \omega\}\subset \Omega(M)
\ee
are the $\cinf$-modules of {\em twisted selfdual} and {\em twisted
anti-selfdual} inhomogeneous differential forms of $(M,g)$,
respectively (see \cite{ga1} for details).  We have the direct sum
decompositions:
\be
\wedge T^\ast M=\wedge^+ T^\ast M\oplus \wedge^- T^\ast M~~,~~\Omega(M)=\Omega^+(M)\oplus \Omega^-(M)~~,
\ee
which corresponds to the complementary projectors: 
\be
P_\pm\eqdef \frac{1}{2}(1\pm {\tilde \ast})~~,~~\Omega^\pm(M)=P_\pm(\Omega(M))~~,
\ee
where ${\tilde \ast}$ is the twisted Hodge operator of \cite{ga1}:
\ben
\label{TwistedHodge}
{\tilde \ast}\omega\eqdef \omega\diamond \nu~~,~~\forall \omega\in \Omega(M)~~.
\een
The latter is related to the ordinary Hodge operator of $(M,g)$ though \cite{ga1}:
\ben
\label{HodgeRel}
{\tilde \ast}=\ast \circ \tau~~.
\een

\paragraph{Image, kernel and signature of $\gamma$.} Well-known facts from the representation theory of Clifford algebras imply the following:

\paragraph{Proposition.}  

\begin{enumerate}
\item The image of the bundle morphism $\gamma$ coincides with the sub-bundle $\End_{\Sigma_\gamma}(S)$. This gives: 
\be
\gamma(\Omega(M))=\Gamma(M,\End_{\Sigma_\gamma}(S))~~.
\ee
\item $\gamma$ is fiberwise injective iff. $\Cl(p,q)$ is simple as an associative
$\R$-algebra, i.e. iff. $p-q\not \equiv_8 1,5$ (the so-called {\em simple case}). 
\item When $\gamma$ fails to be fiberwise injective (i.e. when $p-q\equiv_8
1,5$, the so-called {\em non-simple case}), we have:
\be
\gamma(\nu)=\epsilon_\gamma \id_S~~,
\ee
where the sign factor $\epsilon_\gamma\in \{-1,1\}$ is called the {\em
signature} of $\gamma$. The two choices for this sign factor lead to
two inequivalent choices for $\gamma$.  In this case,
$\wedge^{\epsilon_\gamma} T^\ast M$ is a sub-bundle of algebras of the
\KA bundle of $(M,g)$ (see \cite{ga1}). Moreover, the kernel of the
bundle morphism $\gamma$ equals $\wedge^{-\epsilon_\gamma} T^\ast M$,
which implies the following relation for the $\cinf$-modules of global sections:
\be
\cK(\gamma)=\Omega^{-\epsilon_\gamma}(M)~~.
\ee
Furthermore, the restriction of $\gamma$ to its so-called {\em effective
domain} $\wedge^{\epsilon_\gamma} T^\ast M$ gives an isomorphism of
bundles of algebras between $(\wedge^{\epsilon_\gamma} T^\ast
M,\diamond)$ and $(\End_{\Sigma_\gamma}(S),\circ)$.
\end{enumerate}
The fiberwise injectivity and surjectivity properties of $\gamma$ are summarized in Table \ref{table:InjSurj}. 
\begin{table}[tt]
\centering
\begin{tabular}{|c|c|c|}
\hline
~ & injective & non-injective \\
\hline\hline
surjective     & $ \mathbin{\textcolor{ForestGreen}{\mathbf{0(\R)}}}, \mathbf{2(\R)}$ & \cellcolor{cyan} $\mathbf{1(\R)}$ \\
\hline
non-surjective & $\mathbf{3(\C)}, \mathbin{\textcolor{ForestGreen}{\mathbf{7(\C), 4(\H), 6(\H)}}}$ & \cellcolor{cyan} $\mathbf{5(\H)}$ \\
\hline
\end{tabular}
\caption{Fiberwise character of real pin representations $\gamma$.  At
the intersection of each row and column, we indicate the values of
$p-q~({\rm mod}~8)$ for which the map induced by $\gamma$ on each
fiber of the \KA algebra has the corresponding properties. In
parentheses, we also indicate the Schur algebra $\S$ of
$\gamma$ for that value of $p-q~({\rm mod}~8)$. Note that $\gamma$ is
fiberwise surjective exactly for the normal case, i.e. when the Schur
algebra is isomorphic with $\R$. Also notice that $\gamma$ fails to be fiberwise injective precisely in the 
non-simple case $p-q\equiv_8 1, 5$, which we indicate through the blue shading of the corresponding table cells.}
\label{table:InjSurj}
\end{table}

\subsection{The effective domain and the partial inverse of $\gamma$}
\label{sec:pininverse}

The following notation allows one to treat fiberwise injective and non-injective cases simultaneously: 
\ben
\label{OmGammaDef}
\Omega^{\gamma}(M)\eqdef \twopartdef{\Omega(M)~~,~~}{~\gamma~~\mathrm{is~fiberwise~injective~(simple~case)}~,}{\Omega^{\epsilon_\gamma}(M)~~,~~}
{~\gamma~~\mathrm{is~not~fiberwise~injective~(non-simple~case)}~}~~
\een
and: 
\ben
\label{OmMinusGammaDef}
\Omega^{-\gamma}(M)\eqdef \twopartdef{0~~,~~}{~\gamma~~\mathrm{is~fiberwise~injective~(simple~case)}~,}
{\Omega^{-\epsilon_\gamma}(M)~~,~~}{~\gamma~~\mathrm{is~not~fiberwise~injective~(non-simple~case)}~.}~~
\een
With this notation, we always have: 
\be
\cK(\gamma)=\Omega^{-\gamma}(M)~~,~~\Omega(M)=\Omega^{\gamma}(M)\oplus \Omega^{-\gamma}(M)~~
\ee
and the restriction of the (map on sections induced by) $\gamma$ to its {\em effective domain}
$\Omega^\gamma(M)$ is injective. Since $\gamma(\Omega^\gamma(M))=\Gamma(M,\End_{\Sigma_\gamma}(S))$, we can define the
{\em partial inverse} $\gamma^{-1}:\End_{\Sigma_\gamma}(S)\rightarrow \wedge^\gamma T^\ast M$ of $\gamma$ to be the (map of bundles which induces the) partial inverse of this restriction: 
\ben
\label{GammaInvDef}
\gamma^{-1}\eqdef \left(\gamma|_{\Omega^\gamma(M)}^{\Gamma(M,\End_{\Sigma_\gamma}(S))}\right)^{-1}:\Gamma(M,\End_{\Sigma_\gamma}(S))~\longrightarrow~ \Omega^\gamma(M)~~.
\een
Notice the relations: 
\be
\gamma\circ \gamma^{-1}=\id_{\End_{\Sigma_\gamma}(S)}~~,~~\gamma^{-1}\circ \gamma=P_\gamma|^{\wedge^\gamma T^\ast M}\eqdef P_{\epsilon_\gamma}|^{\wedge^{\epsilon_\gamma} T^\ast M}~~,
\ee
where in the right hand side we indicate explicitly the appropriate co-restriction.

\paragraph{Local expressions.} Let $e^m$ be a pseudo-orthonormal local coframe of $M$ defined above an open subset $U\subset M$. 
For later reference, we define: 
\ben
\label{emgamma}
\boxed{e^m_\gamma\eqdef \gamma^{-1}(\gamma^m)=P_\gamma(e^m)\in \Omega^\gamma(U)~~,~~\forall m=1\ldots d}~~,
\een
as well as:
\ben
\label{eAgamma}
\boxed{e^A_\gamma\eqdef \gamma^{-1}(\gamma^A)=\gamma^{-1}(\gamma^{m_1})\diamond 
\ldots \diamond \gamma^{-1}(\gamma^{m_k})=e^{m_1}_\gamma\diamond \ldots \diamond e^{m_k}_\gamma\in \Omega^\gamma(U)}
\een
for any increasingly-ordered $k$-uple $A=(m_1,\ldots,m_k)$. 

\subsection{Representation types} 
\label{sec:pinreptypes}

\begin{table}[tt]
\centering
\begin{tabular}{|c|c|c|}
\hline
Type &  $\pqarray$ & $\S$\\
\hline\hline
normal & $\mathbin{\textcolor{ForestGreen}{\mathbf{0}}},\mathbin{\textcolor{cyan}{\mathbf{1}}},\mathbf{2}$ & $\R$\\
\hline
almost complex & $\mathbf{3},\mathbin{\textcolor{ForestGreen}{\mathbf{7}}}$ & $\C$ \\
\hline
quaternionic & $\mathbin{\textcolor{ForestGreen}{\mathbf{4}}},\mathbin{\textcolor{cyan}{\mathbf{5}}},\mathbin{\textcolor{ForestGreen}{\mathbf{6}}}$ & $\H$\\
\hline
\end{tabular}
\caption{Type of the pin bundle of $(M,g)$ according to the mod $8$
reduction of $p-q$. The pin bundle $S$ is called {\em normal}, {\em
almost complex} or {\em quaternionic} depending on whether its Schur
algebra is isomorphic with $\R$, $\C$ or $\H$. The non-simple
sub-cases are indicated in blue, while those cases when a spin
operator can be defined are indicated in green.}
\label{table:types}
\vspace{5mm}
\end{table}
Recall that the Schur algebra $\S$ is isomorphic with $\R$ (the {\em normal case}), $\C$ (the
{\em almost complex case}) or $\H$ (the {\em quaternionic case}),
where the terminology in brackets is due to \cite{Okubo1, Okubo2, Okubo3}.
The results of \cite{Okubo1, Okubo2, Okubo3} imply that the three types of
real pin bundles occur according to the mod $8$ reduction of the
difference $p-q$ as shown in Table \ref{table:types} (see also Table \ref{table:types_extended}).
In this section, we summarize some properties of the three types of real pin bundles. 
\hyphenation{faith-ful-ness}
\begin{table}[tt]
\centering
\begin{tabular}{|c|c|c|c|c|c|c|c|}
\hline
$\S$ & $\begin{array}{c} p-q\\ {\rm mod}~8 \end{array}$ 
& $\begin{array}{c}\wedge T^\ast_x M\\\approx \Cl(p,q)\end{array}$ & $\Delta$ & $N$ & 
$\begin{array}{c} \mathrm{Number~of}\\ \mathrm{choices}\\ \mathrm{for}~\gamma \end{array}$   
& $\gamma_x(\wedge T^\ast_x M)$ & $\begin{array}{c} \mathrm{Fiberwise}\\ \mathrm{injectivity}\\ \mathrm{of}~\gamma\end{array}$\\
\hline\hline
$\R$ & $\mathbin{\textcolor{ForestGreen}{\mathbf{0}}}, \mathbf{2}$ & $\Mat(\Delta,\R)$ & $2^{[\frac{d}{2}]}=2^{\frac{d}{2}}$ & $2^{[\frac{d}{2}]}$ & $1$ & $\Mat(\Delta,\R)$   & injective \\
\hline
$\H$ & $\mathbin{\textcolor{ForestGreen}{\mathbf{4, 6}}}$ & $\Mat(\Delta,\H)$ & $2^{[\frac{d}{2}]-1}=2^{\frac{d}{2}-1}$ & $2^{[\frac{d}{2}]+1}$ & $1$ & $\Mat(\Delta,\H)$   & injective\\
\hline
$\C$ & $\mathbf{3}, \mathbin{\textcolor{ForestGreen}{\mathbf{7}}}$ & $\Mat(\Delta,\C)$ & $2^{[\frac{d}{2}]}=2^{\frac{d-1}{2}}$ &$2^{[\frac{d}{2}]+1}$& 1 & $\Mat(\Delta,\C)$    & injective\\
\hline
$\H$ & $\mathbin{\textcolor{cyan}{\mathbf{5}}}$ & $\Mat(\Delta,\H)^{\oplus 2}$ & $2^{[\frac{d}{2}]-1}=2^{\frac{d-3}{2}}$ &$2^{[\frac{d}{2}]+1}$&$2$ $(\epsilon_\gamma=\pm 1)$&$\Mat(\Delta,\H)$& 
\cellcolor{cyan}non-injective \\
\hline
$\R$ & $\mathbin{\textcolor{cyan}{\mathbf{1}}}$ & $\Mat(\Delta,\R)^{\oplus 2}$ & $2^{[\frac{d}{2}]}=2^{\frac{d-1}{2}}$ & $2^{[\frac{d}{2}]}$& $2$ $(\epsilon_\gamma=\pm 1)$ & $\Mat(\Delta,\R)$ & 
\cellcolor{cyan} non-injective \\
\hline 
\end{tabular}
\caption{Summary of pin bundle types. The non-negative integer
$N\eqdef \rk_\R S$ is the real rank of $S$ while $\Delta\eqdef
\rk_{\Sigma_\gamma}S$ is the Schur rank of $S$. The non-simple cases are indicated by 
the blue shading of the corresponding table cells.}
\label{table:types_extended}
\end{table}

\subsection{The normal case} 
\label{sec:pincasesnormal}

This occurs when $\S\approx \R$, which happens for $p-q\equiv_8
0,1,2$. We have $N=\Delta=2^{\left[\frac{d}{2}\right]}$ and the Schur bundle
$\Sigma_\gamma\approx \cO_\R$ is the trivial real line bundle generated by the
identity section of $\End(S)$. This implies:
\be 
\Gamma(M,\Sigma_\gamma)=\cinf~\id_S=\{f~\id_S~|~f\in \cinf\}\approx \cinf~~,
\ee
where $\approx$ denotes the obvious isomorphism of $\R$-algebras. We have
$\nu\diamond\nu=+1$ for $p-q\equiv_8 0, 1$ and $\nu\diamond\nu=-1$ for
$p-q\equiv_8 2$. Furthermore, $\nu$ is central in the \KA algebra of $(M,g)$ iff.  $p-q\equiv_8 1$. 

\subsubsection{Injectivity and surjectivity} 
The morphism $\gamma$ is always fiberwise surjective. It is fiberwise injective for $p-q\equiv_8 0,2$ (the
{\em normal simple case}) but fails to be fiberwise injective when
$p-q\equiv_8 1$ (the {\em normal non-simple case}). When $p-q\equiv_8 1$, we
have $\gamma(\nu)=\epsilon_\gamma \id_S$, where $\epsilon_\gamma\in \{-1,+1\}$
is the signature of $\gamma$. The two choices of signature correspond to
different fiberwise representations of the real Clifford algebra $\Cl(p,q)$
and of the pin group $\Pin(p,q)\subset \Cl(p,q)$, but induce equivalent
representations of the spin group $\Spin(p,q)\subset \Pin(p,q)$.

\subsubsection{Spin projectors} 
In the normal case, the restriction $\gamma_\ev$ is fiberwise reducible iff.  $p-q\equiv_8 0$, in which
case spin projectors can be constructed from the product structure
$\cR=\gamma(\nu)$, which determines the eigen sub-bundles $S^\pm$ of {\em
  Majorana-Weyl spinors}, on which $\gamma(\nu)$ has eigenvalues $\pm
1$. Sections of $S^+$ or $S^-$ are called Majorana-Weyl spinors (of positive
and negative chiralities, respectively). We have:
\be
\Gamma(M,S^\pm)=\{\xi\in \Gamma(M,S)~|~\gamma(\nu)\xi=\pm \xi\}~~.
\ee
The situation is summarized in Table \ref{table:NormalCase}. 
\begin{table}[H]
\centering
\begin{tabular}{|c|c|c|c|c|c|c|c|}
\hline
$\pqarray$ & $\Cl(p,q)$ & $\begin{array}{c}\gamma\\\mathrm{is~injective}\end{array}$ & $\epsilon_\gamma$ &  $\cR$ (real spinors) & $\nu
\diamond \nu$ & $\begin{array}{c}\nu\\ \mathrm{is~central}\end{array}$ \\
\hline\hline
$ \mathbin{\textcolor{ForestGreen}{\mathbf{0}}}$ & simple & Yes & N/A &
\cellcolor{ForestGreen} $\gamma(\nu)$ (Majorana-Weyl) & $+1$ & No \\
\hline
$ \mathbin{\textcolor{cyan}{\mathbf{1}}}$ & non-simple & No &
\cellcolor{cyan} $\pm 1$ & N/A & $+1$ & Yes\\
\hline
$ \mathbf{2}$ & simple & Yes &  N/A & N/A & $-1$ & No \\
\hline
\end{tabular}
\caption{Summary of subcases of the normal case.}
\label{table:NormalCase}
\end{table} 

\subsection{The almost complex case}
\label{sec:pincasesac}

This case occurs when $\S\approx \C$, which happens for $p-q\equiv_8
3,7$. In this case, $d$ is odd and we have
$N=2\Delta=2^{[\frac{d}{2}]+1}$. We also have $\nu\diamond \nu=-1$ and
$\nu$ is always central in the \KA algebra.

\subsubsection{Complex structures and the endomorphism $D$} 
There exist two complex structures on the bundle $S$ which
lie in the commutant of the image of $\gamma$, i.e. two {\em
globally-defined} endomorphisms $J\in \Gamma(M,\End(S))$ which
satisfy:
\ben
\label{CJconds}
J^2=-\id_S~~\mathrm{and}~~[J, \gamma(\omega)]_{-,\circ}=0~~,~~\forall \omega\in \Omega(M)~~.
\een
The two solutions $J$ of \eqref{CJconds} differ by a sign factor:
\be
J\rightarrow -J~~
\ee
and are given by:
\be
J=\pm \gamma(\nu)~~.
\ee
In this case, the Schur bundle $\Sigma_\gamma$ is the trivial rank two
real vector bundle spanned by $\id_S$ and by any of these two choices
of $J$. In particular, we have:
\be
\Gamma(M,\Sigma_\gamma)=\cinf \id_S\oplus\cinf J=\{f\id_S+gJ~|~f,g\in \cinf\}~~.
\ee
Each of the two choices of $J$ determines an isomorphism of
$\cinf$-algebras from $\cC^\infty(M,\C)$ to $\Gamma(M,\Sigma_\gamma)$:
\be
\varphi_J(f+i g)=f+g J \in \Gamma(M,\Sigma_\gamma)~~,~~\forall f, g\in \cinf~~,
\ee
so $\varphi_J$ and $\varphi_{-J}$ are related through complex
conjugation, which is an $\cinf$-linear involutive automorphism of
$\cC^\infty(M,\C)$:
\be
\varphi_{-J}(u)=\varphi_J(\bar{u})~~,~~\forall u\in \cC^\infty(M,\C)~~.
\ee
A choice for $J$ makes the Schur bundle into a trivial complex line
bundle, the two opposite choices being related through complex
conjugation.  When viewing $S$ as a bundle of $\Sigma_\gamma$-modules, 
opposite choices for $J$ correspond to two choices of complex structure
on $S$, which are again related through complex conjugation \footnote{The two choices for $J$ are exchanged when changing the orientation of $M$, since 
this maps the volume form $\nu$ into its opposite.}. The results of
\cite{Okubo1} imply that there exists a globally-defined
endomorphism $D\in \Gamma(M,\End(S))$ which satisfies:
\beqan
&& D\circ \gamma(\omega)=\gamma(\pi(\omega))\circ D~~,~~\forall \omega\in \Omega(M)~~, \label{Dgamma_acomm}\\
&& D^2=(-1)^{\frac{p-q+1}{4}}\id_S =\twopartdef{-\id_S~,~}{p-q\equiv_8 3~,}{+\id_S~,~}{p-q\equiv_8 7~,}~~\label{Dsquare}\\
&& [J,D]_{+,\circ}=0~~\label{DJacomm}
\eeqan
and which is determined by these properties up to a sign ambiguity $D\rightarrow
-D$. 

\subsubsection{Injectivity and surjectivity}

In this case, $\gamma$ is always fiberwise injective but
non-surjective. The image of $\gamma$ coincides with the sub-bundle
$\End_\C(S)\eqdef \End_{\Sigma_\gamma}(S)$ of $\End(S)$, which in turn
is the bundle of complex-linear endomorphisms of $S$, when the latter
is viewed as a $\C$-vector bundle upon using the complex structure
$J$. This sub-bundle of $\End(S)$ is isomorphic through $\gamma$ (as a
bundle of $\R$-algebras) with the \KA bundle of $(M,g)$. The situation
is summarized in Table \ref{table:ComplexCase}.  
\begin{table}[t]
\centering
\begin{tabular}{|c|c|c|c|c|c|c|c|}
\hline
$\pqarray$ & $\Cl(p,q)$ & $\begin{array}{c}\gamma\\ \mathrm{is~injective}\end{array}$ & $\epsilon_\gamma$ &  $D^2$ & $\cR$ (real spinors) & $\nu
\diamond \nu$ & $\begin{array}{c}\nu\\ \mathrm{is~central}\end{array}$ \\
\hline\hline
$ \mathbf{3}$ & simple & Yes  & N/A & $-\id_S$ &N/A& $-1$ & Yes \\
\hline
$ \mathbin{\textcolor{ForestGreen}{\mathbf{7}}}$ & simple & Yes & N/A & $+\id_S$ &
\cellcolor{ForestGreen} $D$ (Majorana) & $-1$ & Yes\\
\hline
\end{tabular}
\caption{Summary of subcases of the almost complex case. In this case,
$\gamma(\nu)$ defines a complex structure $J$ on $S$ and we have $\im
\gamma=\End_\C(S)$. We also have an endomorphism $D$ of $S$ which
anticommutes with $J$ (thus giving a complex conjugation on $S$, when the
latter is viewed as a complex vector bundle) and satisfies
$[D,\gamma^m]_{+,\circ}=0$. The two subcases $p-q\equiv_8 3$ and $p-q\equiv_8
7$ are distinguished by whether $D^2$ equals $-\id_S$ or $+\id_S$. In both
cases, $\gamma$ can be viewed as an isomorphism of bundles of $\R$-algebras from the \KA
bundle $(\wedge T^\ast M,\diamond)$ to $(\End_\C(S),\circ)$, while its
complexification $\gamma_\C$ gives an isomorphism of bundles of $\C$-algebras
from the complexified \KA bundle $(\wedge T^\ast_\C M,\diamond)$ to
$(\End_\C(S),\circ)$. When $p-q\equiv_8 7$, $D$ is a real structure which can be used to identify $S$ with
the complexification $(S_+)_\C\eqdef S_+\otimes \cO_\C$ of the real bundle $S_+\subset S$ of
{\em Majorana spinors}. When $p-q\equiv_8 3$, $D$ is a second complex
structure on $S$, which anticommutes with the complex structure
$J=\gamma(\nu)$. In that case, the operators $J,D$ and $J\circ D$ define a
global quaternionic structure on $S$ --- which, however, is not compatible
with $\gamma$ since $D$ {\em anti}commutes with $\gamma^m$.}
\label{table:ComplexCase}
\end{table} 

\subsubsection{Spin projectors} The restriction
$\gamma_\ev$ is fiberwise reducible iff. $p-q\equiv_8 7$, so spin projectors
can only be defined when this condition holds, which we assume
to be the case for the remainder of this sub-subsection. When $p-q\equiv_8 7$,
the spin endomorphism in  \eqref{Rprops} is given by $\cR=D$.  Since $D$ and
$J$ anticommute (see \eqref{DJacomm}), $D$ can be viewed as
a real structure (complex conjugation operation) on the complex vector bundle
obtained by endowing $S$ with the complex structure $J$. We then have 
$J(\xi)=i\xi$ for all $\xi \in \Gamma(M,S)$ and $\gamma(\omega)$ is an
endomorphism of $S$ as a {\em complex} vector bundle:
\be
\gamma(\omega)\in \Gamma(M,\End_\C(S))~~,~~\forall \omega\in \Omega(M)~~,
\ee
because $\gamma(\omega)$ commutes with $J$. Since $\gamma$ is fiberwise
injective in this case, we can thus view $\gamma$ as an
{\em isomorphism} of bundles of algebras from the {\em real} \KA bundle  of $(M,g)$ to
the bundle $(\End_\C(S),\circ)$, where the latter is viewed as a 
bundle of $\R$-subalgebras of the bundle of $\R$-algebras $(\End(S),\circ)$:
\be
\gamma:(\wedge T^\ast M,\diamond)\stackrel{\sim}{\rightarrow} (\End_\C(S),\circ)~~.
\ee
When this interpretation is used, we can denote the action of $D$ by an overline, defining the {\em
complex conjugate} of a section of $S$ through:
\be
{\bar \xi}\eqdef D(\xi)~~,~~\forall \xi\in \Gamma(M,S)~~.
\ee
The spin projectors $\cP_\pm=\frac{1}{2}(\id_S \pm D)$ relate to taking the real and imaginary parts:
\be
\xi_\pm\eqdef \cP_\pm(\xi)=\frac{1}{2}(\xi\pm {\bar \xi})~~,~~\forall \xi\in \Gamma(M,S)~~, 
\ee
with:
\be
\Re \xi\eqdef \xi_+~~,~~\Im \xi\eqdef -J(\xi_-)~\Longleftrightarrow~
\xi_+=\Re\xi~~,~~\xi_-= J(\Im \xi)=i\Im \xi~~.
\ee
The $\R$-vector bundles $S_\pm\eqdef \cP_\pm(S)$ give the decomposition
$S=S_+\oplus S_-$, while the fact that $J$ anticommutes with $D$ implies:
\be
J(S_\pm)=S_\mp~\Longrightarrow~ S=S_+\oplus J(S_+)~~.
\ee
Sections of $S_\pm$ can be characterized through:
\be
\Gamma(M,S_\pm)=\{\xi\in \Gamma(M,S)~|~D(\xi)=\pm \xi\Leftrightarrow {\bar \xi}=\pm \xi \} \subset \Gamma(M,S)~~
\ee
and are called {\em real} and {\em purely imaginary},
respectively. In the physics literature, real sections of $S$  (i.e. sections
$\xi$ of $S_+$) are also called {\em Majorana spinors}. Since $S$ can be
identified with the complexification $(S_+)_\C\eqdef S_+\otimes \cO_\C$ of $S_+$, this means that the bundle $S$
of real pinors can be viewed as the underlying $\R$-vector bundle of the
complex vector bundle of complexified Majorana spinors. Since the latter are
usually called complex spinors, it follows that real pinors in this case are simply 
complex spinors for which one has forgotten the complex structure $J$. We also have: 
\be
\Gamma(M,S_+)=\{\xi\in \Gamma(M,S)~|~\Im \xi=0\}~~,~~\Gamma(M,S_-)=\{\xi\in \Gamma(M,S)~|~\Re \xi=0\}~~.
\ee
In fact, the restrictions $J_\pm \eqdef J|_{S_\pm}^{S_\mp}$ give
isomorphisms of $\R$-vector bundles: 
\be
J_\pm:S_\pm \stackrel{\sim}{\longrightarrow} S_\mp~~
\ee
which satisfy $J_\mp \circ J_\pm=\id_{S^\pm}$, so any section $\xi \in \Gamma(M,S)$ decomposes {\em uniquely} as: 
\ben
\label{xiReImDec}
\xi=\xi_++\xi_-=\xi_R +i \xi_I~~{\rm with}~~\xi_\pm\in \Gamma(M,S_\pm)~,~~\xi_R=\Re\xi\in \Gamma(M,S_+)~,~~\xi_I=\Im\xi \in \Gamma(M,S_+)~~.
\een
As mentioned above, this decomposition corresponds to a bundle isomorphism which identifies the
complexification $(S_+)_\C$ with the complex vector bundle $S$, mapping sections of the former to sections of the
latter via:
\be
\Gamma(M,(S_+)_\C)\approx \Gamma(M,S_+)\oplus i \Gamma(M,S_+) \ni \xi_R+i\xi_I
\longrightarrow \xi=\xi_R+J(\xi_I)\in \Gamma(M,S)~,~~\forall
\xi_R,\xi_I\in \Gamma(M,S_+)~.
\ee
Since $D$ anticommutes with $J$, we have: 
\be
{\bar \xi}=\xi_R - J(\xi_I)=\xi_R-i\xi_I ~\Longleftrightarrow~ \Re(\bar{\xi})=\Re(\xi)~~,~~\Im(\bar{\xi})=-\Im(\xi)~~,
\ee
so $D$ coincides with the complex conjugation induced by this presentation of
$S$ as the complexification of $S_+$. We define the {\em complex conjugate} $\bar{T}$ of any endomorphism $T\in \Gamma(M,\End(S))$ through: 
\be
\bar{T}\eqdef D\circ T\circ D\in \Gamma(M,\End(S))~~,~~{\rm so}~~{\bar
  T}(\xi)=\overline{T({\overline \xi})}~~,~~\forall \xi \in \Gamma(M,S)~~,
\ee
thus obtaining an antilinear involutive automorphism $T\rightarrow
\bar{T}$ of $\Gamma(M,\End(S))$. We say that $T$ is {\em real} or {\em imaginary} if
$\bar{T}=T$ or $\bar{T}=-T$, respectively. We have:
\be
T(S_\pm)\subset \twopartdef{S_\pm ~,~}{T=\mathrm{real}~,}{S_\mp~,~}{T=\mathrm{imaginary}~.}~~
\ee
For example, notice that $D$ is real while $J$ is imaginary. Also notice that the product of two endomorphisms is real when both of them are either real or imaginary and 
imaginary when one of them is real and the other is imaginary. Relation \eqref{Dgamma_acomm} takes the form: 
\ben
\label{bar_gamma}
\overline{\gamma(\omega)}=\gamma(\pi(\omega))~~,~~\forall \omega\in \Omega(M)~~,
\een
and hence:
\ben
\label{goreality}
\gamma(\omega)=\twopartdef{\mathrm{real}~,~}{\omega\in \Omega^\ev(M)~,}{\mathrm{imaginary}~,~}{\omega\in \Omega^\odd(M)~.}~~
\een
\paragraph{Local expressions for $p-q\equiv_8 7$.} Let $U\subset M$ be an open subset supporting
both a local pseudo-orthonormal coframe $(e^a)_{a=1\ldots d}$ of $(M,g)$ and a
local frame $(\epsilon_\alpha)_{\alpha=1\ldots \Delta}$ of $S_+$. Then
$(\epsilon_\alpha,J(\epsilon_\alpha))$ is a local frame of $S$ above
$U$. Setting $\gamma^a\eqdef \gamma(e^a)\in \Gamma(U,\End(S))$, relation
\eqref{goreality} shows that $\gamma^a$ are imaginary:
\be
\overline{\gamma^a}=-\gamma^a~\Longleftrightarrow~ D\circ \gamma^a\circ D=-\gamma^a
\ee
while $\gamma^A\eqdef \gamma^{a_1}\circ \ldots \circ \gamma^{a_k}\in
\Gamma(U,\End(S))$ for an ordered index set $A=(a_1\ldots a_k)$ are real or
imaginary according to whether the length $k=|A|$ of $A$ is even or odd:
\be
\gamma^A=\twopartdef{\mathrm{real}~,~}{|A|=\mathrm{even}~,}{\mathrm{imaginary}~,~}{|A|=\odd~.}~~
\ee
In particular, we have: 
\be
\gamma^a(S_\pm)=S_\mp~~,~~\gamma^A(S_\pm)\subset \twopartdef{S_\pm ~,~}{|A|=\mathrm{even}~,}{S_\mp~,~}{|A|=\odd~.}~~
\ee
The matrix ${\hat \Gamma}^a=({\hat \Gamma}^a_{i,j=1\ldots N})$ of $\gamma^a$
with respect to the local frame $(\epsilon_\alpha,J(\epsilon_\alpha))$ of the
$\R$-vector bundle $S$ (a square matrix of size $N\times N$ with entries in
$\cC^\infty(U,\R)$) is given by the expansions:
\beqa
\gamma^a(\epsilon_\alpha)&=&\sum_{\beta=1}^\Delta \gamma^a_{\beta\alpha}\epsilon_\beta
=\sum_{\beta=1}^\Delta \left[\gamma^a_{R, \beta\alpha}\epsilon_\beta+\gamma^a_{I, \beta\alpha}J(\epsilon_\beta)\right]~~,\\
\gamma^a(J(\epsilon_\alpha))&=&J(\gamma^a(\epsilon_\alpha))=\sum_{\beta=1}^\Delta \gamma^a_{\beta\alpha}J(\epsilon_\beta) 
=\sum_{\beta=1}^\Delta \left[\gamma^a_{R, \beta\alpha}J(\epsilon_\beta)-\gamma^a_{I, \beta\alpha} \epsilon_\beta\right]~~,
\eeqa
where we decomposed the complex-valued functions $\gamma^a_{\beta\alpha}\in \cC^\infty(U,\C)$ into their real and imaginary parts: 
\be
\gamma^a_{\beta\alpha}=\gamma^a_{R, \beta\alpha}+i\gamma^a_{I, \beta\alpha}~~,~~{\rm with}~~\gamma^a_{R, \beta\alpha},\gamma^a_{I, \beta\alpha}\in \cC^\infty(U,\R)
\ee
and used the facts that $J$ and $\gamma^a$ commute and that the complex structure on $S$ is defined through $i\xi=J(\xi)$. It is convenient to encode the coefficients of 
the expansions above into the following square matrices of size $\Delta\times \Delta$ with entries in $\cC^\infty(U,\C)$:
\be
{\hat \gamma}^a\eqdef (\gamma^a_{\alpha\beta})_{\alpha,\beta=1\ldots \Delta}~~,
\ee
which are called the {\em complex} gamma matrices defined by the local coframe
$(e^a)_{a=1\ldots d}$ of $M$ with respect to the local frame
$(\epsilon_\alpha)_{\alpha=1\ldots \Delta}$ of $S_+$. 
Since $(\epsilon_\alpha)_{\alpha=1\ldots \Delta}$ is also a local frame of the
{\em complex} vector bundle $(S,J)$, these are just the gamma matrices of
$e^a$ with respect to this local frame of $S$, when the latter is viewed as a
{\em complex} vector bundle.  With the notations above, the 
{\em real} gamma matrices defined by  $(e^a)_{a=1\ldots d}$ with respect to the local frame $(\epsilon_\alpha, J(\epsilon_\alpha))_{\alpha=1\ldots \Delta}$ of $S$ are given by: 
\be
{\hat \Gamma}^a=\left[\begin{array}{cc}{\hat \gamma}^a_R & {\hat \gamma}^a_I\\ -{\hat \gamma}^a_I & {\hat \gamma}^a_R\end{array}\right]~~, 
\ee
where ${\hat \gamma}^a_R, {\hat \gamma}^a_I\in \Mat(\Delta, \cC^\infty(U,\R))$ are the real and imaginary parts of ${\hat \gamma}^a$: 
\be
{\hat \gamma}^a={\hat \gamma}^a_R+i {\hat \gamma}^a_I~~,~~{\hat \gamma}^a_R=({\hat \gamma}^a_{R,\alpha \beta})_{\alpha,\beta=1\ldots \Delta}~~,
~~{\hat \gamma}^a_I=({\hat \gamma}^a_{I,\alpha \beta})_{\alpha,\beta=1\ldots \Delta}~~.
\ee
A local section $\xi\in \Gamma(U,S)$ expands as: 
\ben
\label{xiExpansion}
\xi=\sum_{\alpha=1}^\Delta \xi^\alpha\epsilon_\alpha=\sum_{\alpha=1}^\Delta \left[\xi^\alpha_R\epsilon_\alpha+\xi^\alpha_I J(\epsilon_\alpha)\right]~~,
\een
where: 
\be
\xi^\alpha=\xi^\alpha_R +i\xi^\alpha_I\in \cC^\infty(U,\C)~~,~~{\rm with}~~\xi^\alpha_R, \xi^\alpha_I\in \cC^\infty(U,\R)~~.
\ee
We let ${\hat \xi}$ denote the $\cC^\infty(U,\C)$-valued column matrix
of size $\Delta$ with entries $\xi^\alpha$. Since $\gamma^A$ commute
with $J$ (i.e. are complex-linear), we have
$\gamma^A\xi=\sum_{\alpha=1}^\Delta
(\gamma^A\xi)^\alpha\epsilon_\alpha$, the complex coefficient functions $
(\gamma^A\xi)^\alpha\in \cC^\infty(U,\C)$ being given by the entries of the matrix ${\hat
\gamma}^A{\hat \xi}$, where:
\be
{\hat \gamma}^A=\widehat{\gamma^A}={\hat \gamma}^{a_1}\circ\ldots\circ {\hat \gamma}^{a_k}\in \Mat(\Delta, \cC^\infty(U,\C))~~,~~\forall~\mathrm{ordered}~~A=(a_1\ldots a_k)~~.
\ee
Notice that $\xi$ is real iff. the matrix ${\hat \xi}$ is real in the sense
that its entries are real-valued functions, i.e. ${\hat \xi}\in
\Mat(U,\cC^\infty(U,\R))$. Furthermore, ${\hat \gamma}^A$ belongs to
$\Mat(U,\cC^\infty(U,\R))$ when $|A|$ is even and to
$\Mat(U,\cC^\infty(U,i\R))$ when $|A|$ is odd. In the physics literature, one
often finds expressions written locally in terms of ${\hat \xi}$ and ${\hat
\gamma}^a$; such expressions should be understood in the sense explained
above.

\subsection{The quaternionic case} 
\label{sec:pincasesquat}

This occurs for $\S\approx \H$, which happens for $p-q\equiv_8
4,5,6$. Then $N=4\Delta=2^{[\frac{d}{2}]+1}$ and $\Sigma_\gamma$ is a
bundle of quaternion algebras (the topology of such bundles was
studied in \cite{Cadek}).  We have $\nu \diamond \nu =+1$ when $p-q\equiv_8
4,5$ and $\nu\diamond \nu=-1$ for $p-q\equiv_8 6$. Furthermore, $\nu$ is
central in the \KA algebra iff. $p-q\equiv_8 5$. When $p-q\equiv_8 5$, we have
$\gamma(\nu)=\epsilon_\gamma\id_S$, where $\epsilon_\gamma\in \{-1,1\}$ is the
signature of $\gamma$, which can only be defined in this subcase (known as the
{\em quaternionic non-simple case}). 

\subsubsection{The quaternionic structure of $S$} 

For any sufficiently small open subset $U\subset M$, there exist three local sections $J_j\in
\Gamma(U,\Sigma_\gamma)\subset \Gamma(U,\End(S))$ $(j=1\ldots 3)$ lying in the commutant of the
subset $\gamma(\Omega(U))\subset \Gamma(U,\End(S))$, which satisfy the
algebra of quaternion relations:
\ben
\label{HJconds}
[J_i,\gamma(\omega)]_{-,\circ}=0~~,~~\forall \omega\in \Omega(U)~~,~~ J_i \circ J_j=-\delta_{i j}\id_S+\sum_{k=1}^3 \epsilon_{i j k} J_k~~,~~\forall i,j =1\ldots 3 ~~,
\een
where $\epsilon_{ijk}$ is the totally anti-symmetric Levi-Civita
symbol. In particular, we have
$J_i^2=-\id_S|_U$.  Setting $J_0\eqdef \id_S|_U\in \Gamma(U,\End(S))$,
we have:
\be
\Gamma(U,\Sigma_\gamma)=\oplus_{\alpha=0}^3\cC^\infty(U,\R) J_\alpha=\{\sum_{\alpha=0}^3 f_\alpha J_\alpha~|~f_\alpha\in \cC^\infty(U,\R)\}~~.
\ee
In particular, the restriction $\Sigma_\gamma|_U$ is topologically
trivial as a bundle of algebras. The group $\cC^\infty(U,\O(3,\R))$ of
$\O(3,\R)$-valued functions defined on $U$ acts transitively on the space of solutions $J_i\in \Gamma(U,\End(S))$ to
\eqref{HJconds} through:
\ben
\label{O3Raction}
J_i\rightarrow \sum_{j=1}^3 R_{ij} J_j~~\mathrm{where}~~R=(R_{ij})_{i,j=1\ldots 3}\in \cC^\infty(U,\O(3,\R))~~.
\een
The most general complex structure $J\in  \Gamma(U,\Sigma_\gamma)$ on the restricted bundle
$S|_U$ which lies in the commutant of $\gamma(\Omega(U))$ takes the
form:
\ben
\label{quatJ}
J=\sum_{i=1}^3 f_i J_i~~\mathrm{where}~~f_i\in \cC^\infty(U,\R) ~~\mathrm{with}~~\sum_{i=1}^3 (f_i)^2=1~~.
\een
Equation \eqref{quatJ} says that $J$ is a local section (defined above
$U$) of the {\em twistor bundle $\cU_\gamma$ of $(S,\gamma)$}, which
is the $S^2$-sub-bundle of $\Sigma_\gamma$ consisting of the imaginary 
units of the fibers of $\Sigma_\gamma$:
\ben
\label{U_gamma}
\cU_\gamma\eqdef\{(x,J_x)~|~J_x\in \Sigma_{\gamma,x}~~\&~~ J_x^2=-1\}\subset \Sigma_\gamma~~.
\een
Let $\j_\alpha$ $(\alpha=0\ldots 3$) be the canonical units of $\H$, where
$\j_0=1$ and $\j_i$ $(i=1\ldots 3$) are the canonical imaginary
units. For any choice of solution $\vec{J}=(J_1,J_2,J_3)$ to
\eqref{HJconds} defined above $U$, the morphism of $\R$-vector
bundles:
\be
\varphi_{{\vec J}}: \H\otimes\cO_\R|_U\rightarrow \Sigma_\gamma|_U~~
\ee
which acts on sections through: 
\be
\varphi_{{\vec J}}(\sum_{\alpha=0}^3 f_\alpha \j_\alpha)\eqdef \sum_{\alpha=0}^3 f_\alpha J_\alpha
\ee
is an isomorphism of bundles of $\R$-algebras from $\H\otimes \cO_U$
to the restricted Schur bundle $\Sigma_\gamma|_U$, which provides a
local trivialization of $\Sigma_\gamma$ (as a bundle of $\R$-algebras)
above $U$. Here, $\cO_\R|_U$ stands for the restriction of $\cO_\R$ to $U$.

\subsubsection{Injectivity and surjectivity} 
The bundle morphism $\gamma$ is fiberwise injective iff. $p-q\equiv_8 4,6$. It
is always fiberwise non-surjective, with image equal to the sub-bundle
$\End_\H(S)\eqdef \End_{\Sigma_\gamma}(S)$ of $\End(S)$. This sub-bundle of $\End(S)$ is isomorphic through
$\gamma$ (as a bundle of algebras) with the \KA bundle of $(M,g)$ when
$p-q\equiv_8 4,6$ and with the sub-bundle $\wedge^{\epsilon_\gamma} T^\ast M$ when
$p-q\equiv_8 5$. The situation is summarized in Table \ref{table:QuaternionicCase}. 
\begin{table}[t]
\centering
\begin{tabular}{|c|c|c|c|c|c|c|c|}
\hline
$\pqarray$ & $\Cl(p,q)$ & $\begin{array}{c}\gamma\\ \mathrm{is~injective}\end{array}$ & $\epsilon_\gamma$ &  $\cR$ (real
spinors) & $\nu \diamond \nu$ & $\begin{array}{c}\nu\\ \mathrm{\tiny is~central}\end{array}$ \\
\hline\hline
$ \mathbin{\textcolor{ForestGreen}{\mathbf{4}}}$ & simple & Yes & N/A &
\cellcolor{ForestGreen} $\gamma(\nu)$ (sympl. Majorana-Weyl) & $+1$ & No \\
\hline
$ \mathbin{\textcolor{cyan}{\mathbf{5}}}$ & non-simple & No & \cellcolor{cyan} $\pm 1$ & N/A & $+1$ & Yes\\
\hline
$ \mathbin{\textcolor{ForestGreen}{\mathbf{6}}}$ & simple & Yes &  N/A &
\cellcolor{ForestGreen} $\gamma(\nu)\circ J$ (sympl. Majorana) & $-1$ & No \\
\hline
\end{tabular}
\caption{Summary of subcases of the quaternionic case. $J$ denotes any of the
  complex structures induced on $S$ by the quaternionic structure. }
\label{table:QuaternionicCase}
\end{table} 

\subsubsection{Spin projectors}
In the quaternionic case, the various possibilities for spin projectors are as follows:
\begin{itemize}
\item If $p-q\equiv_8 4$, then the restriction $\gamma_\ev$ is
fiberwise reducible and we can use the spin projectors defined by the product
structure $\cR=\gamma(\nu)$, which lies in the commutant of $\Gamma(M,\Sigma_\gamma)$.
Sections of the sub-bundles $S^\pm=\ker(\gamma(\nu)\mp \id_S)\subset S$  are called {\em symplectic Majorana-Weyl}
spinors of positive and negative chiralities, respectively. They can be viewed
as those sections $\xi$ of $S$ which satisfy the conditions: 
\be
\xi\in \Gamma(M,S_\pm)~\Longrightarrow~ \gamma(\nu)\xi=\pm \xi~~.
\ee
We have $S=S^+\oplus S^-$, so any section $\xi\in \Gamma(M,S)$ decomposes
uniquely as: 
\be
\xi=\xi^++\xi^-~~{\rm with}~~\xi^\pm \in \Gamma(M,S^\pm)~~.
\ee
Since $J_\alpha$ commute with $\gamma(\nu)$, we have $J_\alpha(S^\pm)\subset S^\pm$ and
hence the sub-bundles $S^\pm$ of symplectic Majorana-Weyl spinors inherit from $S$ the
structure of bundles of free modules over the Schur bundle
$\Sigma_\gamma$ --- in particular, each of $S^\pm$ carries a quaternionic structure. The restricted morphism $\gamma_\ev$ can be viewed as an
isomorphism of bundles of algebras from the even \KA algebra of $M$ to the sub-bundle of algebras
$\End_\H(S^+)\oplus \End_\H(S^-)$ of $(\End_\H(S),\circ)$: 
\be
\gamma_\ev:(\wedge^\ev T^\ast M,\diamond)\stackrel{\sim}{\longrightarrow} (\End_\H(S^+),\circ) \oplus (\End_\H(S^-),\circ)~~.
\ee

\item If $p-q\equiv_8 5$ (the quaternionic non-simple case), then the restriction $\gamma_\ev$ is
fiberwise irreducible, so spin projectors cannot be defined. This is also the
only quaternionic non-simple subcase, i.e. the only quaternionic subcase for
which $\gamma$ fails to be fiberwise injective. 
 
\item If $p-q\equiv_8 6$, then the restriction $\gamma_\ev$ is
fiberwise reducible and we can define {\em symplectic Majorana spinors} using the
product structure $\cR^J=\gamma(\nu)\circ J$ induced by any {\em global}
section  $J\in \Gamma(M,\cU_\gamma)$ of the twistor bundle $\cU_\gamma$ of $(S,\gamma)$.  
\end{itemize}

\subsubsection{The biquaternion formalism}
Real pinors in the quaternionic case can be described indirectly by first
complexifying the real vector bundle $S$ and then recovering it from its
complexification by using the appropriate real structure. This approach is
common in the physics literature, though its precise relation with the direct
construction of real pinors (on which we rely) is rarely clarified.  

\paragraph{The complexified pin bundle and its biquaternionic structure.}

Recall that the algebra of biquaternions (or complexified quaternions) is the
$\C$-algebra $\C \otimes _\R \H$, which is isomorphic with the $\C$-algebra
$\Mat(2,\C)\approx \Cl_\C(2)$ through the map:
\be
\label{BiquatMatIsom}
\C\otimes \H \ni w_0\j_0+w_1\j_1+w_2\j_2+w_3\j_3\longrightarrow \left[\begin{array}{cc} w_0+iw_1~ & ~w_2+i w_3~ \\ -w_2+iw_3~ & ~w_0-i w_1 \end{array}\right]\in \Mat(2,\C)~~,
\ee 
where $w_\alpha\in \C$. This well-known isomorphism maps the imaginary quaternion units into $i$ times the Pauli matrices: 
\beqa
\j_1\rightarrow  i \left[\begin{array}{cc} 1 & 0 \\ 0 & -1 \end{array}\right]=i\boldsymbol{\sigma}_3 ~~,~~\j_2\rightarrow  \left[\begin{array}{cc} 0 & 1 \\ -1 & 0 \end{array}\right]=i\boldsymbol{\sigma}_2~~,~~
\j_3\rightarrow  i \left[\begin{array}{cc} 0 & 1 \\ 1 & 0 \end{array}\right]=i\boldsymbol{\sigma}_1~~,
\eeqa
where $i$ is, as usual, the imaginary unit of the field $\C$ of complex numbers. The {\em complexified Schur bundle}: 
\be
(\Sigma_\gamma)_\C\eqdef \Sigma_\gamma\otimes \cO_\C~~
\ee
is a bundle of biquaternion algebras over $M$. We let $S_\C\eqdef S\otimes
\cO_\C$ be the {\em complexified pin bundle}, i.e. the complexification of the $\R$-vector bundle $S$, which is a
bundle of free modules over the complexified Schur bundle and thus carries a
biquaternionic structure. Notice that $S_\C$
is also a bundle of modules over the complexified \KA bundle $\wedge T^\ast_\C
M=\wedge T^\ast M\otimes \cO_\C$ of $(M,g)$, through the morphism of bundles
of algebras given by the complexification $\gamma_\C$ of $\gamma$, which
commutes with the biquaternionic structure of $S_\C$. Using the
$(\Sigma_\gamma)_\C$-module structure, left multiplication by the complex
imaginary unit $i$ gives a globally-defined endomorphism $\fI\in \Gamma(M,\End(S_\C))$ whose
action on sections is given by:
\be
\fI(\xi)\eqdef i \xi~~,~~\forall \xi\in \Gamma(M,S_\C)~~,
\ee 
while the complexifications of $J_k$ (which we denote $\fJ_k$)
define fiberwise $\C$-linear endomorphisms of $S_\C|_U$ which satisfy  the
quaternion relations. In particular, we have: 
\be
[\fJ_k,\fI]_{-,\circ}=0~~
\ee
and $\fI, \fJ_1,\fJ_2,\fJ_3$ form a local frame of the underlying $\R$-vector bundle
of $(\Sigma_\gamma)_\C$ above $U$. In every fiber of $S_\C|_U$, the endomorphisms  $\fI,
\fJ_1,\fJ_2,\fJ_3$ represent the canonical basis $i, \j_1,\j_2,\j_3$ of the
biquaternion algebra, when the latter is viewed as an eight-dimensional
associative algebra over the real numbers. The complexification $S_\C$ comes endowed with a natural real structure --- a
globally-defined endomorphism $\fZ\in \Gamma(M,\End(S_\C))$, which is $\C$-antilinear:
\be
[\fZ,\fI]_{+,\circ}=0~~,
\ee
and satisfies:
\be
\Gamma(M,S)=\{\xi\in \Gamma(M,S_\C)~|~\fZ(\xi)=\xi\}~~,~~~i\Gamma(M,S)=\fI(\Gamma(M,S))=\{\xi\in
\Gamma(M,S_\C)~|~\fZ(\xi)=-\xi\}~.
\ee
We have: 
\ben
\label{fDJkComm}
[\fZ|_U,\fJ_k]_{-,\circ}=0~~,
\een
since the locally-defined endomorphisms $\fJ_k$ are obtained through complexification.  

\paragraph{The product structure induced by $\fJ_1$. }
Since $\fI$ and $\fJ_1$ commute and since both of them square to $-\id_{S_\C}|_U$,
the opposite of their composition: 
\be
\fR\eqdef -\fI\circ \fJ_1=-i\fJ_1~~
\ee
is a {\em $\C$-linear} product structure on $S_\C|_U$ which anticommutes with $\fZ|_U$:
\be
\fR^2=+\id_{S_\C}|_U~~,~~[\fR,\fZ|_U]_{+,\circ}=0~~ 
\ee
and hence the $\C$-linear endomorphisms of $S_\C|_U$ defined through: 
\be
\fP_\pm=\frac{1}{2}(\id_{S_\C}|_U\pm \fR)
\ee
are complementary in $\End_\C(S_\C|_U)$. It follows that the complex sub-bundles $S_\C|_U^\pm$ of $S_\C|_U$ 
consisting of eigenvectors of $\fR$ corresponding to the eigenvalues $\pm 1$ give a direct sum decomposition: 
\be
S_\C|_U=S_\C|_U^+\oplus S_\C|_U^-~~.
\ee
Notice the equalities: 
\be
\fI(S_\C|_U^\pm)= S_\C|_U^\pm~~, ~~\fZ(S_\C|_U^\pm)= S_\C|_U^\mp
\ee 
and the fact that $S_\C|_U^\pm$ are the eigen sub-bundles of $S_\C|_U$ determined by the eigenvalues $\pm i$ of the complexified
endomorphism $\fJ_1$. In particular, we have: 
\be
\Gamma(U,S_\C|_U^\pm)=\{\xi\in \Gamma(U,S_\C|_U)~|~\fJ_1(\xi)=\pm i\xi\}~~.
\ee 
Any $\xi\in \Gamma(U,S_\C)$ decomposes uniquely as: 
\be
\xi=\xi^+ +\xi^-~~{\rm with}~~\xi^\pm \in \Gamma(U,S_\C|_U^\pm)~~
\ee
and we have: 
\be
\fJ_1(\xi)^\pm=\pm i \xi^\pm~~.
\ee
Since $\fJ_2$ anticommutes with $\fJ_1$,  it induces an isomorphism of $\C$-vector bundles:
\be
\fJ_2|_{S_\C|_U^+}^{S_\C|_U^-}: S_\C|_U^+\stackrel{\sim}{\longrightarrow} S_\C|_U^- ~~,
\ee
which can be used to identify $S_\C|_U^-$ with $S_\C|_U^+$. Using this isomorphism,
we find that any $\xi$ also decomposes uniquely as: 
\be
\xi=\xi^1 -\fJ_2(\xi^2)~~,~~{\rm where} ~~ \xi^1\eqdef \xi^+\in \Gamma(U,S_\C|_U^+)~~{\rm
  and}~~\xi^2= \fJ_2(\xi^-)\in \Gamma(U,S_\C|_U^+)~~.
\ee
We can thus describe $\xi$ through the column matrix: 
\ben
\label{ximatrix}
{\hat \xi}\eqdef \left[\begin{array}{c} \xi^1\\ \xi^2\end{array}\right]\in \Mat(2,1;\Gamma(U,S_\C|_U^+))~~
\een
and describe any endomorphism $T\in \Gamma(U,\End(S_\C|_U))$ through the matrix
${\hat T}\in \Mat(2,2; \Gamma(U,\End(S_\C|_U^+))$ defined by: 
\ben
\label{Tmatrix}
\widehat{T\xi}={\hat T}{\hat \xi}~~,~~\forall \xi \in \Gamma(U,S_\C)~~,
\een
where juxtaposition in the right hand side stands for matrix multiplication. 
Since $\fJ_1(\xi^\pm)=\pm i \xi^\pm$ and since $\fJ_2$
is $\C$-linear, we have $\fJ_1(\xi)^1=i\xi^1$ and $\fJ_1(\xi)^2=-i \xi^2$. On the
other hand, an easy computation gives:
\be
\fJ_2(\xi)^1= \xi^2~~,~~\fJ_2(\xi)^2=-\xi^1 ~\Longrightarrow~ \fJ_3(\xi)^1= i\xi^2~~,~~\fJ_3(\xi)^2= i\xi^1~~,
\ee
where the implication displayed follows from the relation $\fJ_3=\fJ_1\circ
\fJ_2$. In terms of description \eqref{Tmatrix}, this amounts to: 
\be
\widehat{\fJ_1}= i\boldsymbol{\sigma}_3= \left[\begin{array}{cc} i & 0 \\ 0 & -i \end{array}\right]~~,~~
\widehat{\fJ_2}=  i\boldsymbol{\sigma}_2=\left[\begin{array}{cc} 0 &1 \\ -1 & 0 \end{array}\right]~~,~~
\widehat{\fJ_3}=  i\boldsymbol{\sigma}_1=\left[\begin{array}{cc} 0 & i \\ i & 0 \end{array}\right]~~
\ee 
and implies:
\be
{\hat \fR}= \boldsymbol{\sigma}_3~\Longrightarrow~ {\hat \fP}_\pm =\frac{1}{2}(1 \pm \boldsymbol{\sigma}_3)~~.
\ee
We have:
\be
\fZ(\xi)^1=-\fZ_0 (\xi^2)~~,~~\fZ(\xi)^2=\fZ_0 (\xi^1)~\Longleftrightarrow~ \widehat{\fZ(\xi)}= -i\sigma_2(\fZ_0\xi)=
\left[\begin{array}{cc} 0 & -\fZ_0  \\  \fZ_0 &
    0  \end{array}\right]\left[\begin{array}{c} \xi^1 \\ \xi^2 \end{array}
\right]~~,~~\forall \xi\in \Gamma(U,S_\C)~~,
\ee
where $\fZ_0=\left[\fJ_2\circ \fZ\right]|_{S_\C|_U^+}^{S_\C|_U^-}=\left[i\fZ\circ
  \fJ_2 \right] |_{S_\C|_U^+}^{S_\C|_U^+} \in \Gamma(U,\End_\R(S_\C|_U^+))$ is an
antilinear endomorphism of $S_\C|_U^+$ which squares to $-\id_{S_\C|_U^+}$.
Hence the $\fZ$-reality condition for sections of $S_\C$ takes the form:
\be
\xi\in \Gamma(U,S)~\Longleftrightarrow~ \fZ(\xi)=\xi~\Longleftrightarrow~
\xi^1=-\fZ_0(\xi^2)~\Longleftrightarrow~ \xi^2=\fZ_0(\xi^1)~~,
\ee
which can be viewed as a {\em `generalized symplectic Majorana condition'}
(see the example in subsection \ref{sec:examplesquaternionic}) on the pair of sections (which are complex pinors) $\xi_1,\xi_2\in \Gamma(U,S_\C|_U^+)$.

\paragraph{The complex pin bundle and complex pinors of spin $1/2$.}
Local sections of $S_\C$ form  a sheaf of left modules over the sheaf of
sections of $(\Sigma_\gamma)_\C$. In particular, the space $\Gamma(U,S_\C)$ is a left module over the non-commutative ring $
\Gamma(U,(\Sigma_\gamma)_\C)$ for any open subset $U$ of $M$ which supports a
local frame of imaginary unit sections of $\Sigma_\gamma$. The discussion above shows that
this module structure is given by: 
\ben
\label{BiquatMult}
\widehat{\f \xi}= \left[\begin{array}{cc} f_0+if_1 ~&~ f_2+i f_3 \\ -f_2+if_3 ~&~
    f_0-i f_1 \end{array}\right] {\hat \xi}~~,~~\forall f\in
\Gamma(U,(\Sigma_\gamma)_\C)~~,~~\forall  \xi\in \Gamma(U,S_\C)~~,
\een
where we used the decomposition:
\be
\f=f_0 \id_{S_\C|_U}+f_1 \fJ_1+f_2 \fJ_2+f_3\fJ_3\in \Gamma(U,(\Sigma_\gamma)_\C)~~,~~{\rm with}~~ f_\alpha\in \cC^\infty(U,\C)~~.
\ee
In equation \eqref{BiquatMult}, any complex-valued smooth function $g=g_R+i g_I\in \cC^\infty(U,\C)$ (with $g_{R,I}\in \cC^\infty(U,\R)$) acts on 
$\xi\in \Gamma(U,S_\C)$ via the complex structure $\fI$:
\be
g\xi =g_R \xi + g_I \fI(\xi)~~,~~\forall \xi\in \Gamma(U,S_\C)~~
\ee 
and we have $g \Gamma(U,S_\C|_U^+)\subset \Gamma(U,S_\C|_U^+)$.  

Since the complexified operators $\gamma(\omega)$ commute with $\fI$ and $\fJ_k$, we have: 
\be
[\gamma(\omega),\fR]_{-,\circ}=0~\Longleftrightarrow~
\gamma(\omega)(\Gamma(U,S_\C^\pm))\subset \Gamma(U,S_\C^\pm)~~{\rm for} ~~\omega\in \Omega(U)
\ee
and $\fJ_2\circ \gamma(\omega)\circ \fJ_2^{-1}=\gamma(\omega) $. Defining
$\gamma_\pm(\omega)\eqdef \gamma(\omega)|_{S_\C|_U^\pm}^{S_\C|_U^\pm}$ and
$\gamma_1(\omega)\eqdef \gamma_+(\omega)$, $\gamma_2(\omega)\eqdef \fJ_2|_{S_\C|_U^-}^{S_\C|_U^+}\circ
\gamma_-(\omega)\circ \left(\fJ_2|_{S_\C|_U^-}^{S_\C|_U^+}\right)^{-1}$, the last
relation implies $\gamma_2(\omega)=\gamma_1(\omega)\eqdef \Gamma(\omega)$. We
thus have a decomposition:
\be
\gamma=\gamma_+\oplus \gamma_-\Longleftrightarrow \gamma=
\Gamma\oplus \left[\left(\fJ_2|_{S_\C|_U^-}^{S_\C|_U^+}\right)^{-1}\circ
\Gamma \circ \fJ_2|_{S_\C|_U^-}^{S_\C|_U^+}\right]~~.
\ee
where: 
\be
\Gamma:(\wedge T^\ast U,\diamond) \rightarrow (\End_\C(S_\C|_U^+),\circ)~~
\ee
is a morphism of bundles of $\R$-algebras. The morphism $\Gamma$ complexifies to a morphism of bundles of $\C$-algebras: 
\be
\Gamma_\C :(\wedge T^\ast_\C U,\diamond) \rightarrow (\End_\C(S_\C|_U^+),\circ)~~
\ee
whose fibers are irreducible representations of the complexified Clifford
algebra $\Cl_\C(p,q)\approx \Cl_\C(d)$. This complexified morphism $\Gamma_\C$
is always fiberwise surjective, being fiberwise injective iff. $d$ is even,
i.e. iff. $p-q\equiv_8 4, 6$.  Therefore, sections of $\Gamma(U,S_\C|_U^+)$ are
locally-defined {\em complex  pinors}\footnote{These are often called complex {\em s}pinors in the
  physics literature, which is justified since any pinor is also a spinor.} of spin $1/2$.
We have:
\be
\widehat{\gamma(\omega)}= \left[\begin{array}{cc} \Gamma(\omega) & 0 \\ 0 & \Gamma(\omega)  \end{array}\right]~~,~~ \forall \omega\in \Omega(U)~~.
\ee
and hence $(S_\C|_U,\gamma)$ is equivalent with the direct sum
$(S_\C|_U^+,\Gamma)\oplus (S_\C|_U^+,\Gamma)$ as a bundle of modules over the \KA
bundle of $(M,g)$. Similarly, $(S_\C|_U,\gamma_\C)$ is equivalent with
$(S_\C|_U^+,\Gamma_\C)\oplus (S_\C|_U^+,\Gamma_\C)$ as a bundle of modules over the
complexified \KA bundle of $(U,g|_U)$. In much of the supergravity literature, one constructs the bundle
$S$ of real pinors by starting from such a direct sum of two copies of the
bundle $S_\C$ of complex pinors and imposing the reality
(`generalized symplectic Majorana') condition $\fZ(\xi)=\xi$. The discussion above shows that this construction can always be applied locally.

\paragraph{Local expressions.} If  $e^a$ is a local pseudo-orthonormal coframe of $(M,g)$ defined above $U$, we set
$\gamma^a\eqdef \gamma_\C(e^a)\in \Gamma(U,\End_\C(S_\C))$ and $\Gamma^a\eqdef
\Gamma(e^a)\in \Gamma(U,\End_\C(S_\C^+))$. We also set
$\gamma^A=\gamma^{a_1}\circ \ldots \circ \gamma^{a_k}\in \Gamma(U,\End_\C(S_\C))$
and $\Gamma^A=\Gamma^{a_1}\circ \ldots \circ \Gamma^{a_k}\in
\Gamma(U,\End_\C(S_\C^+))$ for any ordered index set $A=(a_1,\ldots, a_k)$. The relation above gives:
\be
{\widehat \gamma}^a= \left[\begin{array}{cc} \Gamma^a & 0 \\ 0 &
    \Gamma^a  \end{array}\right]~~,~~
{\widehat \gamma}^A={\hat \gamma}^{a_1}\circ \ldots \circ {\hat \gamma}^{a_k}= \left[\begin{array}{cc} \Gamma^A & 0 \\ 0 & \Gamma^A  \end{array}\right]~~.
\ee

\subsection{Summary of spin projectors}
\label{sec:pinprojsummary}

The spin projectors in the three cases are summarized in Table \ref{table:chiral_projectors}.
\begin{table}[H]
\centering
\begin{tabular}{|c|c|c|c|c|c|c|}
\hline
$\S$ & $\pqarray$ & $\Cl(p,q)$ & $\cR$ & $\begin{array}{c} \mathrm{Terminology}\\\mathrm{for~real~spinors}\end{array}$\\
\hline\hline
$\R$ & $\mathbin{\textcolor{ForestGreen}{\mathbf{0}}}$ & simple & $\gamma(\nu)$ & Majorana-Weyl\\
\hline
$\C$ & $\mathbin{\textcolor{ForestGreen}{\mathbf{7}}}$ & simple & $D$ &  Majorana\\
\hline
$\H$ &  $\mathbin{\textcolor{ForestGreen}{\mathbf{4}}}$& simple & $\gamma(\nu)$ & symplectic Majorana-Weyl\\
\hline
$\H$ &  $\mathbin{\textcolor{ForestGreen}{\mathbf{6}}}$ & simple &
$\gamma(\nu)\circ J$ & symplectic Majorana \\
\hline
\end{tabular}
\caption{The product structure $\cR$ used in the construction of the
spin projectors $\cP^\cR_\pm\eqdef \frac{1}{2}(1\pm \cR)$ for those
cases when they can be defined and the corresponding terminology for
real spinors. When $p-q\equiv_8 6$, the locally-defined endomorphism $J\in
\Gamma(U,\End(S))$ appearing in the expression for $\cR$ is any of the
complex structures associated with the quaternionic structure
of $S$. Notice that $\Cl(p,q)$ is always simple as an $\R$-algebra (and hence
$\gamma$ is fiberwise injective) in those
cases when spin projectors can be defined. }
\label{table:chiral_projectors}
\end{table}

\subsection{Relation to pin bundles over the complexified \KA bundle of $(M,g)$}
\label{sec:pincomplex}

\subsubsection{General remarks}

Let $T^\ast_\C M$ be the complexified cotangent bundle of $M$, endowed
with the nondegenerate fiberwise $\C$-bilinear pairing induced by the
complexification of $g$. The complexified exterior bundle $\wedge
T^\ast_\C M$ carries a structure of bundle of algebras whose product
(which we again denote by $\diamond$) is obtained by complexfiying the
geometric product induced on $\wedge T^\ast M$ by $g$. The bundle
$(\wedge T^\ast_\C M,\diamond)$ of unital associative algebras over
$\C$ is called the {\em complexified \KA bundle of $(M,g)$}; it
coincides with the complexification of the real \KA bundle as a bundle
of algebras. Its fibers are isomorphic with $\Cl_\C(p,q)\approx \C
\otimes_\R \Cl(p,q)\approx \Cl_\C(d,0)$, the complexification of the
real Clifford algebra $\Cl(p,q)$.  It is natural to consider {\em
bundles of complex pinors}, i.e. bundles $S$ of modules over $(\wedge
T^\ast_\C M,\diamond)$; these are $\C$-vector bundles $S$ over $M$
endowed with a morphism:
\ben
\label{gamma_C}
\gamma_\C: (\wedge T^\ast_\C M,\diamond)\rightarrow (\End_\C(S),\circ)
\een
of bundles of $\C$-algebras.  A $\C$-vector bundle $S$ over $M$ can be
identified with the pair $(S,J)$, where $S$ is the underlying
$\R$-vector bundle while $J\in \Gamma(M,\End_\R(S))$ is the
globally-defined endomorphism given by multiplication with the
imaginary unit in each fiber. This satisfies $J^2=-\id_S$, being the
complex structure on $S$ defining its original $\C$-vector bundle
structure. The bundle $\End_\C(S)$ of complex-linear endomorphisms of
$S$ identifies with the commutant of $J$ in $\End_\R(S)$, being a
bundle of $\R$-subalgebras of the latter. On the other hand, the
complexified \KA bundle can be written $\wedge T^\ast_\C
M=\cO_\C\otimes \wedge T^\ast M $, where $\cO_\C$ is the trivial
complex line bundle on $M$. It is now easy to check that there exists
a bijection between morphisms \eqref{gamma_C} of bundles of
$\C$-algebras and morphisms:
\ben
\label{gamma}
\gamma:(\wedge T^\ast M,\diamond)\rightarrow (\End_\C(S),\circ)
\een
of bundles of $\R$-algebras, where $\gamma_\C$ is recovered from
$\gamma$ through {\em $J$-complexification}, an operation which takes the
following form when applied to global sections\footnote{Notice that
$\Gamma(M,\wedge T^\ast_\C M)=\Omega_\C(M)=
\cC^\infty(M,\C)\otimes_{\cinf}\Omega(M)=\Omega(M)\otimes_\R \C$ is
the $\cC^\infty(M,\C)$-module of complex-valued forms defined on
$M$.}:
\ben
\label{gamma_Cgamma}
\gamma_\C((f+i g)\otimes \omega)=(f\id_S+g J)\circ \gamma(\omega)=\gamma(\omega)\circ (f\id_S+g J)~~,
~~\forall \omega\in \Omega(M)~~,~~\forall f, g\in \cinf~~.
\een
We can thus view a bundle of complex pinors as a pair $(S,J)$ where
$J$ is a complex structure on $S$ and $S$ is a bundle of real pinors
whose underlying morphism $\gamma$ has the property that its image
lies in the commutant of $J$. It is quite obvious that fiberwise irreducibility of $\gamma$ implies 
fiberwise irreducibility (over $\C$) of $\gamma_\C$. Well-known results from the representation theory of complex Clifford algebras 
imply that the converse is also true, i.e. we have: 

\paragraph{Proposition.} Let $\gamma$, $\gamma_\C$ be related through \eqref{gamma_Cgamma} as above. 
Then $\gamma$ is fiberwise irreducible (over $\R$) iff. $\gamma_\C$ is fiberwise irreducible (over $\C$). 

\

\noindent As in \cite{ga1}, it is convenient to consider the complex-valued volume form: 
\ben
\label{nuC}
\nu_\C\eqdef i^{q+\left[\frac{d}{2}\right]}\nu
\een
When fiberwise irreducibility holds (i.e. when $(S,\gamma)$ is a real
{\em pin} bundle and thus $(S,\gamma_\C)$ is a complex pin bundle),
the Schur algebra $\S$ must equal $\C$ or $\H$ (since $J$
belongs to the commutant of the image of $\gamma$ and thus $J$ is a
section of the Schur bundle $\Sigma_\gamma$ of $\gamma$), so the real
rank $N$ of $S$ equals $2^{\left[\frac{d}{2}\right]+1}$ and hence its
complex rank equals $2^{\left[\frac{d}{2}\right]}$; this agrees with a
well-known fact from the representation theory of complex Clifford
algebras. Let us consider the two cases in turn:

\subsubsection{The case $\S=\C$} 

In this case, we have two choices for $J$, namely $J=\pm \gamma(\nu)$, leading through \eqref{gamma_Cgamma} to the two
$J$-complexifications:
\be
\gamma_\C^\pm ((f+i g)\otimes \omega)=(f\id_S\pm g \gamma(\nu))\circ \gamma(\omega)~~,~~\forall \omega\in \Omega(M)~~,~~\forall f, g\in \cinf~~,
\ee
whose fiberwise representations are complex-conjugate to each
other. They are distinguished by the property:
\be
\gamma_\C^\pm (i\nu)=\mp \id_S ~\Longleftrightarrow~ \gamma_\C^\pm(\nu_\C) =\epsilon_{\gamma_\C^\pm}\id_S~~
~~,
\ee
where we introduced the signature (see \cite{ga1}) of
$\gamma_\C^{\pm}$ (as morphisms of bundles of $\C$-algebras):
\be
\epsilon_{\gamma_\C^\pm}\eqdef \pm (-1)^{\frac{1}{2}(1+q+\left[\frac{d}{2}\right])}=\pm \twopartdef{(-1)^{q+1}~,~}{p-q\equiv_8 3}{(-1)^q~,~}{p-q\equiv_8 7}~~
\ee
and we used the fact that $\gamma_\C(\nu)=\gamma(\nu)=\pm J$ as well as the congruence:
\be
1+q+\left[\frac{d}{2}\right]\equiv_4 \twopartdef{2(q+1)~,~}{p-q\equiv_8 3}{2q~,~}{p-q\equiv_8 7}~~.
\ee 
Notice that $(S,J=\gamma(\nu), \gamma_\C^+)$ and $(S,J=-\gamma(\nu), \gamma_\C^-)$ are inequivalent but mutually complex-conjugate 
complex pin bundles over $(M,g)$ --- this corresponds to the well-known fact that the complexified Clifford algebra
$\Cl_\C(p,q)\approx \Cl_\C(d,0)$ has two inequivalent (but mutually complex-conjugate) irreducible
$\C$-representations when $d$ is odd. 

\subsubsection{The case $\S=\H$} 

In this case, let us assume for simplicity that we are given a global section $J\in \Gamma(M,\cU_\gamma)$ of the twistor bundle \eqref{U_gamma} of $(S,\gamma)$ (with slight adaptations of notations, the 
discussion generalizes when such a section is given only locally). We then have a corresponding morphism (acting on sections as in \eqref{gamma_Cgamma}) of bundles of $\C$-algebras, which we denote through 
$\gamma_J$. This satisfies:
\ben
\label{gamma_Jnu}
\gamma_J(\nu_\C)=(-1)^{q+1}\twopartdef{\gamma(\nu)~,~}{p-q\equiv_8 4,5}{\gamma(\nu)\circ J~,~}{p-q\equiv_8 6}~~,
\een
where we used the congruence: 
\be
q+\left[\frac{d}{2}\right]\equiv_4\twopartdef{2(q+1)~~~,~}{p-q\equiv_8 4,5}
{2q+3~,~}{p-q\equiv_8 6}~~.
\ee
It is convenient to distinguish the following subcases:
\begin{itemize}
\item When $p-q\equiv_8 5$ (so that $d$ is odd), we have
$\gamma(\nu)=\epsilon_\gamma \id_S$, which gives
$\gamma_J(\nu_\C)=\epsilon_{\gamma_J}\id_S$, where
$\epsilon_{\gamma_J}=(-1)^{q+1}\epsilon_\gamma$. The two choices for
the sign factor $\epsilon_{\gamma_J}$ correspond to the two
inequivalent irreducible $\C$-representations of the complexified
Clifford algebra $\Cl_\C(p,q)\approx \Cl_\C(d,0)$ and lead to
inequivalent complex pin bundles over $(M,g)$.
\item When $p-q\equiv_8 4, 6$ (so that $d$ is even), equation
\eqref{gamma_Jnu} gives $\gamma_J(\nu_\C)=(-1)^{q+1}\cR$, where $\cR$
is the corresponding spin endomorphism, which obviously commutes with
$J$. Hence $\gamma_J(\nu_\C)$ is a globally-defined $\C$-linear
endomorphism of $S$, when the latter is endowed with the complex
structure induced by $J$ --- it plays the
role of a spin endomorphism acting on complex pinors given by
sections of $S$. 
\end{itemize}

\section{Admissible bilinear pairings on the pin bundle}
\label{sec:pairings}

\subsection{Basics}
\label{sec:pairingsbasics}

\paragraph{The $\cB$-transpose.}

For any non-degenerate fiberwise bilinear pairing $\cB$ on $S$, we let
$T^t$ denote the transpose of $T\in \Gamma(M,\End(S))$ with respect to
$\cB$, which is defined through:
\ben
\label{transpose}
\cB(T\xi,\xi')=\cB(\xi,T^t\xi')~~,~~\forall \xi,\xi'\in \Gamma(M,S)~~.
\een
This operation satisfies $(T^t)^t=T$ and $(\id_S)^t=\id_S$.  The
operation $T\rightarrow T^t$ of taking the $\cB$-transpose defines a
$\cinf$-linear anti-automorphism of the algebra
$(\Gamma(M,\End(S)),\circ)$. 

\paragraph{Remark.} Consider a fiberwise non-degenerate bilinear pairing ${\tilde \cB}$ on
$S$ which is related to $\cB$ through: 
\ben
\label{BtildeB}
{\tilde \cB}=\cB\circ (\id_S\otimes A)~\Longleftrightarrow~ {\tilde \cB}(\xi,
\xi')=\cB(\xi, A\xi')~~,~~\forall \xi,\xi'\in \Gamma(M,S)~~,
\een
where $A\in\Gamma(M, \Aut(S))$ is a globally-defined automorphism of $S$. Then
the transpose $T^{{\tilde t}}$ of a section $T\in \Gamma(M,\End(S))$ with respect to ${\tilde
  \cB}$ is related to the  transpose $T^t$ of $T$ with respect to $\cB$
through: 
\ben
\label{TranspRel}
T^{{\tilde t}}=A^{-t}\circ T^t \circ A^t~~.
\een
Let us assume further that $\cB(\xi,\xi')=\sigma\cB(\xi',\xi)$ for all $\xi, \xi'\in \Gamma(M,S)$,
with $\sigma\in \{-1,1\}$ and that $A^t=\eta A$ for some $\eta\in
\{-1,1\}$. Then  \eqref{BtildeB} implies ${\tilde
  \cB}(\xi,\xi')={\tilde \sigma}{\tilde \cB}(\xi',\xi)$, with ${\tilde
  \sigma}=\eta \sigma $. 

\paragraph{Admissible pairings on $S$.}

Recall from \cite{AC1, AC2} that a nondegenerate bilinear pairing $\cB$ on the pin bundle $S$ is called {\em
admissible} if it has the following properties:

\begin{enumerate}

\item $\cB$ is either symmetric or skew-symmetric:
\ben
\label{Bsymmetry}
\cB(\xi,\xi')=\sigma_\cB \cB(\xi', \xi)~~,~~\forall \xi,\xi' \in \Gamma(M,S)~~,
\een
where the sign factor $\sigma_\cB =\pm1$ is called the {\em symmetry} of $\cB$;

\item For any $\omega\in \Omega(M)$, we have:
\ben
\label{Btype}
\gamma(\omega)^t=\gamma(\tau_\cB(\omega))~\Longleftrightarrow ~
\cB(\gamma(\omega)\xi,\xi')= \cB(\xi,\gamma(\tau_\cB(\omega)) \xi')~~,~~\forall \xi,\xi' \in \Gamma(M,S)~~,
\een
where: 
\ben
\label{tauBdef}
\tau_\cB\eqdef \tau\circ \pi^{\frac{1-\epsilon_\cB}{2}}=\twopartdef{\tau~,~}{\epsilon_\cB=+1}{\tau\circ \pi~,~}{\epsilon_\cB=-1}
\een
is the {\em $\cB$-modified reversion} (an anti-automorphism of the
\KA algebra of $(M,g)$) and the sign factor $\epsilon_\cB \in
\{-1,1\}$ is called the {\em type} of $\cB$;
 
\item If $p-q\equiv_8 0,4,6,7$ (thus $S=S^+ \oplus S^-$ where $S^\pm\subset S$ are the real spin bundles), then $S^+$ and $S^-$ are either $\cB$-orthogonal 
to each other or $\cB$-isotropic. The {\em isotropy} of $\cB$ is the sign factor $\iota_\cB\in \{-1,1\}$ defined through:
\ben
\label{Bisotropy}
 \iota_\cB \eqdef \twopartdef{+1~,~}{\cB(S^+,S^-)=0}{-1~,~}{\cB(S^\pm,S^\pm)=0}~~.
\een
When $p-q\not \equiv 0,4,6,7$, then $\iota_\cB$ is undefined. 
\end{enumerate}

\noindent When $p-q\equiv_8 0,4,6,7$, we have $\iota_\cB=+1$
iff. $(S^\pm)^\perp=S^\mp$, where $~^\perp$ denotes the
$\cB$-orthogonal complement of a sub-bundle of $S$. This case can
occur for any symmetry $\sigma_\cB$ of $\cB$. The case $\iota_\cB=-1$
can occur only when $\sigma_\cB=-1$, in which case $\cB$ is a
symplectic pairing on $S$. In this case, the condition $\iota_\cB=-1$
implies that $S^+$ and $S^-$ are Lagrangian sub-bundles of the
symplectic vector bundle $(S,\cB)$.

\paragraph{Remark.} Recall that (when $\gamma$ is reducible on the
fibers) the spin projectors are given by $\cP^\cR_\pm
=\frac{1}{2}(1\pm \cR)$, where $\cR$ is a product structure on $S$. It is
easy to see that $\cR$ satisfies the following relation, where $^t$
stands for the transpose with respect to an admissible pairing $\cB$
on $S$:
\ben
\label{Rtranspose}
\cR^t=\iota_\cB \cR~\Longleftrightarrow~ (\cP^\cR_\pm )^t=
\cP^\cR_{\pm \iota_\cB}~~.
\een
In fact, this relation can be used as an alternative definition of
$\iota_\cB$. In particular, an admissible pairing on $S$ has the
property that $\cR$ is either self-adjoint $(\iota_\cB=+1$) or
anti-selfadjoint ($\iota_\cB=-1$) with respect to $\cB$.  For later
reference, we also note that \eqref{Btype} implies the relation:
\ben
\label{voltranspose}
\gamma(\nu)^t=(-1)^{\left[\frac{d}{2}\right]}\epsilon_\cB^d \gamma(\nu)~~,
\een
where we noticed that $\tau(\nu)=(-1)^{\left[\frac{d}{2}\right]}\nu$
and $\pi^{\frac{1-\epsilon_\cB}{2}}(\nu)=\epsilon_\cB^d\nu$~. 
  
\paragraph{Local expressions.} Let $e^m$ be a pseudo-orthonormal local coframe of $(M,g)$
defined above an open subset $U\subset M$. Then property \eqref{Btype}
amounts to:
\ben
\label{type1}
(\gamma^m)^t=\epsilon_\cB\gamma^m~\Longleftrightarrow~  \cB(\gamma^m\xi,\xi')=\epsilon_\cB \cB(\xi,\gamma^m\xi')~~,~~\forall m=1\ldots d~~,
\een
which in turn implies:
\ben
\label{gammaAtranspose}
\boxed{(\gamma^A)^t=\epsilon_\cB^{|A|} (-1)^{\frac{|A|(|A|-1)}{2}}\gamma^{A}}
\een
and: 
\ben
\label{typeA}
\boxed{(\gamma^A)^{-t}=\epsilon_\cB^{|A|}\gamma_A~\Longleftrightarrow ~
\cB((\gamma^A)^{-1}\xi,\xi')=\epsilon_\cB^{|A|} \cB(\xi,\gamma_A\xi')~~,
~~\forall \xi,\xi'\in \Gamma(M,S)}~~,
\een
where $\epsilon^{|A|}_{\cB}\eqdef (\epsilon_\cB)^{|A|}$. If
$(\varepsilon_i)_{i=1\ldots N}$ is an arbitrary local frame of $S$
defined above $U$ (with dual local coframe of $S^\ast$ denoted by
$(\varepsilon^i)_{i=1\ldots N}$), then any $T\in \Gamma(M,\End(S))$
acts through:
\be
T|_U\varepsilon_i=\sum_{j=1}^N T_{ji}\varepsilon^j=\sum_{j=1}^N \varepsilon^j(T\varepsilon_i)\epsilon_j \nn
\ee 
where $T_{ij}\eqdef \varepsilon^i(T|_U\varepsilon_j)\in \cC^\infty(U,\R)$. This gives:
\beqa
T(\xi) =_{_U} \sum_{j=1}^N \varepsilon^j (T\xi|_U)\varepsilon_j~~,
~~\forall \xi\in \Gamma(M,S)~~\mathrm{and}~~\tr(T|_U) &=& \sum_{i=1}^N T_{ii}=\sum_{i=1}^N
\varepsilon^i(T|_U\varepsilon_i)~~.
\eeqa
In particular, we have: 
\be
\gamma^A_{ij}\eqdef \varepsilon^i(\gamma^A\varepsilon_j)\in \cC^\infty(U,\R)~~
{\rm and}
~~\gamma^A(\varepsilon_i)=\sum_{j=1}^N (\gamma^A)_{ji}\varepsilon^j~~,
\ee
with similar relations for $\gamma_A^{-1}$. The matrices
$(\gamma^m_{ij})_{i,j=1 \ldots N}$ are the {\em gamma matrices} of
$e^m$ (in fact, matrix-valued {\em functions}) with respect to the
local frame $\varepsilon_i$ of $S$.

Let us give more detail on the admissible bilinear pairings $\cB$ for
each of the three types of real representations. If one drops the
non-degeneracy assumptions, bilinear pairings on $S$ satisfying
properties $1.-3.$ above for any fixed $\sigma,\epsilon$ and $\iota$ form a free $\cinf$-module whose rank
depends on $p$ and $q$ (see \cite{AC1, AC2}). A convenient basis of
this free module is provided by certain admissible bilinear pairings
(determined up to multiplication with a nowhere vanishing function)
which were constructed in loc. cit.  Below, we give more detail on the
properties of such admissible pairings.

\subsection{Normal representation ($p-q\equiv_8 0,1,2$)} 
\label{sec:pairingscasesnormal}

The spin projection exists only when $p-q\equiv_8 0$, with
$\cR=\gamma(\nu)$. The situation for admissible pairings is as
follows.

\paragraph{When $p-q\equiv_8 0,2$, i.e. the normal simple case.} 
Up to multiplication by elements of $\cinf$, there are two admissible pairings $\cB_+$ and $\cB_-$,
which are distinguished by the value $\epsilon\in \{-1,1\}$ of their
type. Up to multiplication with a nowhere-vanishing function, 
we can take\footnote{Relation \eqref{Bpmrel} implies $(\gamma^m)^{t_{-}}=\gamma(\nu)^{-t_+}\circ
  (\gamma^m)^{t_+}\circ \gamma(\nu)^{t_+}$, where
$t_\pm$ denotes the transpose taken with respect to the pairing
$\cB_\pm$.  Since $d$ is even in this case, it follows that $\gamma(\nu)$ anticommutes
with all $\gamma^m$ and thus $(\gamma^m)^{t_{-}}=-(\gamma^m)^{t_+}$, so 
pairings related by \eqref{Bpmrel} have opposite type, i.e. $\epsilon_{-}=-\epsilon_{+}$. Furthermore, we have
$\gamma(\nu)^{t_+}=(-1)^{\frac{d}{2}}\gamma(\nu)$, which implies
via relation \eqref{Bpmrel}  that the symmetries $\sigma_+$ and $\sigma_-$ are
related through $\sigma_+=(-1)^{\frac{d}{2}} \sigma_-$. }: 
\ben
\label{Bpmrel}
\cB_+=\cB_-\circ (\id_S\otimes \gamma(\nu))\Leftrightarrow \cB_-=(-1)^{\frac{p-q}{2}}\cB_+\circ (\id_S\otimes \gamma(\nu))=\twopartdef{+ \cB_+\circ (\id_S\otimes
\gamma(\nu))~,~}{p-q\equiv_8 0}{- \cB_+\circ (\id_S\otimes
\gamma(\nu))~,~}{p-q\equiv_8 2}~~
\een
where we used that $d$ is even in this case and the relation
$\gamma(\nu)^2=(-1)^{q+\frac{d}{2}}\id_S=(-1)^{\frac{p-q}{2}}\id_S$ (which holds for
$d=$even).  The symmetry $\sigma(\epsilon, d)$ of $\cB_\epsilon$ is uniquely determined by
$\epsilon$ and by the mod $8$ reduction of $d$ according to the table:
\begin{center}
\begin{tabular}{|c|c|c|c|c|}
\hline
 $\begin{array}{c}d\\({\rm mod}~8)\end{array}$ &  $0$ & $2$ & $4$ & $6$ \\
\hline
 $\sigma(\epsilon, d)$ &  $+1$ & $+\epsilon$ & $-1$& $-\epsilon$\\
\hline
\end{tabular}
\end{center}
\noindent The isotropy type $\iota$ of $\cB_\epsilon$ depends only
on $p$ and $q$ and is as follows:
\begin{itemize}
\item When $p-q\equiv_8 0$, the restriction $\gamma_\ev$ is fiberwise
reducible and $\iota=(-1)^{\frac{d}{2}}=\twopartdef{+1~,~}{d=0,4~,}{-1~,~}{d=2,6~.~}$  Indeed, we
have $\cR=\gamma(\nu)$. Since $d$ is even for such $p,q$,
$\cR$ satisfies $\cR^t=(-1)^{\frac{d}{2}} \cR$ for any
admissible pairing $\cB$ on $S$ due to \eqref{voltranspose}. Hence
$\iota=(-1)^{\frac{d}{2}}$ (see \eqref{Rtranspose}) for any
admissible $\cB$.  Together with \eqref{Bpmrel}, this gives: 
\begin{enumerate}
\item When $d\equiv_8 0, 4$: $\cB_+|_{S_\pm \otimes S_\mp}=\cB_-|_{S_\pm \otimes S_\mp}=0~~,~~\cB_+|_{S_\pm
  \otimes S_\pm}=\pm \cB_-|_{S_\pm \otimes S_\pm}$
\item When $d\equiv_8 2, 6$: $\cB_+|_{S_\pm \otimes S_\pm}=\cB_-|_{S_\pm \otimes S_\pm}=0~~,~~\cB_+|_{S_\pm
  \otimes S_\mp}=\mp \cB_-|_{S_\pm \otimes S_\mp}$, where we used the fact that $\gamma(\nu)|_{S_\pm}=\pm \id_{S^\pm}$.
\end{enumerate}
\item When $p-q\equiv_8 2$ , the restriction $\gamma_\ev$ is fiberwise
irreducible, so $\iota$ is not defined.
\end{itemize}
The pairing $\cB_+$ (which is uniquely determined up to multiplication by a nowhere-vanishing smooth function) will be called the {\em
  basic admissible pairing} and will also be denoted $\cB_0$. 

\paragraph{When $p-q\equiv_8 1$, i.e. the normal non-simple case.} Then the admissible bilinear pairing $\cB$ is
unique up to multiplication by a nowhere vanishing smooth
function\footnote{Recall that $\gamma(\nu)=\epsilon_\gamma\id_S$ in this case, so $\cB\circ
(\id_S\otimes \gamma(\nu))=\epsilon_\gamma \cB$ is proportional to $\cB$.} and $\gamma_\ev$ is fiberwise irreducible, so $\iota_\cB$ is not
defined.  The values of $\sigma_\cB$ and $\epsilon_\cB$ are determined
by the mod $8$ reduction of $d$ as shown in the table below:
\begin{center}
\begin{tabular}{|c|c|c|c|c|} \hline $\begin{array}{c}d\\({\rm mod}~8)\end{array}$ & $1$ & $3$ & $5$ &
$7$\\ \hline\hline $\sigma_\cB$ & $+1$ & $-1$ & $-1$ & $+1$ \\ \hline
$\epsilon_\cB$ & $+1$ & $-1$ & $+1$ & $-1$ \\ \hline
\end{tabular}
\end{center}

\subsection{Almost complex representation ($p-q\equiv_8 3,7$)}
\label{sec:pairingscasesac}

In this case, there are four independent choices $\cB_0,\cB_1,\cB_2$ and $\cB_3$ for the nondegenerate admissible
pairing, which we can take to be related through:
\ben
\label{Cpairings}
\cB_1=\cB_0\circ (\id_S\otimes J)~~,~~\cB_2=\cB_0\circ (\id_S\otimes D)~~,~~\cB_3=\cB_0\circ [\id_S\otimes (D\circ J)]~~.
\een
Here, $\cB_0$ (which we shall call the {\em basic admissible pairing})
is a particular choice of admissible pairing (determined up to
multiplication by a nowhere-vanishing smooth real-valued function), with
type $\epsilon_0=-1$, whose symmetry $\sigma_0$ is given by:
\begin{center}
\begin{tabular}{|c|c|c|c|c|} \hline $\begin{array}{c}d\\({\rm mod}~8)\end{array}$ & $1$ & $3$ & $5$ &
$7$ \\ \hline $\sigma_0$ & $+1$ & $-1$ & $-1$ & $+1$ \\ \hline
\end{tabular}
\end{center}
\noindent and whose isotropy $\iota_0$ is as follows \cite{AC2}:
\begin{itemize}
\item {\bf When $p-q\equiv_8 3$}, the restriction $\gamma_\ev$ is
fiberwise irreducible so $\iota_0$ is not defined.
\item {\bf When $p-q\equiv_8 7$}, the restriction $\gamma_\ev$ is
fiberwise reducible and we have $\iota_0=1$, i.e. $\cB_0|_{S_\pm\otimes S_\mp}=0$.
\end{itemize} The symmetry $\sigma_k$, type $\epsilon_k$ and isotropy $\iota_k$ (the latter being
defined iff. $p-q\equiv_8 7$) of $\cB_k$ for $k=1\ldots 3$ are
related to those of $\cB_0$ as shown in the table below.
\begin{table}[H]
\centering
\begin{tabular}{|c|c|c|c|} \hline $k$ & $1$ & $2$ & $3$ \\
\hline\hline $\sigma_k/\sigma_0$ & $-(-1)^{\left[\frac{d}{2}\right]}$
& $ +1$ & $(-1)^{\left[\frac{d}{2}\right]}$ \\ \hline $\epsilon_k/\epsilon_0$
& $+1$ & $-1$ & $-1$ \\ \hline $\iota_k/\iota_0$ & $-1$ & $+1$ & $-1$
\\ \hline
\end{tabular}
\caption{Characteristics of admissible bilinear forms in the almost complex
  case.}
\label{table:CBilChars}
\end{table}

\paragraph{$\cB_0$-symmetry properties of $J$ and $D$.} 
In all subcases of the almost complex case, the globally-defined endomorphisms $J=\gamma(\nu)$ and $D$ satisfy
\cite{Okubo1,AC2}:
\ben
\label{JDtranspose}
\boxed{J^t=(-1)^{\frac{d(d+1)}{2}}J~~,~~D^t = D~\Longrightarrow~ (D\circ
  J)^t=(-1)^{\frac{d(d-1)}{2}} D\circ J}~~,
\een
where $^t$ denotes the transpose taken {\em with respect to the basic pairing $\cB_0$}.  In particular, $D$ is
$\cB_0$-selfadjoint. It is easy to check that these relations agree with the
first row of Table \ref{table:CBilChars} if one uses the congruence: 
\be
\frac{d(d-1)}{2}\equiv_2 \left[\frac{d}{2}\right]~\Longrightarrow~
\frac{d(d+1)}{2}\equiv_2 1+\left[\frac{d}{2}\right]~~.
\ee
Relations \eqref{Dsquare} and \eqref{JDtranspose} imply:
\ben
\label{BD}
\boxed{D^{-t}=(-1)^{\frac{p-q+1}{4}}D~\Longleftrightarrow ~
\cB_0(D^{-1}\xi,\xi')=(-1)^{\frac{p-q+1}{4}}\cB_0(\xi,D\xi')}~~.
\een

\subsubsection{$\cB_k$-symmetry properties} 
The symmetry properties of various endomorphisms of $S$ with respect to the other pairings
$\cB_k$ ($k=1\ldots 3$) can be obtained using \eqref{Cpairings}. Applying \eqref{TranspRel}, we find that the transpositions $T^{t_k}$ of $T\in \Gamma(M,\End(S))$ with
respect to the admissible pairings $\cB_k$ ($k=1\ldots 3$) are related to the
transposition $T^t$ of $T$ with respect to $\cB_0$ through: 
\beqan
\label{TransposeComplex}
T^{t_1}&=&J^{-t}\circ T^t\circ J^t=-J\circ T^t\circ J~~,\nn\\
T^{t_2}&=&D^{-t}\circ T^t\circ D^t=(-1)^{\frac{p-q+1}{4}} D\circ T^t \circ D~~,\\
T^{t_3}&=&D^{-t}\circ J^{-t}\circ T^t\circ J^t\circ
D^t=-(-1)^{\frac{p-q+1}{4}} D\circ J\circ T^t \circ J \circ D~~.\nn
\eeqan
Relations \eqref{TransposeComplex} simplify further when $T$ is $\C$-linear or $\C$-antilinear, i.e. if $T$
commutes or anticommutes with $J$. In this case, we define:
\be
\lambda_T\eqdef\twopartdef{+1~,~}{[T,J]_{-,\circ}=0~,}{-1~,~}{[T,J]_{+,\circ}=0~~}~\Longleftrightarrow~
T\circ J=\lambda_TJ\circ T~~
\ee
and find: 
\be
T^{t_1}=\lambda_T T^t~~,~~T^{t_2}=(-1)^{\frac{p-q+1}{4}} D\circ T^t \circ D~~,~T^{t_3}=-(-1)^{\frac{p-q+1}{4}} \lambda_T D\circ T^t  \circ D~~,\nn
\ee
where we noticed that $T^t$ has the same (anti-)commutation properties with
$D$ as $T$.  Similarly, we find simplifications when $T$ commutes or
anticommutes with $D$, in which case we define:
\be
\delta_T\eqdef\twopartdef{+1~,~}{[T,D]_{-,\circ}=0~}{-1~,~}{[T,D]_{+,\circ}=0~~}~\Longleftrightarrow~
T\circ D=\delta_T D\circ T~~
\ee
and find: 
\be
T^{t_1}=-J\circ T^t\circ J~~,~T^{t_2}=\delta_T T^t ~~,~~T^{t_3}=- \delta_T J\circ T^t \circ J ~~,
\ee
noticing now that $T^t$ has the same (anti-)commutation properties with
$J$ as $T$. A case often encountered in applications is when $T$ is both
$\C$-linear/antilinear and commutes/anti-commutes with $D$. In this situation,
both $\lambda_T$ and $\delta_T$ are defined and we find:
\be
T^{t_1}=\lambda_T T^t ~~,~~T^{t_2}= \delta_T T^t ~~,~~T^{t_3}= \lambda_T \delta_T T^t  ~~.\nn
\ee
\paragraph{The $\cB_\alpha$-transpose of $\gamma^{A}$. }
A typical example of the last type mentioned in the previous paragraph is $T=\gamma^A$ for some ordered
multi-index $A=(a_1,\ldots,a_k)$ of length $|A|=k$. Then $\lambda_{\gamma^A}=+1$ (since
$\gamma^a$ commute with $J$) and $\delta_{\gamma^A}=(-1)^{|A|}$ (since $\gamma^a$
anti-commute with $D$). Since $\epsilon_0=-1$, equation \eqref{gammaAtranspose} becomes: 
\be
(\gamma^A)^t=(-1)^{\frac{|A|(|A|+1)}{2}}\gamma^A
\ee 
and the relations above give: 
\beqan
\label{CGammaTransposes}
(\gamma^A)^{t_1}&=& (-1)^{\frac{|A|(|A|+1)}{2}} \gamma^A ~\Longrightarrow~ (\gamma^a)^{t_1}=-\gamma^a~~,\nn\\
(\gamma^A)^{t_2}&=& (-1)^{\frac{|A|(|A|-1)}{2}} \gamma^A ~\Longrightarrow~ (\gamma^a)^{t_2}=+\gamma^a~~,\\
(\gamma^A)^{t_3}&=& (-1)^{\frac{|A|(|A|-1)}{2}} \gamma^A  ~\Longrightarrow~ (\gamma^a)^{t_3}=+\gamma^a~~.\nn
\eeqan
These identities agree with the second row of Table \ref{table:CBilChars}. Relations \eqref{CGammaTransposes} imply: 
\beqan
\label{CBilkSym}
\cB_0(\xi, \gamma^A\xi')&=& \sigma_0(-1)^{\frac{|A|(|A|+1)}{2}} \cB_0(\xi', \gamma^A\xi) ~~,\nn\\
\cB_1(\xi, \gamma^A\xi')&=& \sigma_1(-1)^{\frac{|A|(|A|+1)}{2}} \cB_1(\xi', \gamma^A\xi) ~~,\nn\\
\cB_2(\xi, \gamma^A\xi')&=& \sigma_2(-1)^{\frac{|A|(|A|-1)}{2}} \cB_2(\xi', \gamma^A\xi) ~~,\\
\cB_3(\xi, \gamma^A\xi')&=& \sigma_3 (-1)^{\frac{|A|(|A|-1)}{2}} \cB_3(\xi', \gamma^A\xi)  ~~.\nn
\eeqan

\subsubsection{The case $p-q\equiv_8 7$} Let us consider this situation
in more detail, given that a spin projection can be defined in this
case (leading to Majorana spinors). Viewing $S$ as a complex vector bundle (with complex structure
given by $J$), recall that in this case $D$ is a real structure
(complex conjugation) on $S$, which we also denote by an overline. For
$p-q\equiv_8 7$, we always have $\iota_0=1$, which means that the
basic admissible pairing $\cB_0$ satisfies:
\ben
\label{Bres}
\cB_0|_{S_\pm\otimes S_\mp}=0~\Longrightarrow~ \cB_0(\xi_\pm,\xi'_\mp)=0~~,~~\forall \xi, \xi'\in \Gamma(M,S)~~
\een
and hence it is determined by its restrictions to $S_+\otimes S_+$ and to
$S_-\otimes S_-$. In turn, the second of these restrictions is determined by
the first due to  the first of identities \eqref{JDtranspose}, which imply: 
{\small
\be
\cB_0(J\xi,J\xi')=(-1)^{\frac{d(d-1)}{2}} \cB_0(\xi,\xi')~~,~~\forall
\xi,\xi'\in \Gamma(M,S)~\Longrightarrow~ \cB_0|_{S_-\otimes
  S_-}=(-1)^{\frac{d(d-1)}{2}} \cB_0|_{S_+\otimes S_+}\circ (J\otimes
J)|_{S_-\otimes S_-}
\ee}
Property \eqref{Bres} implies:
\ben
\label{B0exp}
\cB_0(\xi,\xi')=\cB_0(\xi_+,\xi'_+)+\cB_0(\xi_-,\xi'_-)=\cB_0(\xi_R,\xi'_R)-(-1)^{\frac{d(d+1)}{2}}\cB_0(\xi_I,\xi'_I)~~,~~\forall \xi,\xi'\in \Gamma(M,S)~~,
\een
where we used decomposition \eqref{xiReImDec} and the first of identities \eqref{JDtranspose}. Together with \eqref{bar_gamma}, this gives: 
\ben
\label{CB0Gamma}
\cB_0(\xi,\gamma(\omega)\xi') =\twopartdef{\cB_0(\xi_R,\gamma(\omega)\xi'_R)-(-1)^{\frac{d(d+1)}{2}}\cB_0(\xi_I,\gamma(\omega)\xi'_I)~,~}{\omega\in \Omega^\ev(M)~,}
{\cB_0(\xi_R,(J\circ \gamma(\omega))\xi'_I)+(-1)^{\frac{d(d+1)}{2}}\cB_0(\xi_I,(J\circ \gamma(\omega))\xi'_R)~,~}{\omega\in \Omega^\odd(M)~,~}~~
\een
where we used \eqref{JDtranspose} and the fact that $J$ and $\gamma(\omega)$ commute. On the other hand, we always have $\epsilon_0=-1$, which gives: 
\be
\gamma(\omega)^t=\gamma(\tau(\pi(\omega)))=\twopartdef{+\gamma(\tau(\omega))~,~}{\omega\in \Omega^\ev(M)~,}{-\gamma(\tau(\omega))~,~}{\omega\in \Omega^\odd(M)~,~}~~
\ee
thereby implying: 
\ben
\label{CB0GammaSymm}
\cB_0(\xi, \gamma(\omega)\xi')=\sigma_0\cB_0(\xi',\gamma((\tau\circ
\pi)(\omega))\xi)~~,~~\forall \xi,\xi'\in \Gamma(M,S)~~, 
\een
which in turn gives: 
\ben
\label{CB0GammaComp}
\cB_0(\xi, \gamma^A\xi')=\sigma_0(-1)^{\frac{|A|(|A|+1)}{2}}\cB_0(\xi',\gamma^A\xi)~~,~~\forall \xi,\xi'\in \Gamma(M,S)~~.
\een
The other admissible pairings \eqref{Cpairings} can be expressed as: 
\beqan
\label{B123exp}
\cB_1(\xi,\xi')&=&\cB_0(\xi, J\xi')=-\cB_0(\xi_R, \xi'_I)-(-1)^{\frac{d(d+1)}{2}}\cB_0(\xi_I, \xi'_R)~~,\nn\\
\cB_2(\xi,\xi')&=&\cB_0(\xi, \overline{\xi'})=\cB_0(\xi_R, \xi'_R)+(-1)^{\frac{d(d+1)}{2}}\cB_0(\xi_I, \xi'_I)~~,\\
\cB_3(\xi,\xi')&=&-\cB_0(\xi, J(\overline{\xi'}))=-\cB_0(\xi_R, \xi'_I)+(-1)^{\frac{d(d+1)}{2}}\cB_0(\xi_I, \xi'_R)~~,\nn
\eeqan
so in particular we have: 
\beqa
\cB_1|_{S_\pm\otimes S_\pm}&=&0~\Longrightarrow~ \iota_1=-1~~,\nn\\
\cB_2|_{S_\pm\otimes S_\mp}&=&0~\Longrightarrow~ \iota_2=+1~~,\nn\\
\cB_3|_{S_\pm\otimes S_\pm}&=&0~\Longrightarrow~ \iota_3=-1~~,\nn
\eeqa
which agrees with the third row of Table \ref{table:CBilChars}. Also note the relations: 
\beqa
\cB_0|_{S_+\otimes S_+}=\cB_2|_{S_+\otimes  S_+}~~&,&~~\cB_0|_{S_-\otimes S_-}=-\cB_2|_{S_-\otimes S_-}~~,\nn\\
\cB_1|_{S_+\otimes S_-}=\cB_3|_{S_+\otimes S_-}~~&,&~~\cB_1|_{S_-\otimes S_+}=-\cB_3|_{S_-\otimes S_+}~~.\nn
\eeqa
When combined with \eqref{B0exp}, identities \eqref{B123exp} give the following expressions
which we list for convenience of the reader:
\!\!\!\!\!\!\!\!\!\beqan
\label{CBiotas}
\!\!\!\!\!\!\!\!\!\cB_0(\xi_R,\xi'_R)=\frac{1}{2}[\cB_0(\xi,\xi')+\cB_2(\xi,\xi')]~&,&~\cB_0(\xi_I,\xi'_I)=\frac{(-1)^{\frac{d(d+1)}{2}}}{2}[\cB_2(\xi,\xi')-\cB_0(\xi,\xi')]~,~~~~~\\
\!\!\!\!\!\!\!\!\!\cB_0(\xi_R,\xi'_I)=-\frac{1}{2}[\cB_1(\xi,\xi')+\cB_3(\xi,\xi')]~&,&~\cB_0(\xi_I,\xi'_R)=\frac{(-1)^{\frac{d(d+1)}{2}}}{2}[\cB_3(\xi,\xi')-\cB_1(\xi,\xi')]~.~~~~~
\eeqan
Relations \eqref{B123exp} also imply the following relations which we list for completeness:
\be
\cB_0(\xi, J\circ\gamma(\omega)\xi')=\twopartdef{-\cB_0(\xi_R,\gamma(\omega)\xi'_I)-(-1)^{\frac{d(d+1)}{2}}\cB_0(\xi_I,\gamma(\omega)\xi'_R),~}{\omega\in \Omega^\ev(M)~,~~~}
{\cB_0(\xi_R,(J\circ \gamma(\omega))\xi'_R)-(-1)^{\frac{d(d+1)}{2}}\cB_0(\xi_I,(J\circ \gamma(\omega))\xi'_I),~}{\omega\in \Omega^\odd(M)~,~~~~}~~
\ee

\be
\cB_0(\xi,D\circ\gamma(\omega)\xi')  =\twopartdef{\cB_0(\xi_R,\gamma(\omega)\xi'_R)-(-1)^{\frac{d(d+1)}{2}}\cB_0(\xi_I,\gamma(\omega)\xi'_I)~,~}{\omega\in \Omega^\ev(M)~,}
{\cB_0(\xi_R,(J\circ \gamma(\omega))\xi'_I)-(-1)^{\frac{d(d+1)}{2}}\cB_0(\xi_I,(J\circ \gamma(\omega))\xi'_R)~,~}{\omega\in \Omega^\odd(M)~,~}~~
\ee

\be
\cB_0(\xi, J\circ D\circ\gamma(\omega)\xi')=\twopartdef{\cB_0(\xi_R,\gamma(\omega)\xi'_I)-(-1)^{\frac{d(d+1)}{2}}\cB_0(\xi_I,\gamma(\omega)\xi'_R),~}{\omega\in \Omega^\ev(M)~,~~~}
{-\cB_0(\xi_R,(J\circ \gamma(\omega))\xi'_R)-(-1)^{\frac{d(d+1)}{2}}\cB_0(\xi_I,(J\circ \gamma(\omega))\xi'_I),~}{\omega\in \Omega^\odd(M)~.~~~~}~~
\ee

\paragraph{The case of Majorana spinors.} The expressions above simplify when
$\xi, \xi'$ are Majorana spinors, i.e. sections of $S_+$ (that is, real sections of
$S$ when the latter is viewed as the complexification of $S_+$), which means
that $\xi_R=\xi,~ \xi'_R=\xi'$ and $\xi_I=\xi'_I=0$,
i.e. $D(\xi)=\bar{\xi}=+\xi$ and $D(\xi')=\overline{\xi'}=+\xi'$. In this
case, identities \eqref{CB0Gamma} become:
\ben
\label{CB0GammaReal}
\cB_0(\xi,\gamma(\omega)\xi') =0~~\mathrm{for}~~\xi,\xi'\in \Gamma(M,S_+)~~\mathrm{and}~~\omega\in \Omega^\odd(M)~~
\een
while \eqref{B123exp} give: 
\be
\cB_1(\xi,\xi')=\cB_3(\xi,\xi')=0~~,~~\cB_2(\xi,\xi')=\cB_0(\xi, \xi')~~,~~\forall \xi,\xi'\in \Gamma(M,S_+)~~.
\ee
Notice that \eqref{CBiotas} and \eqref{bar_gamma} imply the following
relations, which generalize \eqref{CB0GammaReal}: 
\beqan
\label{CSelectionForTwoSpinors}
\cB_0(\xi,\gamma(\omega)\xi')&=&\cB_2(\xi,\gamma(\omega)\xi') =0~~\mathrm{for}~~\xi,\xi'\in \Gamma(M,S_+)~~\mathrm{and}~~\omega\in \Omega^\odd(M)~~,\nn\\
\cB_1(\xi,\gamma(\omega)\xi')&=&\cB_3(\xi,\gamma(\omega)\xi') =0~~\mathrm{for}~~\xi,\xi'\in \Gamma(M,S_+)~~\mathrm{and}~~\omega\in \Omega^\ev(M)~~
\eeqan
and which can be summarized as: 
\be
\cB_k(\xi,\gamma(\omega)\xi')=0~~\mathrm{for}~~\xi,\xi'\in \Gamma(M,S_+)~~\mathrm{and}~~\omega\in \Omega^{{\rm opar}(k)}(M)~~,
\ee
where: 
\be
{\rm opar}(k)\eqdef \twopartdef{\ev~,~}{k=\odd~,}{\odd~,~}{k=\even~,}~~
\ee
is the opposite of the parity of $k$. In particular, we have: 
\ben
\label{CShortSelectionForTwoSpinors}
\boxed{\cB_k(\xi,\gamma^A\xi')=0~~\mathrm{for}~~\xi,\xi'\in
\Gamma(M,S_+)~~\mathrm{unless}~~|A|\equiv_2 k}~~.
\een
We also have: 
\beqan
\label{CB0GammaRed}
\cB_0(\xi,\gamma(\omega)\xi')&=&\cB_2(\xi,\gamma(\omega)\xi') ~~\mathrm{for}~~\xi,\xi'\in \Gamma(M,S_+)~~\mathrm{and}~~\omega\in \Omega^\ev(M)~~,\nn\\
\cB_1(\xi,\gamma(\omega)\xi')&=&\cB_3(\xi,\gamma(\omega)\xi') ~~\mathrm{for}~~\xi,\xi'\in \Gamma(M,S_+)~~\mathrm{and}~~\omega\in \Omega^\odd(M)~~.
\eeqan
For a single Majorana spinor $\xi$, relations \eqref{CBilkSym} imply the
following properties, which are often encountered in applications: 
\beqan
\label{CSelectionForOneSpinor}
\cB_0(\xi, \gamma^A\xi)&=& 0~~{\rm unless}~~(-1)^{\frac{|A|(|A|+1)}{2}}=\sigma_0~~, \nn\\
\cB_1(\xi, \gamma^A\xi)&=&0~~{\rm unless}~~(-1)^{\frac{|A|(|A|+1)}{2}} =\sigma_1~~,\nn\\
\cB_2(\xi, \gamma^A\xi)&=&0~~{\rm unless}~~(-1)^{\frac{|A|(|A|-1)}{2}} =\sigma_2 ~~,\\
\cB_3(\xi, \gamma^A\xi)&=&0~~{\rm unless}~~(-1)^{\frac{|A|(|A|-1)}{2}} =\sigma_3  ~~\nn
\eeqan
and which can be summarized as: 
\ben
\label{CShortSelectionForOneSpinor}
\boxed{\cB_k(\xi, \gamma^A\xi)= 0~~{\rm unless}~~(-1)^{\frac{|A|(|A|-1)}{2}}\epsilon_k^{|A|}=\sigma_0}~~.
\een

\subsubsection{Local expressions} 
Consider a pseudo-orthonormal local coframe $(e^a)_{a=1\ldots d}$ of $(M,g)$ and a local frame
$(\epsilon_\alpha)_{\alpha=1\ldots \Delta}$ of $S_+$, both supported above an
open subset $U$ of $M$. Defining:
\be
\cB_{\alpha\beta}\eqdef \cB_0(\epsilon_\alpha,\epsilon_\beta)\in \cC^\infty(U,\R)~~,~~\forall \alpha,\beta=1\ldots \Delta~~,
\ee
we let ${\hat \cB}$ denote the $\cC^\infty(U,\R)$ square matrix of dimension $\Delta$ with entries $\cB_{\alpha\beta}$. The symmetry property of $\cB$ implies: 
\be
\cB_{\beta\alpha}=\sigma_0\cB_{\alpha\beta}~~,~~\forall \alpha,\beta=1\ldots \Delta ~\Longleftrightarrow~ {\hat \cB}^T=\sigma_0 {\hat \cB}~~,
\ee 
where $~^T$ denotes the ordinary transpose of matrices. If $\xi,\xi'\in \Gamma(M,S_+)$ expand as in \eqref{xiExpansion} (with $\xi_I^\alpha=0,~ \xi^\alpha=\xi_R^\alpha\in \cC^\infty(U,\R)$ 
and similarly for $\xi'$), we have: 
\be
\cB_0(\xi,\xi')=_{_U}{\hat \xi}^T {\hat \cB}{\hat \xi}'\in \cC^\infty(U,\R)~~.
\ee

\paragraph{Remark.} As mentioned above, one can view the complex vector bundle
$S$ as the compexification $S\approx S_+ \otimes \cO_\C$ of the real vector
bundle $S_+$, where $\cO_\C$ is the trivial complex line bundle over $M$. With
this interpretation, $D$ is the complex conjugation of this complexified
bundle and we can consider the fiberwise $\C$-bilinear pairing $\beta$ on $S$
obtained by complexification of the restriction $\cB_0|_{S_+\otimes S_+}$:
\ben
\label{BetaDef}
\beta(\xi,\xi')\eqdef \cB_0(\xi_R,\xi'_R)-\cB_0(\xi_I,\xi'_I)+
i\left[\cB_0(\xi_R,\xi'_I)+\cB_0(\xi_I,\xi'_R)\right]\in \cC^\infty(M,\C)~~,~~\forall \xi, \xi'\in \Gamma(M,S)~~,
\een
i.e. (using \eqref{B123exp}): 
\beqa
2\beta(\xi,\xi')=&~&~~~\cB_0(\xi,\xi')+\cB_2(\xi,\xi') +(-1)^{\frac{d(d+1)}{2}}(\cB_0(\xi,\xi')-\cB_2(\xi,\xi')) \\
&-&i\left[\cB_1(\xi,\xi')+\cB_3(\xi,\xi')+(-1)^{\frac{d(d+1)}{2}}(\cB_1(\xi,\xi')-\cB_3(\xi,\xi'))\right]~~.
\eeqa
This fiberwise $\C$-bilinear form satisfies: 
\beqa
&&\beta(J\xi,\xi')=\beta(\xi,J\xi')=i\beta(\xi,\xi')~~,~~\forall \xi, \xi'\in\Gamma(M,S)~~,\nn\\
&&\beta(\xi,\xi')=\cB_0(\xi,\xi')~~,~~\forall \xi,\xi'\in \Gamma(M,S_+)~~.
\eeqa
Using the relations above, it is easy to check we have the following local expression for any pair of pinors
$\xi,\xi'\in \Gamma(M,S)$:
\ben
\label{BetaLocal}
\beta(\xi, \gamma^{A}\xi')={\hat \xi}^T {\hat \gamma}^{A}{\hat \xi'}~~.
\een
We stress that the fiberwise $\C$-bilinear pairing $\beta$ on $S$ is often used in the
physics literature instead of the admissible form $\cB_0$, which is only
fiberwise $\R$-bilinear. The admissible forms $\cB_k$ can be reconstructed from
$\beta$ as follows: 
\beqa
\cB_0(\xi,\xi')&=&\beta(\xi_R,\xi'_R)-(-1)^{\frac{d(d+1)}{2}}\beta(\xi_I,\xi'_I)~~,\nn\\
\cB_1(\xi,\xi')&=&-\beta(\xi_R,\xi'_I)-(-1)^{\frac{d(d+1)}{2}}\beta(\xi_I,\xi'_R)~~,\nn\\
\cB_2(\xi,\xi')&=&\beta(\xi_R,\xi'_R)+(-1)^{\frac{d(d+1)}{2}}\beta(\xi_I,\xi'_I)~~,\nn\\
\cB_3(\xi,\xi')&=&-\beta(\xi_R,\xi'_I)+(-1)^{\frac{d(d+1)}{2}} \beta(\xi_I,\xi'_R)~~.\nn\\
\eeqa

\subsection{Quaternionic representation ($p-q\equiv_8 4,5,6$)}
\label{sec:pairingscasesquat}

In this case, spin projectors can only be defined when $p-q\equiv_8 4,
6$. In order to simplify notation, we shall assume that $M$ is contractible, so that all bundles under consideration are topologically trivial; in particular,
$J_i$ are globally defined on $M$. Given that the discussion below is local, this condition can always be removed by replacing $M$ with a sufficiently small open subset $U$. 
The situation for admissible pairings is summarized below \cite{Okubo1, AC1, AC2}. 

\subsubsection{The quaternionic simple case ($p-q\equiv_8 4,6$)} 
When $p-q\equiv_8 4,6$, the morphism $\gamma$ is fiberwise-injective
and we have eight independent admissible pairings
$\cB^\epsilon_\alpha$ ($\epsilon=\pm 1,~ \alpha=0\ldots 3$), which are
given by \cite{AC2}:
\ben
\label{cBkepsilon}
\cB_k^\epsilon=\cB^\epsilon_0\circ (\id_S\otimes J_k)~~,~~\forall k=1\ldots 3~~,~~\forall \epsilon\in \{-1,+1\}~~,
\een
where $\cB^\epsilon_0$ are the two {\em fundamental admissible pairings} --- which we can take to be related through:
\ben
\label{cBpm}
\cB_0^+=\cB_0^-\circ (\id_S\otimes \gamma(\nu))~\Longleftrightarrow~
\cB_0^-=(-1)^{\frac{p-q}{2}} \cB_0^+\circ (\id_S\otimes
\gamma(\nu))=\twopartdef{+  \cB_0^+\circ (\id_S\otimes
\gamma(\nu))~,~}{p-q\equiv_8 4}{- \cB_0^+\circ (\id_S\otimes
\gamma(\nu))~,~}{p-q\equiv_8 6}
\een
and have type $\epsilon$, symmetry $\sigma_0(\epsilon,d)$ given
in the table below:
\begin{center}
\begin{tabular}{|c|c|c|c|c|}
\hline
 $\begin{array}{c}d\\({\rm mod}~8)\end{array}$ &  $0$ & $2$ & $4$ & $6$ \\
\hline
 $\sigma_0(\epsilon,d)$ &  $-1$ & $-\epsilon$ & $+1$ & $+\epsilon$\\
\hline
\end{tabular}
\end{center}
and isotropy given as follows:
\begin{itemize}
\item If $p-q\equiv_8 4$, then $\gamma_\ev$ is fiberwise reducible and
  $\gamma(\nu)^2=\id_S$. Chiral projectors can be constructed from the product structure
$\cR=\gamma(\nu)$, which satisfies $\cR^t=(-1)^{\frac{d}{2}} \cR$ with
respect to any admissible pairing $\cB$ due to \eqref{voltranspose}
since $d$ is even. Thus $\iota= (-1)^{\frac{d}{2}}$ (see
\eqref{Rtranspose}) for any admissible $\cB$ and in particular
$\iota^\pm_0=(-1)^{\frac{d}{2}}$. In this case, we have $E_k(S_\pm)=S_\pm$,
where $S_\pm$ are the bundles of {\em symplectic Majorana-Weyl spinors} of positive
and negative chiralities (which are defined as the eigenbundles of $\gamma(\nu)$
corresponding to the eigenvalues $\pm 1$ of the latter). 
\item If $p-q\equiv_8 6$, then $\gamma_\ev$ is fiberwise reducible and
  $\gamma(\nu)^2=-\id_S$, so $\gamma(\nu)$ is a globally-defined complex
  structure on $S$. Spin projectors can be constructed using the product structure
$\cR=\gamma(\nu)\circ J$, where $J\in \Gamma(U,\cU_\gamma)$ is any 
locally-defined complex structure associated with the quaternionic structure.  The
eigenbundles of $\cR$ corresponding to the eigenvalues $\pm 1$ are denoted by
$S_\pm$ and $S_+$ is the bundle of {\em symplectic Majorana spinors}. We
have $\cR^t=-(-1)^{\frac{d}{2}}\cR$ where $~^t$ denotes the
$\cB_0^\epsilon$-transpose and hence $\iota_0^\epsilon= -
(-1)^{\frac{d}{2}}$ (independent of $\epsilon$).
\end{itemize} 
The pairing $\cB_0^+$ will also be denoted through $\cB_0$:
\be
\cB_0 \eqdef \cB_0^+~~
\ee
and will be called the {\em basic pairing}. The pairings $\cB^\epsilon_k$ for
$k=1\ldots 3$ will be called {\em the derived pairings}. Notice that relations
\eqref{cBkepsilon} and \eqref{cBpm} imply: 
\ben
\label{cBalphaepsilon}
\boxed{\cB_\alpha^\epsilon=\cB_0\circ (\id_S\otimes
\gamma(\nu)^{\frac{1-\epsilon}{2}}\circ J_\alpha)~~,~~\forall \alpha=0\ldots
3~~,~~\forall \epsilon=\pm 1}~~,
\een
so that it suffices to work with the basic pairing $\cB_0$. 

\paragraph{Properties of derived pairings in the quaternionic simple case.}
The type $\epsilon_k(\epsilon,d)$ and symmetry $\sigma_k(\epsilon,d)$ of $\cB_k^\epsilon$ ($k=1\ldots
3$) satisfy: 
\be
\epsilon_k(\epsilon,d)=\epsilon_0(\epsilon,d)~~,~~\sigma_k(\epsilon,d)=-\sigma_0(\epsilon,d)~~
\ee
while the isotropies of $\cB_k^\epsilon$ are as follows:
\begin{itemize}
\item When $p-q\equiv_8 4$, we have: \be
\iota_k^\epsilon=\iota_0^\epsilon= (-1)^{\frac{d}{2}}~~,~~\forall
k=1\ldots 3~~,~~\forall \epsilon\in \{-1,1\}~~.  \ee
\item When $p-q\equiv_8 6$, we have the following isotropies if we
define spin projectors using the product structure
$\cR=\gamma(\nu)\circ J_1$:
\begin{center}
\begin{tabular}{|c|c|c|c|} \hline $k$ & $1$ & $2$ & $3$ \\ \hline
$\iota_k^\pm/\iota_0^\pm$ & $+1$ & $-1$ & $-1$\\ \hline
\end{tabular}
\end{center}
\end{itemize}
Also notice the relations $J_1(S_\pm)= S_\pm$ and $J_2(S_\pm)= S_\mp$,
$J_3(S_\pm)= S_\mp$. 

\subsubsection{The quaternionic non-simple case ($p-q\equiv_8 5$)} In this
case, $\gamma$ is fiberwise non-injective and we have $\gamma(\nu)=\epsilon_\gamma \id_S$, where $\epsilon_\gamma\in \{-1,1\}$ is
the signature of $\gamma$. We have four admissible nondegenerate bilinear pairings (up to multiplication with a nowhere
vanishing function), given by:
\ben
\label{cBk}
\cB_k=\cB_0\circ (\id_S\otimes J_k)~~,~~\forall k=1\ldots 3~~,
\een
where $\cB_0$ is the {\em basic admissible pairing}, whose type
$\epsilon_0(d)$ and symmetry $\sigma_0(d)$ are given in the table below: 

\begin{center}
\begin{tabular}{|c|c|c|c|c|}
\hline
 $\begin{array}{c}d\\({\rm mod}~8)\end{array}$ &  $1$ & $3$ & $5$ & $7$ \\
\hline\hline
$\epsilon_0(d)$ & $+1$ & $-1$ & $+1$ & $-1$ \\
\hline
 $\sigma_0(d)$ &  $-1$ & $+1$ & $+1$& $-1$\\
\hline
\end{tabular}
\end{center}

\noindent In the quaternionic non-simple case, the restricted bundle morphism
$\gamma_\ev$ is fiberwise irreducible so the isotropy of admissible pairings
on $S$ is not defined. The type $\epsilon_k(d)$ and symmetry $\sigma_k(d)$ of
$\cB_k$ ($k=1\ldots 3$) are given by:
\be
\epsilon_k(d)=\epsilon_0(d)~~,~~\sigma_k(d)=-\sigma_0(d)~~.
\ee
Note that relation \eqref{cBk} implies: 
\ben
\label{cBalpha}
\boxed{\cB_\alpha=\cB_0\circ (\id_S\otimes J_\alpha)~~,~~\forall \alpha=0\ldots 3}~~,
\een
since $J_0=\id_S$. 

\subsubsection{The $\cB_0$-transpose of $J_\alpha$} 
In all sub-cases of the quaternionic case, the globally-defined endomorphisms 
$J_\alpha$ are $\cB_0^\epsilon$-orthogonal (see \cite{Okubo1, AC2}) for $p-q\equiv_8
4,6$ and $\cB_0$-orthogonal for $p-q\equiv_8 5$:  
\ben
\label{BJalpha}
\boxed{(J_\alpha)^{-t}=J_\alpha~\Longleftrightarrow~ 
\cB_0(J^{-1}_\alpha\xi,\xi')=\cB_0(\xi,J_\alpha\xi')~~,~~\forall \alpha=0\ldots 3}~~,
\een
where $~^t$ denotes the $\cB_0^\epsilon$-transpose (for any $\epsilon=\pm
1$) when $p-q\equiv_8 4,6$ or the $\cB_0$-transpose, when $p-q\equiv_8
5$. Since $J_k^2=-\id_S$, we have $J_k^{-1}=-J_k$ and hence:
\ben
\boxed{J_k^t=-J_k~~,~~\forall k=1\ldots 3}~~,
\een
which implies: 
\be
J^t=-J~~,~~\forall J\in \Gamma(U,\cU_\gamma)~~.
\ee
\subsubsection{Admissible pairings in the biquaternion formalism}

Possibly replacing $M$ with a sufficiently small open subset, we assume for simplicity of notation 
that $M$ is contractible and hence $J_i$ are globally defined. Passing to the complexified bundle $S_\C=S\otimes \cO_\C$, the bilinear pairings
$\cB_\alpha^\epsilon$ (for the quaternionic simple case) and $\cB_\alpha$(for
the quaternionic non-simple case) complexify to fiberwise $\C$-bilinear pairings
$\beta_\alpha^\epsilon$ and $\beta_\alpha$ on $S_\C$, respectively. Relations
\eqref{cBalphaepsilon} and \eqref{cBalpha} give: 
\be
\beta_\alpha^\epsilon=\cB_0^\epsilon\circ (\id_{S_\C}\otimes
\fJ_\alpha)~~(\mathrm{for}~ p-q\equiv_8 4, 6)~~\mathrm{and}~~\beta_\alpha=\cB_0\circ
(\id_{S_\C}\otimes \fJ_\alpha)~~(\mathrm{for}~p-q\equiv_8 5)~~.
\ee
Of course, the complexified basic pairing $\beta_0$ has the same symmetry
$\sigma_0$ as $\cB_0$. Using a $\beta_0$-orthogonal local frame of $S_\C$ when
$\sigma_0=+1$ and a  $\beta_0$-symplectic (i.e. Darboux) local frame of $S_\C$
when $\sigma_0=-1$, the various relations given above translate immediately
into matrix identities familiar from the supergravity literature. We refer the
reader to Section \ref{sec:examples} for an example of this. 

\section{Fierz identities for real pinors}
\label{sec:fierz} 

\subsection{Preparations}
\label{sec:fierzpreps}

Given an admissible fiberwise bilinear pairing $\cB$ on $S$, we define
endomorphisms $E_{\xi,\xi'}\in\Gamma(M,\End(S))\approx
\Hom_{\cinf}(\Gamma(M,S),\Gamma(M,S))$ through:
\be
E_{\xi,\xi'}(\xi'')\stackrel{\rm{def}}{=}\cB(\xi'',\xi')\xi~~,~~\forall \xi,\xi'\in \Gamma(M,S)~~.
\ee
It is easy to check the identities: 
\ben
\label{mainid}
E_{\xi_1,\xi_2}\circ E_{\xi_3,\xi_4}=\cB(\xi_3,\xi_2)E_{\xi_1,\xi_4}~~,~~\forall \xi_1,\xi_2,\xi_3,\xi_4 \in \Gamma(M,S)~~,
\een
as well as:
\ben
\label{traceid}
\tr(T\circ E_{\xi,\xi'})=\cB(T\xi,\xi')~~,~~\forall \xi,\xi' \in \Gamma(M,S)~~.
\een
For later reference, note that the bundle $\End(S)$ is endowed with
the natural nondegenerate and symmetric fiberwise bilinear pairing
$\langle~,~\rangle$ whose action on sections is the $\cinf$-bilinear
map given by:
\ben
\label{natbilform}
\langle A,B\rangle \eqdef \tr(A\circ B)\in \cinf~~,~~\forall A,B \in \Gamma(M,\End(S))~~.
\een
Notice that this pairing does not depend on any choice of pairing of
the bundle $S$ itself.  Given $T\in \Gamma(M,\End(S))$, we define the operators of {\em left
and right composition} with $T$ to be the following $\cinf$-linear
operators acting in $\Gamma(M,\End(S))$:
\ben
\label{LTdef}
L_T(A)\eqdef T\circ A~~,~~R_T(A)\eqdef A\circ T~~,~~\forall A\in \Gamma(M,\End(S))~~.
\een
Cyclicity of the trace implies that $L_T$ and
$R_T$ are adjoint with respect to $\langle~,~\rangle$:
\be
\langle L_T(A),B\rangle=\langle A, R_T(B)\rangle~~,~~\forall T, A, B \in \Gamma(M,\End(S))~~.
\ee

\subsection{Fierz identities for the normal case}
\label{sec:fierznormal}

In this subsection, we consider the normal case, which corresponds to
$p-q\equiv_80,1,2$. Since this was already discussed in \cite{ga1} and
\cite{ga2} (see also \cite{Rand, Okubo1, Okubo2, Okubo3}), we shall be
brief.

\subsubsection{The completeness relation}  
We start from the local completeness relation (equation (2.7) of \cite{Okubo2}, see also \cite{Okubo1, ga1, ga2}):
\ben
\label{RealOkuboCompleteness}
\sum_\Aord(\gamma^{-1}_A)_{jk}(\gamma_A)_{lm}=\frac{2^d}{N}\delta_{jm}\delta_{lk}~~,
\een
where $A$ runs over increasingly-ordered tuples of length $|A|=1\ldots
d$. Multiplying both sides of \eqref{RealOkuboCompleteness} by
$T_{kj}$ and summing over $j,k$ gives:

\paragraph{Proposition.} We have the {\em completeness relation for
the normal case}:
\ben
\label{realA}
\boxed{T=_{_U}\frac{N}{2^d}\sum_\Aord   \tr(\gamma^{-1}_A \circ T)\gamma_A~~,~~\forall T\in \Gamma(M,\End(S))}~~.
\een
Setting $T=E_{\xi,\xi'}$ in this relation gives the expansion:
\be
E_{\xi,\xi'}=_{_U}\frac{N}{2^d}\sum_\Aord   \tr(\gamma^{-1}_A \circ E_{\xi,\xi'})\gamma_A~~,
\ee
which also takes the following form upon using identities \eqref{traceid} and \eqref{typeA}:
\ben
\label{RealEExpFinal}
E_{\xi,\xi'}=_{_U}\frac{N}{2^d}\sum_\Aord   \epsilon_\cB^{|A|} \cB (\xi,\gamma^A \xi')\gamma_A~~.
\een

\subsubsection{The geometric Fierz identities} 
Relation \eqref{RealEExpFinal} implies that the inhomogeneous differential forms:
\be
\check{E}_{\xi,\xi'}\eqdef \left(\gamma|_{\Omega^\gamma(M)}\right)^{-1}(E_{\xi,\xi'}) \in \Omega^\gamma(M)~~
\ee
have the expansion:
\ben
\label{Egeomreal}
\boxed{\check{E}_{\xi,\xi'}=_{_U}\frac{N}{2^d}\sum_\Aord   \epsilon_\cB^{|A|} \cB (\xi,\gamma_A \xi')e^A_\gamma~~,
~~\forall \xi,\xi' \in\Gamma(M, S)}~~,
\een
where we used \eqref{eAgamma}. We shall use the notation:
\ben
\label{notationreal}
\check{E}_{\xi,\xi'}=\frac{N}{2^d}\sum_{k=0}^d \bcE^{(k)}_{\xi,\xi'}~~,
\een
where:
\be 
\bcE^{(k)}_{\xi,\xi'}\eqdef_{_U}\frac{1}{k!}\epsilon_{\cB}^k \cB(\xi,\gamma_{a_1...a_k}\xi')e_{\gamma}^{a_1...a_k} ~ \in \Omega^k(M)\cap \Omega^\gamma(M)~~.
\ee
The geometric Fierz identities amount to:
\ben
\label{Fierzreal}
\boxed{\check{E}_{\xi_1,\xi_2} \diamond \check{E}_{\xi_3,\xi_4} = 
\cB (\xi_3,\xi_2) \check{E}_{\xi_1,\xi_4}~~,~~\forall \xi_1,\xi_2,\xi_3,\xi_4 \in \Gamma(M,S)}~~,
\een
an equality which holds in $\Omega^\gamma(M)$. We refer the reader to \cite{ga1,ga2} for further details regarding this case. 

\subsection{Fierz identities for the almost complex case}
\label{sec:fierzalmostcomplex}

This is the case $p-q\equiv_8 3,7$~~ (which implies that $d=p+q$ is
odd). In this situation, we always have $\epsilon_{\cB_0}=-1$ (see
Section \ref{sec:pairings}) and :
\ben
\label{CN}
N=2\Delta=2^{\left[\frac{d}{2}\right]+1}~~.
\een
The Schur algebra $\S$ is isomorphic with the
$\R$-algebra $\C$ of complex numbers, while the $\cinf$-algebra
$\Gamma(M,\Sigma_\gamma)$ is spanned by the operators $\id_S$ and
$J\in \Gamma(M,\End(S))$, where $J$ is a complex structure on
$S$ which lies in the commutant of the image of $\gamma$:
\ben
\label{Jprops}
J^2=-1~~,~~[J,\gamma_m]_{-,\circ}=0~~,~~\forall m=1\ldots d ~\Longrightarrow~ [J,\gamma_A]_{-,\circ}=0~~{\rm for~all~}A~~.
\een
We have: 
\ben
\label{Jnu}
J=\gamma(\nu)=\gamma^{(d+1)}
\een
and $\nu\diamond \nu=-1$ (which, of course, implies the property
$J^2=-\id_S$ listed above). As shown in \cite{Okubo1}, there always exists a globally-defined
endomorphism $D\in \Gamma(M,\End(S))$ which satisfies
\eqref{Dgamma_acomm}, \eqref{Dsquare} and \eqref{DJacomm}.    
The space $\Gamma(M,S)$ admits a structure of $\cC^\infty(M,\C)$-module defined through:
\be
(\alpha+i\beta)\xi\eqdef \alpha\xi+\beta J(\xi)~~,~~\forall \alpha,\beta\in \cC^\infty(M,\C)~~,~~\forall \xi\in \Gamma(M,S)~~.
\ee
The module $\Gamma(M,\End_\C(S))$ of $\cC^\infty(M,\C)$-linear operators can be identified with the submodule of $\Gamma(M,\End(S))$
consisting of those $\cinf$-linear operators which commute with $J$:
\ben
\label{CEnd}
\Gamma(M,\End_\C(S))\equiv \{T\in \Gamma(M,\End(S))~|~[J,T]_{-,\circ}=0\}~~.
\een
The map: 
\be
\gamma: \Omega(M) \longrightarrow~\Gamma(M,\End(S))
\ee
is always injective, with image:
\be
\gamma(\Omega(M))=\Gamma(M,\End_\C(S))~~.
\ee
Its partial inverse is given by: 
\ben
\label{CPartialInverse}
\gamma^{-1}\eqdef (\gamma|^{\Gamma(M,\End_\C(S))})^{-1}:\Gamma(M,\End_\C(S))\rightarrow \Omega(M)~~.
\een

\subsubsection{Preparations} 
Consider the following $\cinf$-linear operator acting in $\Gamma(M,\End(S))$:
\be
\cJ(T)\eqdef J\circ T\circ J~~,~~\forall T\in \Gamma(M,\End(S))~~.
\ee
The identity $J^2=-\id_S$ implies: 
\be
\cJ^2=\id_{\Gamma(M,\End(S))}~~,
\ee
which shows that $\cJ$ is a product structure on the $\cinf$-module $\Gamma(M,\End(S))$.
On the other hand, direct computation using cyclicity of the trace
shows that $\cJ$ is self-adjoint with respect to the natural pairing
\eqref{natbilform} on $\End(S)$:
\be
\langle \cJ(A),B\rangle=\langle A,\cJ(B)\rangle~~,~~\forall A,B\in \Gamma(M,\End(S))~~.
\ee
It follows that the operators:
\be
\Pi_\pm\eqdef \frac{1}{2}(\id_{\Gamma(M,\End(S))}\pm \cJ)~~
\ee
are complementary $\langle~,~\rangle$-orthoprojectors on the space $\Gamma(M,\End(S))$. In particular, we have: 
\be
\Pi_\pm^2=\Pi_\pm~~,~~\Pi_+\circ \Pi_-=\Pi_-\circ \Pi_+~~,~~\Pi_++\Pi_-=\id_{\Gamma(M,\End(S))}~~
\ee
and the submodules: 
\ben
\label{Endpm}
\Gamma(M,\End(S))^\pm\eqdef \Pi_\pm(\Gamma(M,\End(S)))=\{T\in \Gamma(M,\End(S))~|~\cJ(T)=\pm T\}\subset \Gamma(M,\End(S))~~
\een
provide a $\langle~,~\rangle$-orthogonal direct sum decomposition:
\be
\Gamma(M,\End(S))=\Gamma(M,\End(S))^+\oplus \Gamma(M,\End(S))^- ~~.
\ee

\paragraph{Corollary.} Every $T\in \Gamma(M,\End(S))$ decomposes uniquely as: 
\ben
\label{CTdec}
T=T_+ + T_-~~{\rm with}~~T_\pm=\Pi_\pm(T)=\frac{1}{2}(T\pm \cJ(T))\in \Gamma(M,\End(S))^\pm~~.
\een

\paragraph{Proposition.} We have: 
\beqan
\Gamma(M,\End(S))^-&=&\{T\in \Gamma(M,\End(S))|[J,T]_{-,\circ}=0\}\equiv \Gamma(M,\End_\C(S))~~,\\
\Gamma(M,\End(S))^+&=&\{T\in \Gamma(M,\End(S))|[J,T]_{+,\circ}=0\}~~,\nn
\eeqan
where, in the first relation,  we used the identification \eqref{CEnd}. 

\

\noindent {\bf Proof.} It suffices to prove the first equality, since the second follows similarly. For this, consider the two inclusions in turn:

\

\noindent ($\subset$) If $[J,T]_{-,\circ}=0$ then direct computation shows that $\cJ(T)=-T$, so that $T\in \Gamma(M,\End(S))^-$ (see \eqref{Endpm}). 

\

\noindent ($\supset$) If $T\in \Gamma(M,\End(S))^-$, the relation $\cJ(T)=-T$ (see \eqref{Endpm}) reads:
\be
J\circ T\circ J=-T~~,
\ee
which implies $T\circ J=J\circ T$ upon composing with $J$ and using $J^2=-\id_S$. Thus $[J,T]_{-,\circ}=0$.

\paragraph{Proposition.} We have: 
\ben
\label{LDEnd}
L_D(\Gamma(M,\End(S))^\pm)= \Gamma(M,\End(S))^\mp~~,
\een
where $L_D$ is the $\cinf$-linear operator of left composition with $T$ (see \eqref{LTdef}). 

\

\noindent{\bf Proof.} Relation \eqref{DJacomm} implies: 
\be
L_D\circ \Pi_\pm=\Pi_\mp\circ L_D~~,
\ee
which immediately gives the conclusion. 

\

\noindent Since $\id_S\in \Gamma(M,\End(S))^-$ and $L_D(\id_S)=D$, the Proposition gives the following: 
\paragraph{Corollary.} We have:
\be
D\in\Gamma(M, \End_\C(S))^+~~,
\ee
as well as: 
\paragraph{Corollary.} Every $T\in \Gamma(M,\End(S))$ decomposes uniquely as:
\ben
\label{TDdec}
T=T_0+D \circ T_1~~{\rm with}~~T_0,T_1\in \Gamma(M,\End_\C(S))~~,
\een
where we used the identification \eqref{CEnd}~~.

\

\noindent {\bf Proof.} Follows immediately from \eqref{CTdec} and \eqref{LDEnd}.  

\

\paragraph{Proposition.} The following identities hold:
\ben
\label{gammaDTtrace}
\tr(\gamma^{-1}_A \circ D^{-1} \circ T)=0~~,~~\forall A, T\in \Gamma(M, \End_\C(S))~~.
\een

\noindent{\bf Proof.} Since $J$ commutes with $\gamma_A$ (see
\eqref{Jprops}), we have $\gamma_A\in \Gamma(M,\End(S))^-$. On the
other hand, we have $D^{-1}\circ T\in L_D(\Gamma(M,\End_\C(S)))\equiv
L_D(\Gamma(M,\End(S))^-)=\Gamma(M,\End(S))^+$ by \eqref{LDEnd}. The
conclusion now follows from the fact that $\Gamma(M,\End(S))^-$ and
$\Gamma(M,\End(S))^+$ are mutually orthogonal with respect to the
natural pairing \eqref{natbilform}.

\paragraph{Proposition.} For all $T\in \Gamma(M,\End(S))$ decomposed as in \eqref{TDdec}, we have:
\ben
\label{CtrgammaAT}
\tr(\gamma^{-1}_A \circ T)=\tr(\gamma^{-1}_A \circ T_0) ~~{\rm and}~~\tr(\gamma^{-1}_A \circ D^{-1} \circ T)=\tr(\gamma^{-1}_A \circ T_1)~~.
\een

\

\noindent {\bf Proof.} Follows immediately from \eqref{gammaDTtrace}.  

\

\noindent Notice the relations:
\ben
\label{DTacomm}
D \circ T =\pi(T)\circ D
\een
where $\pi$ is the parity signature acting in the $\Z_2$-graded $\cinf$-module
$\Gamma(M,\End_\C(S))=\gamma(\Omega(M))$ --- the $\Z_2$-grading on the later being defined by transport from
$\Omega(M)$ through the isomorphism $\gamma:\Omega(M)\stackrel{\sim}{\longrightarrow} \Gamma(M,\End_\C(S))$. 
Consider the unital associative and commutative (but {\em not} graded-commutative) $\Z_2$-graded algebra $\A=\{\alpha+\beta e|\alpha,\beta\in \R\}$ generated over $\R$ by an
odd element $\e$ which satisfies the single relation (cf. \eqref{Dsquare}):
\be
\e^2=(-1)^{\frac{p-q+1}{4}} 1 =\twopartdef{-1~,~}{p-q\equiv_8 3}{+1~,~}{p-q\equiv_8 7}~~.
\ee
Since $D$ satisfies \eqref{Dsquare}, the results above show that $\Gamma(M,\End(S))$ is a
{\em free} left $\Z_2$-graded module over the $\Z_2$-graded ring $\A\otimes_\R \cinf$ of $\A$-valued smooth functions defined on $M$, 
where left multiplication with $\e$ is given by $L_D$: 
\be
\e T\eqdef L_D(T)=D \circ T~~,~~\forall T\in \Gamma(M,\End(S))~~
\ee
and the $\Z_2$-grading is given by the decomposition: 
\be
\Gamma(M,\End(S))=\Gamma(M,\End(S))^\ev\oplus \Gamma(M,\End(S))^\odd~~,
\ee
with:
\beqa
\Gamma(M,\End(S))^\ev~&\eqdef& \Gamma(M,\End_\C(S))^\ev \oplus L_D(\Gamma(M,\End_\C(S))^\odd)~~,~~\\
\Gamma(M,\End(S))^\odd&\eqdef& \Gamma(M,\End_\C(S))^\odd\oplus L_D(\Gamma(M,\End_\C(S))^\ev)~~.
\eeqa
In fact, relation \eqref{DTacomm} implies that $(\Gamma(M,\End(S)),\circ)$ it is a unital $\Z_2$-graded $\A\otimes_\R\cinf$-algebra with internal multiplication given
by the composition $\circ$ of $\cinf$-linear operators acting in $\Gamma(M,S)$.  We have a unital  isomorphism of $\A\otimes_\R\cinf$-algebras (which maps $D$ into $\e{\hat \otimes}_\R\id_S$): 
\be
\Gamma(M,\End(S))\approx \A{\hat \otimes}_\R \Gamma(M,\End_\C(S))~~,
\ee
Here, $\Gamma(M,\End_\C(S))$ is viewed as a $\Z_2$-graded unital associative algebra with the $\Z_2$-grading induced via $\gamma$ from that of $(\Omega(M),\diamond)$ and
${\hat \otimes}_\R$ is the graded tensor product of $\Z_2$-graded $\R$-algebras. In particular, an $\R$-linear endomorphism $T\in \Gamma(M,\End(S))$ can be identified with: 
\be
T\equiv T_0+ \e \otimes T_1\in \A{\hat \otimes}_\R \Gamma(M,\End_\C(S))~~,
\ee
where $T_0, T_1\in \Gamma(M,\End_\C(S))$ are the components appearing in the decomposition
\eqref{TDdec}.

\subsubsection{The partial and full completeness relations}

The two identities below (which hold above any open subset $U\subset M$ carrying a local orthonormal coframe $e^m$ of $M$ and a local frame $\varepsilon^i$ of 
$S$) follow immediately from the results proved in \cite{Okubo1}:
\ben
\label{cx1}
\sum_\Aord(\gamma^{-1}_A)_{jk}(\gamma_A)_{lm} 
 =\frac{2^d}{N} \left( \delta_{jm}\delta_{lk} - J_{jm} J_{lk} \right)~~
\een
and:
\ben
\label{cx2}
\sum_\Aord\left[(\gamma^{-1}_A)_{jk}(\gamma_A)_{lm} + \gamma^{-1}_A (D^{-1})_{jk}(D\gamma_A)_{lm} \right]
 =\frac{2^{d+1}}{N}\delta_{jm}\delta_{lk}~~.
\een
These identities imply: 

\paragraph{Proposition.} We have the {\em partial completeness relation for the almost complex case}:
\ben
\label{cx1trace}
\boxed{\frac{2^{d+1}}{N}T=\frac{2^d}{\Delta}T=_{_U}\sum_\Aord   \tr(\gamma^{-1}_A \circ T)\gamma_A   ~~,~~\forall T\in \Gamma(M,\End_\C(S))}~~.
\een
and the {\em full completeness relation for the almost complex case}:
\ben
\label{cx2trace}
\boxed{\frac{2^{d+1}}{N}T =\frac{2^d}{\Delta} T=_{_U} \sum_\Aord   \left[ \tr(\gamma^{-1}_A \circ T)\gamma_A +  \tr(\gamma^{-1}_A \circ D^{-1} \circ T) D\circ \gamma_A \right]
~~,~~\forall T\in \Gamma(M,\End(S))}~~.
\een

\noindent{\bf Proof.} Multiplying both sides of \eqref{cx2} with $T_{jk}$ and summing over $j,k$ gives: 
\ben
\label{cx1bis}
\frac{2^d}{\Delta}\Pi_-(T)=\frac{2^d}{N}(T-\cJ(T))=_{_U}\sum_\Aord    \tr(\gamma^{-1}_A \circ T)\gamma_A~~,~~\forall T\in \Gamma(M,\End(S))~~,
\een
which implies \eqref{cx1trace} for $T\in \Gamma(M,\End_\C(S))\equiv \Gamma(M,\End(S))^-$. On the other hand, multiplying 
both sides of \eqref{cx2} with $T_{jk}$ and summing over $j,k$ gives \eqref{cx2trace}. 

\paragraph{Corollary.} For any $T\in \Gamma(M,\End(S))$~, we have:
\beqan
\label{T01Exp}
T_0 &=_{_U}&\frac{\Delta}{2^d}\sum_\Aord   \tr(\gamma^{-1}_A \circ T)\gamma_A=\frac{\Delta}{2^d}\sum_\Aord   \tr(\gamma^{-1}_A \circ T_0)\gamma_A ~~, \nn\\
T_1 &=_{_U}& \frac{\Delta}{2^d}\sum_\Aord   \tr(\gamma^{-1}_A \circ D^{-1}\circ T)\gamma_A=\frac{\Delta}{2^d}\sum_\Aord   \tr(\gamma^{-1}_A \circ T_1)\gamma_A ~~,
\eeqan
where $T_0$ and $T_1$ are defined as in \eqref{TDdec}.  

\

\noindent{\bf Proof.} The equalities follow immediately from \eqref{cx2trace} and from the decomposition \eqref{TDdec}. 
In the second form of the expansions, we used \eqref{CtrgammaAT}.

\subsubsection{The Fierz identities}
Relations \eqref{TDdec} show that we can decompose $E_{\xi,\xi'}$ uniquely as:
\be
E_{\xi,\xi'}=E^{(0)}_{\xi,\xi'}+D\circ E^{(1)}_{\xi,\xi'}~~,
\ee
where:
\beqan
\label{E01expansion}
E^{(0)}_{\xi,\xi'} &=_{_U}& \frac{\Delta}{2^d}\sum_\Aord   \tr(\gamma^{-1}_A \circ E_{\xi,\xi'} )\gamma_A \in \Gamma(M,\End_\C(S))~~, \\
E^{(1)}_{\xi,\xi'} &=_{_U}& \frac{\Delta}{2^d}\sum_\Aord   \tr(\gamma^{-1}_A \circ
D^{-1} \circ E_{\xi,\xi'})\gamma_A \in \Gamma(M,\End_\C(S))~~ \nn
\eeqan
and (since $E^{(\alpha)}_{\xi,\xi'}\in \Gamma(M,\End_\C(S))$):
\ben
\label{DEacomm}
D\circ E^{(\alpha)}_{\xi,\xi'}=\pi(E^{(\alpha)}_{\xi,\xi'})\circ D~~,~~\forall\alpha=0, 1~~,
\een
where we used \eqref{DTacomm}. Hence identity \eqref{mainid} takes the form: 
\be
\left(E^{(0)}_{\xi_1,\xi_2}+D \circ E^{(1)}_{\xi_1,\xi_2}\right)\circ \left(E^{(0)}_{\xi_3,\xi_4}+D \circ E^{(1)}_{\xi_3,\xi_4}\right)
=\cB_0(\xi_3,\xi_2)\left(E^{(0)}_{\xi_1,\xi_4}+D \circ E^{(1)}_{\xi_1,\xi_4}\right)~~.\nn
\ee
Using \eqref{DEacomm} and \eqref{Dsquare}, this becomes: 
\beqa
E^{(0)}_{\xi_1,\xi_2}\circ E^{(0)}_{\xi_3,\xi_4} + (-1)^{\frac{p-q+1}{4}} \pi(E^{(1)}_{\xi_1,\xi_2})\circ E^{(1)}_{\xi_3,\xi_4} &+&
D \circ \left( \pi(E^{(0)}_{\xi_1,\xi_2})\circ E^{(1)}_{\xi_3,\xi_4} + E^{(1)}_{\xi_1,\xi_2}\circ E^{(0)}_{\xi_3,\xi_4} \right)=\nn\\
&=&\cB_0(\xi_3,\xi_2)\left(E^{(0)}_{\xi_1,\xi_4}+D \circ E^{(1)}_{\xi_1,\xi_4}\right)~~\nn
\eeqa
Separating components according to the decomposition \eqref{TDdec} gives:
\beqan
\label{Randalcx}
E^{(0)}_{\xi_1,\xi_2}\circ E^{(0)}_{\xi_3,\xi_4} + (-1)^{\frac{p-q+1}{4}} \pi(E^{(1)}_{\xi_1,\xi_2})\circ E^{(1)}_{\xi_3,\xi_4} &=& \cB_0(\xi_3,\xi_2)E^{(0)}_{\xi_1,\xi_4}~~,\nn\\
\pi(E^{(0)}_{\xi_1,\xi_2})\circ E^{(1)}_{\xi_3,\xi_4} + E^{(1)}_{\xi_1,\xi_2}\circ E^{(0)}_{\xi_3,\xi_4} &=& \cB_0(\xi_3,\xi_2)E^{(1)}_{\xi_1,\xi_4}~~.
\eeqan
Using \eqref{traceid}, \eqref{typeA}, \eqref{BD} and the fact that $\epsilon_0=-1$, we compute:
\beqa
\tr(\gamma^{-1}_A \circ E_{\xi,\xi'})=\cB_0(\gamma^{-1}_A \xi,\xi') &=& (-1)^{|A|}\cB_0(\xi,\gamma^A\xi')~~,\nn\\
\tr(\gamma^{-1}_A \circ D^{-1} \circ E_{\xi,\xi'})=\cB_0((\gamma^{-1}_A \circ D^{-1}) \xi,\xi') &=& (-1)^{|A|}\cB_0(D^{-1}\xi,\gamma^A\xi')= \nn\\
&=&
(-1)^{\frac{p-q+1}{4}} (-1)^{|A|}\cB_0(\xi,(D\circ \gamma^A)\xi')~~,\nn
\eeqa
which allows us to rewrite \eqref{E01expansion} as: 
\beqan
\label{E01final}
E^{(0)}_{\xi,\xi'} &=_{_U}& \frac{\Delta}{2^d}\sum_\Aord   (-1)^{|A|}\cB_0(\xi,\gamma^A\xi')\gamma_A~~, \nn\\
E^{(1)}_{\xi,\xi'} &=_{_U}& \frac{\Delta}{2^d}\sum_\Aord   (-1)^{\frac{p-q+1}{4}}(-1)^{|A|}\cB_0(\xi,D\circ\gamma^A\xi')\gamma_A ~~. 
\eeqan

\subsubsection{Geometric algebra formulation}
Using the partial inverse \eqref{CPartialInverse}, we define:
\be
\check{E}^{(\alpha)}_{\xi,\xi'}\eqdef \gamma^{-1}(E^{(\alpha)}_{\xi,\xi'})~\in\Omega(M)~~,~~\forall \alpha=0,1~~.\nn
\ee
Applying $\gamma^{-1}$ shows that 
identities \eqref{Randalcx} are equivalent with the {\em geometric algebra form of the Fierz identities in the almost complex case}:
\ben
\label{cRandalcx}
\addtolength{\fboxsep}{3pt}
\boxed{
\begin{split}
& \check{E}^{(0)}_{\xi_1,\xi_2}\diamond \check{E}^{(0)}_{\xi_3,\xi_4} + (-1)^{\frac{p-q+1}{4}} \pi(\check{E}^{(1)}_{\xi_1,\xi_2})\diamond \check{E}^{(1)}_{\xi_3,\xi_4} 
= \cB_0(\xi_3,\xi_2)\check{E}^{(0)}_{\xi_1,\xi_4}~~,\\
& \pi(\check{E}^{(0)}_{\xi_1,\xi_2})\diamond \check{E}^{(1)}_{\xi_3,\xi_4} +  \check{E}^{(1)}_{\xi_1,\xi_2}\diamond \check{E}^{(0)}_{\xi_3,\xi_4} 
= \cB_0(\xi_3,\xi_2)\check{E}^{(1)}_{\xi_1,\xi_4}~~,
\end{split}}
\een
while \eqref{E01final} give the expansions: 
\ben
\label{E01geomfinal}
\addtolength{\fboxsep}{2pt}
\boxed{
\begin{split}
\check{E}^{(0)}_{\xi,\xi'} &=_{_U} \frac{\Delta}{2^d}\sum_\Aord   (-1)^{|A|}\cB_0(\xi, \gamma_A\xi')e^A~, \\
\check{E}^{(1)}_{\xi,\xi'} &=_{_U} \frac{\Delta}{2^d}\sum_\Aord   (-1)^{\frac{p-q+1}{4}} (-1)^{|A|}\cB_0(\xi, (D\circ \gamma_A)\xi')e^A ~. 
\end{split}}
\een
Later on, we shall also use the notations:
\beqa
\bcE^{(0,k)}_{\xi,\xi'}&\eqdef_{_U}& \frac{1}{k!} (-1)^{k} \cB_0(\xi, \gamma_{a_1...a_k}\xi')e^{a_1...a_k} ~ \in \Omega^k(M)\\
\bcE^{(1,k)}_{\xi,\xi'}&\eqdef_{_U}& \frac{1}{k!} (-1)^{\frac{p-q+1}{4}} (-1)^{k}\cB_0(\xi, (D\circ \gamma_{a_1...a_k})\xi')e^{a_1...a_k} ~ \in \Omega^k(M)~~,  
\eeqa
so that: 
\ben
\label{notationcomplex}
\check{E}^{(\alpha)}_{\xi,\xi'} =\frac{\Delta}{2^d}\sum_{k=0}^d \bcE^{(\alpha,k)}_{\xi,\xi'}~~,~~\forall \alpha=0,1~~.
\een

\noindent Consider the unital and associative $\Z_2$-graded $\cC^\infty(M,\R)$-algebra
defined through: 
\be
\Omega_\A(M)\eqdef \A{\hat \otimes}_\R (\Omega(M),\diamond)~~,
\ee
where the \KA algebra of $M$ is viewed as a $\Z_2$-graded algebra. Denoting
the composition of $\Omega_\A(M)$ again by $\diamond$ for simplicity of
notation, equations \eqref{E01geomfinal} can be written in the equivalent
form:
\ben
\label{CFierzGeom}
\check{E}_{\xi_1,\xi_2}\diamond \check{E}_{\xi_3,\xi_4}=\cB_0(\xi_3,\xi_2)\check{E}_{\xi_1,\xi_4}~~, \nn
\een
where we defined: 
\ben
\label{CcheckEgeom}
\check{E}_{\xi,\xi'}\eqdef \check{E}_{\xi,\xi'}^{(0)}+\e\otimes
\check{E}_{\xi,\xi'}^{(1)}\in \Omega_\A(M)~~.
\een

\subsection{Fierz identities for the quaternionic case}
\label{sec:fierzquaternionic}

In this case, the Schur algebra is isomorphic with the $\R$-algebra $\H$ of quaternions, being spanned over $\R$ by
four linearly-independent elements $J_\alpha\in \Gamma(M,\End(S))$
($\alpha=0\ldots 3$), which we can take to correspond to the
quaternion units --- where we once again (possibly replacing $M$ with a sufficiently small open subset --- we assume for simplicity of notation that $M$ is contractible and $J_i$ are globally-defined. 
Hence $J_0=\id_S$ while $J_1,J_2,J_3$ satisfy:
\ben
\label{JiJj}
J_i\circ J_j=-\delta_{ij} J_0+\epsilon_{ijk} J_k~~,~~\forall i,j,k =1\ldots 3~\Longrightarrow~ [J_i,J_j]_{+,\circ}=0~~.
\een
This makes $S$ into a left $\H$-module through:
\be
u\xi=\sum_{\alpha=0}^3 u_\alpha J_\alpha(\xi)~~,~~\forall u=\sum_{\alpha=0}^3 u_\alpha \j_\alpha~\in \H~~,\nn
\ee
where $u_\alpha\in\R$ ~ and $\j_\alpha$~ are quaternion units. The
space $\Gamma(M,\End_\H(S))$ of globally-defined
$\Sigma_\gamma$-linear endomorphisms of this module can be identified
with the subspace of those $\cinf$-linear operators acting in
$\Gamma(M,S)$ which commute with $J_1,J_2$ and $J_3$:
\ben
\label{EndH}
\Gamma(M,\End_\H(S))\equiv\{T\in \Gamma(M,\End(S))~|~[T,J_i]_{-,\circ}=0~~,~~\forall i=1\ldots 3\}~~.
\een

\subsubsection{Preparations}
Consider the following $\cinf$-linear operators $\cJ_\alpha$ acting in the space $\Gamma(M,\End(S))$: 
\be
\cJ_\alpha(T)\eqdef J_\alpha\circ T\circ J_\alpha~~,~~\forall i=1\ldots 3~~.
\ee
We obviously have $\cJ_0=\id_{\Gamma(M,\End(S))}$. Relations \eqref{JiJj} imply: 
\be
\cJ_i^2=\cJ_0~~,~~[\cJ_i,\cJ_j]_{-,\circ}=0~~,~~\forall i\neq j~~
\ee
as well as: 
\be
\cJ_i\circ \cJ_j=\delta_{ij}\cJ_0-|\epsilon_{ijk}|\cJ_k~~\forall i,j=1\ldots 3~~,
\ee
i.e.:
\be
\cJ_1\circ \cJ_2=\cJ_2\circ \cJ_1=-\cJ_3~~,~~\cJ_2\circ \cJ_3=\cJ_3\circ \cJ_2=-\cJ_1~~,~~\cJ_1\circ \cJ_3=\cJ_3\circ \cJ_1=-\cJ_2~~.
\ee
Let us define: 
\be
\cJ\eqdef \sum_{k=1}^3{\cJ_k}~~,~~\cR\eqdef \frac{1}{2}(\cJ_0+\cJ)~~.
\ee
Using the relations above, we compute: 
\be
\cJ^2=3\cJ_0-2\cJ~\Longleftrightarrow~ \cR^2=\cJ_0~~.
\ee
This shows that $\cR$ is a product structure on the $\cinf$-module
$\Gamma(M,\End(S))$. An easy computation using cyclicity of the trace
shows that $\cJ$ is selfadjoint with respect to the pairing
\eqref{natbilform}:
\be
\langle \cJ(A),B\rangle=\langle A, \cJ(B)\rangle~~,~~\forall A,B\in \Gamma(M,\End(S))~~.
\ee
It follows that the $\cinf$-linear operators $\Pi_\pm\in \End_{\cinf}(\Gamma(M,\End(S)))$ defined through:
\be
\Pi_\pm\eqdef \frac{1}{2}(\cJ_0\pm \cR)~\Longleftrightarrow~ \Pi_+=\frac{1}{4}(3\cJ_0+\cJ)~~\mathrm{and}~~\Pi_-=\frac{1}{4}(\cJ_0-\cJ)~~
\ee
are complementary $\langle~,~\rangle$-orthoprojectors. In particular, we have:
\be
\Pi_\pm^2=\Pi_\pm~~,~~\Pi_+\circ \Pi_-=\Pi_-\circ \Pi_+=0~~,~~\Pi_++\Pi_-=\cJ_0~~
\ee
and the $\cinf$-submodules: 
\be
\Gamma(M, \End(S))^\pm\eqdef \Pi_\pm(\Gamma(M,\End(S))=\{T\in \Gamma(M,\End(S))|\cR(T)=\pm T\}\subset \Gamma(M,\End(S))
\ee
give an $\langle~,~\rangle$-orthogonal direct sum decomposition of the $\cinf$-module $\Gamma(M,\End(S))$:
\be
\Gamma(M,\End(S))=\Gamma(M,\End(S))^+\oplus \Gamma(M,\End(S))^-~~.
\ee
Notice the characterizations: 
\beqa
&&\Gamma(M,\End(S))^+=\{T\in \Gamma(M,\End(S))|\cJ(T)=T\}~~,\\
&&\Gamma(M,\End(S))^-=\{T\in \Gamma(M,\End(S))|\cJ(T)=-3T\}~~.
\eeqa
Using identities \eqref{JiJj}, we compute:
\be
\sum_{i=1}^3J_i\circ J_j\circ J_i=-J_j+\epsilon_{ijk}J_k\circ J_i~~,~~\forall i,j=1\ldots 3~~,
\ee
a relation which implies:
\ben
\label{QP1}
\cJ(J_i)=J_i~\Longleftrightarrow~ J_i\in \Gamma(M,\End(S))^+~~,~~\forall i=1\ldots 3~~. 
\een
For any $T\in \Gamma(M,\End(S))$, we define: 
\be
T_\pm\eqdef \Pi_\pm (T)~\Longrightarrow~ T=T_+ + T_-~~.
\ee
Then: 
\ben
\label{Tpm}
T_+=\frac{1}{4}(3T+\cJ(T))~~,~~T_-=\frac{1}{4}(T-\cJ(T))~~.
\een

\paragraph{Proposition.} We have: 
\be
\Gamma(M,\End(S))^-=\{T\in \Gamma(M,\End(S))~|~[T,J_i]_{-,\circ}=0~,~\forall i=1\ldots 3~\}\equiv \Gamma(M,\End_\H(S))~~,
\ee
where we used the identification \eqref{EndH}. 

\

\noindent{\bf Proof.}

\noindent ($\subset$) Direct computation using \eqref{JiJj} gives:
\be
J_i \circ T_- =T_- \circ J_i =\frac{1}{4}([J_i,T]_{+,\circ}+\epsilon_{ijk}J_j \circ T \circ J_k)~~,
~~\forall T\in \Gamma(M,\End(S))~~,~~\forall i,j=1\ldots 3~~,\nn
\ee
which implies:
\ben
\label{QP2}
[J_i,T_-]_{-,\circ}=0~~,~~\forall T\in \Gamma(M,\End(S))~~,~~\forall i=1\ldots 3~~.
\een
For $T\in \Gamma(M,\End(S))^-$, we have $T_-=T$ and \eqref{QP2} shows that $T$ commutes with all $J_i$. 

\

\noindent ($\supset$) Let $T\in \Gamma(M,\End(S))$ satisfy
$[T,J_i]_{-,\circ}=0 ~,~ \forall i=1\ldots 3$. Using these commutation
relations as well as the identities $J_i^2=-\id_S$, equation
\eqref{Tpm} gives $T_-=T$, i.e. $T=\Pi_-(T)\in \Gamma(M,\End(S))^-$.

\

\paragraph{Proposition.} The $\cinf$-submodules $L_{J_i}(\Gamma(M,\End(S)))$ are
contained in $\Gamma(M,\End(S))^+$, have trivial pairwise
intersection and are mutually orthogonal with respect to the pairing
$\langle~,~\rangle$. Hence they give the
$\langle~,~\rangle$-orthogonal direct sum decomposition:
\be
\Gamma(M,\End_\H(S))^+=\oplus_{i=1}^3 L_{J_i}(\Gamma(M,\End_\H(S)))~~,\nn
\ee
where we have identified $\Gamma(M,\End_\H(S))\equiv \Gamma(M,\End(S))^-$. 

\

\noindent{\bf Proof.} 

\noindent ($L_{J_i}(\Gamma(M,\End(S)))$ are contained in $\Gamma(M,\End(S))^+$). 
For any $T\in\Gamma(M,\End_\H(S))$,  we have $[T,J_k]_{-,\circ}=0~,~\forall k=1\ldots 3$, which gives:
\be
(J_i\circ T)_-=\frac{1}{4}(J_i \circ T-\sum_{j=1}^{3} J_j \circ J_i \circ T \circ J_j )=\frac{1}{4}(J_i-\cJ(J_i))\circ T=\Pi_-(J_i)\circ T=0 
~\Longrightarrow~ J_i \circ T \in \Gamma(M,\End_\H(S))^+~~\nn
\ee
where we used \eqref{QP1}, which implies $\Pi_-(J_i)=0$. Thus $L_{J_i}(\Gamma(M,\End(S)))$ are contained in $\Gamma(M,\End(S))^+$:
\be
J_i(\Gamma(M,\End_\H(S)))\subset \Gamma(M,\End_\H(S))^+~~.
\ee
\noindent (trivial mutual intersection). Let $1\leq i\neq j\leq 3$. If ~$T\in L_{J_i}(\Gamma(M,\End_\H(S)))\cap 
L_{J_j}(\Gamma(M,\End_\H(S)))$, then $T=J_i\circ A=J_j\circ B$ for some $A,B\in \Gamma(M,\End_\H(S))$. Composing
with $J_i$ from the left and using $J_i^2=-\id_{S}$ gives:
\be
A=-J_i\circ J_j\circ B=\epsilon_{ijk} J_k\circ B\in \Gamma(M,\End(S))^+~~,
\ee
where we used identities \eqref{JiJj} (with $i\neq j$) as well as property \eqref{QP1}. Since
$A\in \Gamma(M,\End_\H(S))\equiv \Gamma(M,\End(S))^-$, this implies
$A\in \Gamma(M,\End(S))^-\cap \Gamma(M,\End(S))^+=\{0\}$, where we used the fact
that $\Gamma(M,\End(S))^-$ and $\Gamma(M,\End(S))^+$ are complementary submodules of
$\Gamma(M,\End(S))$. It follows that $A=0$ and hence $T=J_i\circ
A=0$. Therefore, we have:
\be
L_{J_i}(\Gamma(M,\End(S)))\cap L_{J_j}(\Gamma(M,\End(S)))=\{0\}~,~~\forall~1\leq i\neq j\leq 3 ~~.
\ee

\

\noindent ($\langle~,~\rangle$-orthogonality).  For any $A,B\in \Gamma(M,\End_\H(S))$, we have 
$\tr(J_i\circ A\circ J_j\circ B)=\tr(J_i\circ J_j\circ A\circ B)$ since $J_j$ and $A$ commute. Using relations \eqref{JiJj}, we 
compute: 
\ben
\label{trJiJjAB}
\tr(J_i\circ J_j \circ A\circ B)=\delta_{ij}\tr(A\circ B)-\epsilon_{ijk}\tr(J_k \circ A\circ B)~~.
\een
On the other hand, the left hand side is symmetric in $i$ and $j$: 
\be
\tr(J_i\circ J_j \circ A\circ B)=\tr(A\circ B\circ J_i\circ J_j)=\tr(J_j\circ A\circ B\circ J_i)=\tr(J_j\circ J_i\circ A\circ B)
\ee
where we used that fact that $A$ and $B$ commute with $J_1, J_2$ and $J_3$ as well as cyclicity of the trace. Hence $\tr(J_i\circ J_j \circ A\circ B)=
\frac{1}{2}(\tr(J_i\circ J_j \circ A\circ B)+\tr(J_j\circ J_i\circ A\circ B))$ and 
\eqref{trJiJjAB} gives: 
\be
\tr(J_i\circ J_j \circ A\circ B)=\delta_{ij}\tr(A\circ B)~\Longrightarrow~ \tr(J_i\circ A\circ J_j\circ B)=\delta_{ij}\tr(A\circ B)~~,
\ee
for all $A,B\in\Gamma(M,\End_\H(S))$ and for all $i,j=1\ldots 3$. 
This shows that the subspaces $L_{J_i}(\Gamma(M, \End(S)))$ are mutually orthogonal with respect to the pairing $\langle~,~\rangle$.

\

\noindent The following consequence of the proposition is obvious. 

\paragraph{Corollary.}  Any $T\in \Gamma(M,\End(S))$ decomposes uniquely as:
\ben
\label{Tdec}
T=\sum_{\alpha=0}^3J_\alpha\circ T_\alpha=T_0 + J_1\circ T_1 + J_2\circ T_2 + J_3\circ T_3 ~~,
\een
where ~$T_\alpha\in\Gamma(M,\End(S))^-\equiv\Gamma(M,\End_\H(S))~,~\forall\alpha=0\ldots 3$. 

\

\noindent Hence $\End(S)$ is (as expected) a bundle of free modules over the Schur algebra, while $\Gamma(M,\End(S))$ is a free left module over the algebra $\Gamma(M,\Sigma_\gamma)\subset \Gamma(M,\End(S))$, 
where left multiplication with  $u\in \Gamma(M,\Sigma_\gamma)$ is given by $L_u$: 
\be
u T\eqdef L_u(T)=u\circ T~~,~~\forall T\in \Gamma(M,\End(S))~~.
\ee
In fact, $(\Gamma(M,\End(S)),\circ)$ is a left $\Gamma(M,\Sigma_\gamma)$-algebra whose internal multiplication is given by
composition of $\R$-linear endomorphisms of $S$.  We have a unital  isomorphism of $\Gamma(M,\Sigma_\gamma)$-algebras: 
\be
\Gamma(M,\End(S))\approx \Gamma(M,\Sigma_\gamma) \otimes_\cinf \Gamma(M,\End_\H(S))~~
\ee
which maps $J_\alpha$ into $J_\alpha\otimes \id_S$. In particular, any endomorphism $T\in \Gamma(M,\End(S))$ can be
identified with: 
\be
T\equiv \sum_{\alpha=0}^3 J_\alpha\otimes T_\alpha\in\Gamma(M,\Sigma_\gamma) \otimes_\cinf \Gamma(M,\End_\H(S)) ~~,
\ee
where $T_\alpha\in \Gamma(M,\End_\H(S))$ are the components appearing in decomposition
\eqref{Tdec}. 

\paragraph{Proposition.} We have: 
\ben
\label{trJiT}
\tr(J_i \circ T)=0~~,~~\forall T\in \Gamma(M,\End_\H(S))\equiv \Gamma(M,\End(S))^- ~~,~~\forall i=1\ldots 3~~.
\een
\paragraph{Proof.} The statement follows immediately from the fact
that $J_i\in \Gamma(M,\End(S))^+$ (see \eqref{QP1}) together with the fact
that $\Gamma(M,\End(S))^+$ and $\Gamma(M,\End_\H(S))\equiv \Gamma(M,\End(S))^-$ are
$\langle~,~\rangle$-orthogonal. The statement can also be proved
directly through the following argument. If $T\in\Gamma(M,\End_\H(S))$, then
$[J_i,T]_{-,\circ}=0~,~\forall i=1\ldots 3$. Relation \eqref{JiJj}
implies:
\be
J_i\circ J_j+J_j\circ J_i=-2\delta_{ij} \id_S~~.\nn
\ee
Multiplying the above with $T$ from the left and taking the trace gives:
\be
\tr(T\circ J_i\circ J_j)+\tr(T\circ J_j\circ J_i)=-2\delta_{ij}\tr(T)~~.\nn
\ee
This implies:
\ben
\label{Q4}
\tr(T\circ J_i\circ J_j)=\tr(T\circ J_j\circ J_i)=-\delta_{ij}\tr(T)~~,~~\forall T\in\Gamma(M,\End_\H(S))~~,
\een
where we noticed the identity $\tr(T\circ J_j\circ J_i)=\tr(J_i\circ
T\circ J_j)=\tr(J_i\circ J_j\circ T)$, which follows from cyclicity of
the trace and from the fact that $T$ commutes with $J_j$. Using
\eqref{JiJj}, we compute:
\be
\tr(T\circ J_i\circ J_j)=-\delta_{ij}\tr(T)+\epsilon_{ijk}\tr(T\circ J_k)~~. 
\ee
Hence $\epsilon_{ijk}\tr(T\circ J_k)=0~,~\forall i,j\in 1\ldots 3$, which implies $\tr(T\circ J_k)=\tr(J_k\circ T)=0$. 

\subsubsection{The partial and full completeness relations}
The two identities below (which hold above any open subset $U$ of $M$
supporting a local pseudo-orthonormal coframe $e^m$ of $(M,g)$ and a local frame $\varepsilon^i$ of $S$) follow
immediately from the results proved in \cite{Okubo1}:
\ben
\label{Q1}
\sum_\Aord (\gamma^{-1}_A)_{jk}(\gamma_A)_{lm}=\frac{2^d}{N}\left(\delta_{jm}\delta_{lk}-\sum_{a=1}^3 (J_a)_{jm}(J_a)_{lk}\right)~~
\een
and
\ben
\label{Q2}
\sum_\Aord(\gamma^{-1}_A)_{jk}(\gamma_A)_{lm}-\sum_{a=1}^3 (\gamma^{-1}_A \circ J_a)_{jk}(J_a\circ \gamma_A)_{lm}=\frac{2^{d+2}}{N}\delta_{jm}\delta_{lk}~~.
\een

\paragraph{Proposition.}
We have the {\em partial completeness relation for the quaternionic case}:
\ben
\label{Q1trace}
\boxed{\frac{2^{d+2}}{N}T=\frac{2^d}{\Delta} T=_{_U}\sum_\Aord   \tr(\gamma^{-1}_A \circ T)\gamma_A~~,~~\forall T\in \Gamma(M,\End_\H(S))~~}
\een
and the {\em full completeness relation for the quaternionic case}: 
\ben
\label{Q2trace}
\boxed{\frac{2^{d+2}}{N}T=\frac{2^d}{\Delta} T=_{_U}\sum_\Aord   \tr(\gamma^{-1}_A\circ T)\gamma_A - \sum_{k=1}^3 \sum_\Aord   \tr(\gamma^{-1}_A \circ J_k \circ T) J_k\circ \gamma_A 
~~,}
\een
for any $ T\in \Gamma(M,\End(S))$.

\noindent{\bf Proof.} Multiplying both sides of \eqref{Q1} with $T_{kj}$ and summing over $j,k$ gives the equivalent form: 
\ben
\label{Q1bis}
\frac{2^d}{\Delta}\Pi_-(T)= \frac{2^d}{N}(T-\cJ(T))=_{_U} \sum_\Aord   \tr\left(\gamma^{-1}_A \circ T\right)\gamma_A~~.
\een
This gives \eqref{Q1trace} for $T\in \Gamma(M,\End_\H(S))$. On the other hand, \eqref{Q2} implies 
\eqref{Q2trace} upon multiplying with the components $T_{kj}$ and summing over $j,k$. 

\paragraph{Corollary.} For ~$\forall\alpha=0\ldots 3$~ we have the expansion: 
\ben
\label{QCor}
T_\alpha=_{_U}\frac{\Delta}{2^d}\sum_\Aord   \tr (\gamma^{-1}_A \circ J^{-1}_\alpha \circ
T)\gamma_A=\frac{\Delta}{2^d} \sum_\Aord   \tr (\gamma^{-1}_A\circ
T_\alpha)\gamma_A~~,~~\forall T\in
\Gamma(M,\End(S))~~.
\een
\noindent{\bf Proof.} Combine the previous proposition with the identity:
\ben
\label{gammaJTtrace}
\tr (\gamma^{-1}_A \circ J_\alpha \circ T)=- \tr (\gamma^{-1}_A \circ T_\alpha)~~,
\een
which follows from \eqref{Tdec} upon using \eqref{JiJj} and property \eqref{trJiT}. 

\subsubsection{The Fierz identities}

Relations \eqref{trJiT} and \eqref{QCor} show that the operators $E_{\xi,\xi'}\in \Gamma(M,\End(S))$ have the unique decompositions:
\be
E_{\xi,\xi'}=\sum_{\alpha=0}^3 J_\alpha\circ E^{(\alpha)}_{\xi,\xi'}~~,
\ee
where $E^{(\alpha)}_{\xi,\xi'}\in\Gamma(M,\End_\H(S))$ for all $\alpha=0\ldots 3$~ and: 
\ben
\label{decomp}
E^{(\alpha)}_{\xi,\xi'}=_{_U}\frac{\Delta}{2^d}\sum_\Aord   \tr(\gamma^{-1}_A\circ J^{-1}_\alpha \circ E_{\xi,\xi'})\gamma_A~~.
\een
Relation \eqref{traceid} gives:
\be
\tr(\gamma^{-1}_A\circ J^{-1}_\alpha \circ E_{\xi,\xi'})=\cB_0((\gamma^{-1}_A\circ J^{-1}_\alpha)\xi,\xi')~~,
\ee
which in turn implies: 
\ben
\label{Ealphaexp}
E^{(\alpha)}_{\xi,\xi'}=_{_U}\frac{\Delta}{2^d}\sum_\Aord   \cB_0((\gamma^{-1}_A\circ J^{-1}_\alpha)\xi,\xi')\gamma_A~~.
\een
Thus:
\beqan
E^{(0)}_{\xi,\xi'} &=_{_U}&\frac{\Delta}{2^d}\sum_\Aord   \cB_0(\gamma^{-1}_A\xi,\xi')\gamma_A~~,\label{QR0}\\
E^{(i)}_{\xi,\xi'} &=_{_U}&\frac{\Delta}{2^d} \sum_\Aord   \cB_0((\gamma^{-1}_A\circ J^{-1}_i)\xi,\xi')\gamma_A~~,
~~\forall i=1\ldots 3~~\label{QRi}
\eeqan
which leads to the expansion: 
\ben
\label{RH}
E_{\xi,\xi'}=_{_U}\sum_{\alpha=0}^3 J_\alpha \circ E^{(\alpha)}_{\xi,\xi'}
=\frac{\Delta}{2^d}\sum_\Aord\sum_{\alpha=0}^3   \cB_0((\gamma^{-1}_A\circ J^{-1}_\alpha)\xi,\xi')J_\alpha\circ\gamma_A~~.
\een
Using \eqref{decomp} in \eqref{mainid} and relations \eqref{JiJj}, we find the {\em Fierz identities for the quaternionic case}: 
\beqan
\label{Rquat}
 E^{(0)}_{\xi_1,\xi_2} \circ E^{(0)}_{\xi_3,\xi_4} - \sum_{i=1}^3 E^{(i)}_{\xi_1,\xi_2} \circ E^{(i)}_{\xi_3,\xi_4} 
&=& \cB_0(\xi_3,\xi_2)E^{(0)}_{\xi_1,\xi_4}~~,\nn\\
 E^{(0)}_{\xi_1,\xi_2} \circ E^{(i)}_{\xi_3,\xi_4} + E^{(i)}_{\xi_1,\xi_2} \circ E^{(0)}_{\xi_3,\xi_4} 
+\sum_{j,k=1}^3 \epsilon_{ijk} E^{(j)}_{\xi_1,\xi_2} \circ E^{(k)}_{\xi_3,\xi_4} &=& \cB_0(\xi_3,\xi_2)E^{(i)}_{\xi_1,\xi_4}~~
~(i=1\ldots 3)~.~~~~~~~~~~~~~
\eeqan

Let us write \eqref{QR0} and \eqref{QRi} in an equivalent form
which is more convenient in applications.  Starting from relations
\eqref{typeA} and \eqref{BJalpha}, we compute:
\be
\cB_0((\gamma^{-1}_A\circ J^{-1}_\alpha)\xi,\xi')=\epsilon^{|A|}_{\cB_0}\cB_0(J^{-1}_\alpha\xi,\gamma^A\xi')
=\epsilon^{|A|}_{\cB_0}\cB_0(\xi,(J_\alpha\circ \gamma^A)\xi')~~,~~\forall \alpha=0\ldots 3~~.  \nn
\ee
This allows us to write \eqref{Ealphaexp} as:
\ben
\label{Ealphaexpfinal}
E^{(\alpha)}_{\xi,\xi'}=_{_U}\frac{\Delta}{2^d}\sum_\Aord   \epsilon^{|A|}_{\cB_0}\cB_0(\xi,(J_\alpha\circ \gamma^A)\xi')\gamma_A~~,
~~\forall \alpha=0\ldots 3~~,~~\forall \xi,\xi'\in \Gamma(M,S)~~.\nn
\een
Thus:
\beqan
\label{E01expfinal}
E^{(0)}_{\xi,\xi'} &=_{_U}& \frac{\Delta}{2^d}\sum_\Aord   \epsilon^{|A|}_{\cB_0}\cB_0(\xi, \gamma^A\xi')\gamma_A~~,\nn\\
E^{(i)}_{\xi,\xi'} &=_{_U}& \frac{\Delta}{2^d}\sum_\Aord   \epsilon^{|A|}_{\cB_0}\cB_0(\xi, (J_i\circ \gamma^A)\xi')\gamma_A~~,
~~\forall i=1\ldots 3 ~~
\eeqan
and we have the expansion:
\ben
\label{Eexpfinal}
E_{\xi,\xi'}=_{_U}\frac{\Delta}{2^d}\sum_\Aord\sum_{\alpha=0}^3   \epsilon^{|A|}_{\cB_0}\cB_0(\xi,(J_\alpha\circ \gamma^A)\xi')J_\alpha\circ \gamma_A~~.\nn
\een

\subsubsection{Geometric algebra formulation}

Since $E^{(\alpha)}_{\xi,\xi'} \in \Gamma(M,\End_\H(S))=\gamma(\Omega(M))$, applying $\gamma^{-1}$ to \eqref{Rquat} gives:
\ben
\label{RquatCheck}
\boxed{
\begin{split}
  \check{E}^{(0)}_{\xi_1,\xi_2} \diamond  \check{E}^{(0)}_{\xi_3,\xi_4} - \sum_{i=1}^3  \check{E}^{(i)}_{\xi_1,\xi_2} \diamond  \check{E}^{(i)}_{\xi_3,\xi_4} 
&= \cB_0(\xi_3,\xi_2) \check{E}^{(0)}_{\xi_1,\xi_4}~~,\nn\\
  \check{E}^{(0)}_{\xi_1,\xi_2} \diamond  \check{E}^{(i)}_{\xi_3,\xi_4} +  \check{E}^{(i)}_{\xi_1,\xi_2} \diamond  \check{E}^{(0)}_{\xi_3,\xi_4} 
+\sum_{j,k=1}^3 \epsilon_{ijk}  \check{E}^{(j)}_{\xi_1,\xi_2} \diamond  \check{E}^{(k)}_{\xi_3,\xi_4} &= \cB_0(\xi_3,\xi_2) \check{E}^{(i)}_{\xi_1,\xi_4}~~
~(i=1\ldots 3)~,
\end{split}
}
\een
while \eqref{Ealphaexpfinal} gives the expansions:
\ben
\label{checkEalphaexpfinal}
\check{E}^{(\alpha)}_{\xi,\xi'}=_{_U}\frac{\Delta}{2^d}\sum_\Aord   \epsilon^{|A|}_{\cB_0}\cB_0(\xi, (J_\alpha\circ \gamma_A)\xi')e^A_\gamma~~,
~~\forall \alpha=0\ldots 3~~,~~\forall \xi,\xi'\in \Gamma(M,S)~~,\nn
\een
i.e.:
\ben
\label{checkE01expfinal}
\boxed{
\begin{split}
\check{E}^{(0)}_{\xi,\xi'} &=_{_U} \frac{\Delta}{2^d} \sum_\Aord   \epsilon^{|A|}_{\cB_0}\cB_0(\xi, \gamma_A\xi')e^A_\gamma~~,~~\forall \xi,\xi'\in \Gamma(M,S)~~,\nn\\
\check{E}^{(i)}_{\xi,\xi'} &=_{_U}  \frac{\Delta}{2^d}\sum_\Aord   \epsilon^{|A|}_{\cB_0}\cB_0(\xi, (J_i\circ \gamma_A)\xi')e^A_\gamma~~,~~\forall i=1\ldots 3 ~~.
\end{split}}
\een
Later on, we shall also use the notation:
\beqa
\bcE^{(0,k)} &\eqdef_{_U}&  \frac{1}{k!} \epsilon^{k}_{\cB_0}\cB_0(\xi, \gamma_{a_1...a_k}\xi')e^{a_1...a_k}_\gamma ~ \in \Omega^k(M)\cap \Omega^\gamma(M)\\
\bcE^{(i,k)} &\eqdef_{_U}&  \frac{1}{k!} \epsilon^{k}_{\cB_0}\cB_0(\xi, (J_i\circ \gamma_{a_1...a_k})\xi')e^{a_1...a_k}_\gamma ~ \in \Omega^k(M)\cap \Omega^\gamma(M)~~,
\eeqa
so that:
\be
\check{E}^{(\alpha)}_{\xi,\xi'}=\frac{\Delta}{2^d}\sum_{k=0}^d \bcE^{(\alpha,k)}_{\xi,\xi'}~~.
\ee
Consider the $\cinf$-module of $\Sigma_\gamma$-valued forms:
\be
\Omega(M,\Sigma_\gamma)\eqdef \Gamma(M,\wedge T^\ast M\otimes \Sigma_\gamma)\approx \Gamma(M,\Sigma_\gamma)\otimes_\cinf \Omega(M)~~,
\ee
which we endow with the noncommutative $\Z_2$-graded algebra structure (with product denoted once again by $\diamond$) induced through the tensor product of algebras from 
the algebra structures of $\Gamma(M,\Sigma_\gamma)$ and of $(\Omega(M),\diamond)$:
\be
(\Omega(M,\Sigma_\gamma),\diamond)\eqdef (\Gamma(M,\Sigma_\gamma),\circ) {\hat \otimes}_\cinf (\Omega(M),\diamond)~~.
\ee
Thus
\be
(u\otimes \omega)\diamond(v\otimes \eta)=(u\circ v)\otimes (\omega\diamond\eta)~~\forall \omega,\eta\in\Omega(M)~,~~\forall u,v\in \Gamma(M,\Sigma_\gamma)~~.
\ee
We define: 
\ben
\label{quatcheckE}
\check{E}_{\xi,\xi'}\eqdef \sum_{\alpha=0}^3 J_\alpha \otimes \check{E}^{(\alpha)}_{\xi,\xi'}\in \Omega(M,\Sigma_\gamma)~~.
\een
Then \eqref{RquatCheck} is equivalent with:
\ben
\label{QuatFierzGeom}
\check{E}_{\xi_1,\xi_2}\diamond \check{E}_{\xi_3,\xi_4}=\cB_0(\xi_3,\xi_2)\check{E}_{\xi_1,\xi_4}~~, \nn
\een

\section{Examples}
\label{sec:examples}

In this section, we illustrate our approach to Fierz identities with three
examples belonging to the normal, real and quaternionic types.  In previous
work, we have already briefly exemplified the simple and non-simple normal cases of our formalism with two
other applications --- namely one real pinor in eight Euclidean dimensions \cite{ga1}
(simple normal type) and two real pinors in nine Euclidean dimensions
\cite{ga2} (non-simple normal type), giving some hints on the computer algebra
implementation used to perform the computations. As in our previous work, the explicit
form of the Fierz identities in the geometric algebra formulation can be extracted and analyzed by using a symbolic computation system such as
{\tt Ricci} \cite{Ricci} \footnote{In contradistinction with the case
studied in \cite{ga2} (where direct computation is prohibitive), some of the examples
discussed below can also be analyzed directly. However, we have used our
implementation in order to speed up and check some of the 
computations. Given the space limitations and the focus of the present paper,
we prefer to discuss the details of the implementation as well as other technical and
mathematical aspects in a separate publication.}.

\subsection{One real pinor in nine Euclidean dimensions (non-simple normal case)}
\label{sec:examplesnormal}

Consider first a single real pinor on a nine-dimensional Riemannian manifold
$(M,g)$, i.e.  $p=d=9,~q=0$. We have $p-q\equiv_8 1$, so this
example belongs to the normal non-simple case, for which the properties of
pinors were treated in Subsections
\ref{sec:pincasesnormal} and \ref{sec:pairingscasesnormal}. This situation arises, for example,
when using the cone or cylinder formalism \cite{ga2} to study ${\cal N}=1$
compactifications of M-theory on an eight-dimensional manifold by lifting the
problem to the nine-dimensional metric cone or metric cylinder $(M,g)$ constructed
over the compactification space\footnote{Such a construction was alluded
to --- though not implemented --- in \cite{Tsimpis}.}.  The pin bundle $S$ is
an $\R$-vector bundle of rank $N=2^{\left[\frac{d}{2}\right]}=16$. The real pin
representation $\gamma:\wedge T^\ast M\rightarrow \End(S)$ of $(M,g)$ is
fiberwise surjective but not fiberwise injective and we have $\gamma(\nu)=\epsilon_\gamma \id_S$, where
$\epsilon_\gamma\in\{-1,+1\}$ is the signature of $\gamma$. We shall choose
$\epsilon_\gamma=+1$. Since $d\equiv_8 1$, the discussion of
Subsection \ref{sec:pairingscasesnormal} shows that --- up to
multiplication by a nowhere-vanishing smooth function --- there is only one admissible
pairing $\cB$ on $S$, which is symmetric ($\sigma_\cB=+1$) and of type
$\epsilon_\cB=+1$. Upon multiplying with an appropriately-chosen nowhere
vanishing function, one can assume without loss of generality that $\cB$ is
positive-definite and thus a scalar product on $S$. We will henceforth denote
the corresponding norm by $||~||$. The isotropy $\iota_\cB$ is not defined, since no spin
projection exists in this case. 

Choosing the pin representation $\gamma$ with signature $\epsilon_\gamma=+ 1$,
we realize the \KA algebra $(\Omega^+(M),\diamond)$ of $(M,g)$ through the truncated model 
\footnote{See \cite{ga1} for a detailed description of truncated models. Here, $\bdiamond_+:\Omega^<(M)\longrightarrow \Omega^<(M)$ denotes the 
reduced geometric product, defined through
$\omega\bdiamond_+\eta=2P_<(P_+(\omega)\diamond P_+(\eta))$, for any inhomogeneous forms
$\omega$ and $\eta$, where $P_+(\omega)=\frac{1}{2}(\omega+\tilde\ast\omega)$ and $P_<(\omega)=\omega_<$.}
$(\Omega^<(M),\bdiamond_+)$, where $\Omega^<(M)=\oplus_{k=0}^4\Omega^k(M)$. We
are interested in pinor bilinears: 
\ben
\label{bE}
\bcE^{(k)}\eqdef
\frac{1}{k!}\cB(\xi,\gamma_{a_1...a_k}\xi')e^{a_1\ldots a_k}\in\Omega^k(M)~~,
~~\forall k\in\overline{0,9}~,
\een
where $\xi,\xi'\in \Gamma(M,S)$.  From a single pinor $\xi\in
\Gamma(M,S)$ (which we normalize through $||\xi||=1$) we can construct --- up
to twisted Hodge duality on $(M,g)$ --- the following nontrivial bilinears (a
scalar, a one-form and a four-form):
\ben
\label{forms9}
\boxed{\cB(\xi,\xi)=1~~,~~V\eqdef \bcE^{(1)}=\cB(\xi,\gamma_a\xi)e^a ~~
,~~\Phi\eqdef \bcE^{(4)}=\frac{1}{24} \cB(\xi,\gamma_{a_1\ldots a_4}\xi) e^{a_1\ldots a_4}~,}
\een
where we used identity \eqref{gammaAtranspose} with $\epsilon_\cB=+1$, which
implies that $ \cB(\xi, \gamma^{a_1\ldots a_k}\xi)$ and thus
$\bcE^{(k)}$ vanish unless
$k(k-1)\equiv_4 0\Leftrightarrow k\equiv_4 0,1 \Leftrightarrow
k=0,1,4,5,8,9$. We have also used the identity
$\gamma(\nu)=\gamma^{(9)}=\gamma^1\circ \ldots \circ \gamma^8=\id_S$ (which
holds since $\epsilon_\gamma=+1$), which implies the relations
$\bcE^{(9-k)}={\tilde \ast} \bcE^{(k)}$ for all $k=0,1,4,5,8,9$, where ${\tilde \ast}$ is the {\em twisted} Hodge operator.  This means that the
inhomogeneous differential form:
\be
\bcE\eqdef \sum_{k=0}^9 \bcE^{(k)}\in \Omega(M)~~
\ee
is twisted selfdual and thus belongs to the effective domain of definition
$\Omega^\gamma(M)=\Omega^+(M)$ of the morphism of $\cinf$-algebras
$\gamma:\Omega(M)\rightarrow \Gamma(M,\End(S))$. The lower truncation of
$\bcE$ is the inhomogeneous differential form:
\be
\bcE_<\eqdef  \sum_{k=0}^4 \bcE^{(k)}=1+V+\Phi \in \Omega^<(M)~~.
\ee
The truncated model of the Fierz algebra \footnote{See \cite{ga1} for the
  definition and properties of the Fierz algebra. This applies here
  since we are in the normal case. } admits a basis consisting of the
single element:
\be
\check{E}_<\eqdef \frac{N}{2^d}\bcE_<=\frac{1}{32}(1+V+\Phi)~~ .\nn
\ee  
We remind the reader of the relations:
\be
\tilde\ast\omega=\omega\diamond\nu~~,~~\ast\omega=\tau(\omega)\diamond\nu~~,~~\forall\omega\in\Omega(M)~~,
\ee
where $\ast$ is the ordinary Hodge operator.  Since in this case the volume
form $\nu$ is central (i.e. $\nu\diamond\omega=\omega\diamond\nu$ 
for any inhomogeneous differential form $\omega$) and since $\nu\diamond\nu=+1_M$, 
 the truncated Fierz identity (where --- for simplicity of notation --- we write $\bdiamond$ instead of $\bdiamond_+$) reads (see \cite{ga1}):
\be
\check{E}_<\bdiamond\check{E}_<=\frac{1}{2}\check{E}_< ~~\left( ~\Longleftrightarrow~ \check{E}\diamond\check{E}=\check{E}~\right)
\ee
and amounts to:
\ben
V\bdiamond V+\Phi\bdiamond\Phi+V\bdiamond\Phi+\Phi\bdiamond V=15+14V+14\Phi ~~.
\label{Fierz9}
\een
Expanding the $\bdiamond$-product into generalized products (see \cite{ga1}
for the definition of the latter), we find:
\beqan
\label{rel1}
  V \bdiamond V & =&||V||^2 ~~, \\
\label{rel2}
  V\bdiamond \Phi &= & \Phi \bdiamond V=\ast(V \wedge\Phi)~~, \\
\label{rel3}
  \Phi \bdiamond \Phi &=& ||\Phi||^2+\ast(\Phi\wedge\Phi)-\Phi\bigtriangleup_2\Phi~~ .
\eeqan
Equality \eqref{rel2} holds since
$\iota_V\Phi$ vanishes \footnote{Indeed, the three-form $\iota_V\Phi$
should be a bilinear in $\xi$ due to the Fierz relations but --- as
shown above --- one cannot construct any nontrivial three-form
bilinear in $\xi$ in this example.}. Using relations
\eqref{rel1}-\eqref{rel3} and separating \eqref{Fierz9} into its rank
components gives:
\ben
\label{Fierz9ranks}
||V||^2+||\Phi||^2=15 ~~,~~\ast(\Phi\wedge\Phi)=14V~~,~~2\ast(V\wedge\Phi)-\Phi\bigtriangleup_2\Phi=14\Phi~~.
\een
Let us define non-negative real-valued functions $a,c\in \cinf$ through: 
\beqan
\label{Rell1}
a \eqdef ||V||^2 ~~,~~ c \eqdef ||\Phi||^2 ~~. 
\eeqan
Since any expression quadrilinear in $\xi$ expands into $\xi$-bilinears and since \eqref{forms9} and their Hodge duals 
generate the $\cinf$-module of globally-defined inhomogeneous differential forms on $M$ which can be constructed as bilinears in $\xi$, 
there must exist functions $b,f,e\in \cinf$ such that the following relations hold:
\beqan
\label{Rell2}
\ast(V \wedge\Phi) &=& b \Phi ~~, \\
\label{Rell3}
\ast(\Phi\wedge\Phi) &=& fV~~ ,\\
\label{Rell4}
\Phi\bigtriangleup_2\Phi &=& e\Phi ~~.
\eeqan
Using equations (\ref{Rell1}-\ref{Rell4}) in \eqref{Fierz9ranks} gives:
\ben
\label{ct1}
a+c=15 ~~,~~f=14~~,~~2b-e=14~~.
\een
In what follows we shall use associativity of the geometric product. Multiplying \eqref{rel1} with $\Phi$ from the right in the truncated
model:
\be
(V\bdiamond V)\bdiamond\Phi=||V||^2\Phi~~
\ee
and using \eqref{Rell1} gives:
\ben
\label{aPhi}
V\bdiamond V\bdiamond\Phi=a\Phi~~.
\een
One the other hand, $\bdiamond$-multiplying \eqref{rel2} with $V$ from the left:
\be
V\bdiamond (V\bdiamond\Phi)=V\bdiamond (\ast(V\wedge\Phi))~~,
\ee
and using \eqref{Rell2} and then again \eqref{rel2} and \eqref{Rell2} gives:
\be
V\bdiamond V\bdiamond\Phi=b V\bdiamond\Phi=b\ast(V\wedge\Phi)=b^2\Phi~~,
\ee
which upon comparing with \eqref{aPhi} lead to:
\ben
\label{ct2}
a= b^2~~. 
\een
Finally, $\bdiamond$-multiplying \eqref{rel2} with $\Phi$ from the right:
\be
(V\bdiamond \Phi)\bdiamond\Phi=\ast(V\wedge\Phi)\bdiamond\Phi~~
\ee
and comparing to the relation obtained through $\bdiamond$-multiplication of \eqref{rel3} with $V$
from the left:
\be
V\bdiamond (\Phi\bdiamond\Phi)=||\Phi||^2 V+V\bdiamond(\ast(\Phi\wedge\Phi))-V\bdiamond(\Phi\bigtriangleup_2\Phi)~~
\ee
gives, upon using \eqref{Rell1}-\eqref{Rell4}:
\ben
\label{ct3}
bf=c~~,~~bc=fa~~. 
\een
Relations \eqref{ct1}-\eqref{ct3} imply a second order equation for $b$:
\be
  b^2 + 14 b - 15 = 0~~ ,
\ee
which has the solutions $b=1$ and $b= -15$. The solution $b=1$ further
gives $a=1$, $e=-12$ and $c=14$, while the second solution $b=-15$
gives $c=-210$,  which cannot be the case since $c$ is the square of the norm of
$\Phi$ and hence must be non-negative. We conclude that the
Fierz relations \eqref{Fierz9} are equivalent with the following system of
conditions on the forms $V$ and $\Phi$:
\beqa
||V||^2=1 ~~&,&~~||\Phi||^2=14~~,\\
\iota_V\Phi=0~~,~~\ast(\Phi\wedge\Phi)=14V ~~&,&~~\ast(V\wedge\Phi)=\Phi~~,~~ \Phi\bigtriangleup_2\Phi=-12\Phi\nn~~.
\eeqa
Via the cone formalism of \cite{ga2}, these relations provide a synthetic
encoding of the Fierz identities used in \cite{MartelliSparks, Tsimpis}.

\subsection {One Majorana spinor in seven Euclidean dimensions (almost complex case)}
\label{sec:examplesalmostcomplex}

Consider a Riemannian seven-manifold $(M,g)$ ($d=p=7,~ q=0$). Since
$p-q\equiv_8 7$, this belongs to the almost complex case, for which the
properties of pinors were treated in Subsections \ref{sec:pincasesac}
and \ref{sec:pairingscasesac}. The case of a single pinor (more
precisely, that of a Majorana spinor) on a Riemannian seven-manifold
arises, for example, in the well-studied case of ${\cal N}=1$
compactifications of $M$-theory on 7-manifolds
\cite{MinasianTomasiello, BehrndtJeschek, ThomasHouse}, which admit a
geometric description through reductions of the structure group of
$TM$ from $O(7)$ to $G_2$.  In this case, the bundle morphism $\gamma$
is fiberwise injective but not fiberwise surjective, having image
equal to $\End_\C(S)\subset \End(S)$. As an $\R$-vector bundle, $S$ has
rank $\rk_\R S=N=2^{\left[\frac{d}{2}\right]+1}=16$, while as a
$\C$-vector bundle (with complex structure given by
$J=\gamma(\nu)$, where $\gamma(\nu)=\gamma^{(8)}=\gamma^1\circ \ldots \circ \gamma^7$) it
has rank $\rk_\C S=\Delta=2^{\left[\frac{d}{2}\right]}=8$.  One has a
spin endomorphism given by $\cR=D$, which is a real structure (complex
conjugation) on $S$ when the latter is viewed as a complex vector
bundle. The $D$-real sections $\xi$ of $S$ (those satisfying
$D(\xi)=\xi$) are {\em Majorana spinors}, while $D$-imaginary sections
can be obtained by applying $J$ on such spinors. Since $S=S_+\oplus
S_-=S_+\oplus J(S_+)$ (where $S_+$ is the sub-bundle of Majorana
spinors), any section $\xi \in \Gamma(M,S)$ decomposes uniquely as
$\xi=\xi_R +i \xi_I=\xi_R+J(\xi_I)$, with $\xi_R=\Re\xi\in
\Gamma(M,S_+)$ and $\xi_I=\Im\xi \in \Gamma(M,S_+)$.

Up to rescaling by nowhere-vanishing smooth functions, there are four
admissible bilinear pairings $\cB_0,\cB_1,\cB_2,\cB_3$ on $S$ (see
Subsection \ref{sec:pairingscasesac}) which can be used to construct
form-valued pinor bilinears. Since all these pairings are related
through the action of $J$ and $D$, we choose to work with the basic
pairing $\cB_0$, which is symmetric ($\sigma_0=+1$) of type $\epsilon_0=-1$ and isotropy $\iota_0=+1$. In
the following, we consider only the particular case when $\xi
\in\Gamma(M,S_+)$ is a {\em Majorana spinor} --- a case which was
studied through less formal methods in \cite{MinasianTomasiello,
BehrndtJeschek, ThomasHouse}.

\paragraph{Form-valued bilinears constructed using $\gamma$ and $D$.}
For $\xi\in \Gamma(M,S_+)$ a Majorana spinor, \eqref{CB0Gamma} reduces to\footnote{Indeed, the
locally-defined endomorphisms $\gamma_a\in \Gamma(U,\End(S))$ are
$D$-imaginary, so $\gamma_{a_1\ldots a_k}$ is $D$-real or $D$-imaginary
according to whether $k$ is even or odd. Since $\iota_0=+1$ and $\xi$ is
$D$-real, this implies that $\bcE^{(k)}$ vanishes unless $k$ is even
(cf. Subsection \ref{sec:pairingscasesac}). }:
\ben
\label{Dselection}
\cB_0(\xi,\gamma_{a_1\ldots a_k}\xi) =0~~{\rm
  unless}~~k=\even~\Longleftrightarrow~ \bcE^{(0,k)}=0~~\mathrm{unless}~~k=\even~~,
\een
where $ \bcE^{(0,k)}$ is defined as in \eqref{E01geomfinal}-\eqref{notationcomplex}. On the other hand, we
have $(\gamma_a)^t=-\gamma_a$ since $\epsilon_0=-1$. This implies:
\be
(\gamma_{a_1\ldots a_k})^t=(-1)^{\frac{k(k+1)}{2}}\gamma_{a_1\ldots a_k}~~. 
\ee
Together with the symmetry of $\cB_0$, this shows $\bcE^{(0,k)}$ vanishes
unless $(-1)^{\frac{k(k+1)}{2}}=\sigma_0=+1$ (cf. \eqref{CBilkSym}),
i.e. unless $k(k+1)\equiv_4 0\Leftrightarrow k=0,3,4,7$.  Combining this with
\eqref{Dselection}, we conclude that --- for $\xi$ a Majorana spinor --- the
forms $\bcE^{(0,k)}$ vanish unless $k=0,4$. When $\xi$ is normalized
through $\cB_0(\xi,\xi)=1$, this gives $\bcE^{(0,0)}=1$ and the form-valued $\xi$-bilinear:
\ben
\label{phiDef}
\boxed{\phi\eqdef \bcE^{(0,4)}=\frac{1}{24}\cB_0(\xi,\gamma_{a_1\ldots a_4}\xi)e^{a_1\ldots a_4}~~}\nn
\een
which in turn form the components of the first generator $\check{E}^{(0)}$ of
the Fierz algebra (cf. expansions \eqref{E01geomfinal}):
\be
\check{E}^{(0)}=\frac{1}{16}(1+\phi) ~~. \nn
\ee
Notice that $\cB_0(\xi,D\circ\gamma_{a_1...a_k}\xi)$ (which equals
$\cB_2(\xi,\gamma_{a_1...a_k}\xi)$ by identity \eqref{B123exp}) coincides in
this case with $(-1)^{k}\cB_0(\xi,\gamma_{a_1...a_k}\xi)$ and hence leads to
the same two bilinears written above. Using \eqref{E01geomfinal}, this shows
that the second generator of the Fierz algebra coincides with the first:
\be
\check{E}^{(1)}=\check{E}^{(0)}~~. 
\ee

\paragraph{The Fierz identities.}
Using the observations above, we find that the Fierz identities \eqref{cRandalcx} are mutually-equivalent
in our case and amount to the following relation:
\be
\check{E}^{(0)}\diamond\check{E}^{(0)}+\pi(\check{E}^{(0)})\diamond\check{E}^{(0)}=\check{E}^{(0)} ~~. \nn\\
\ee
Since $\check{E}^{(0)}$ has only components of even rank, we have
$\pi(\check{E}^{(0)})=\check{E}^{(0)}$ and hence the Fierz relations further
reduce to:
\ben
\label{F7d}
2\check{E}^{(0)}\diamond\check{E}^{(0)}=\check{E}^{(0)} ~~\Longleftrightarrow~~(1+\phi)\diamond(1+\phi)=8(1+\phi)~~. 
\een
Expanding the geometric product into generalized products, we find: 
\ben
\label{phidiamondphi}
\phi\diamond\phi= ||\phi||^2-\phi\bigtriangleup_2\phi~~,
\een 
where we used the identity \footnote{One has 
$\omega\bigtriangleup_k\omega=||\omega||^2$ 
for all $\omega\in \Omega^k(M)$ and any $k=0\ldots d$.}
$\phi\bigtriangleup_4\phi=||\phi||^2$. Substituting \eqref{phidiamondphi} 
into \eqref{F7d} and separarting into rank components shows that the Fierz identities
are equivalent with the following two conditions:
\ben
\label{Fierz7phi}
\boxed{||\phi||^2=7 ~~~\mathrm{and}~~~ \phi\bigtriangleup_2\phi=-6\phi}~~.
\een

\paragraph{Form-valued bilinears constructed using $\gamma$ and $J$ or $J\circ D$.}
As usual in the almost complex case, one can construct further form-valued $\xi$-bilinears using
both $\gamma$ and either $J$ or $J\circ D$. As we shall see below, these
bilinears contain no new information in our example, but it is
instructive to discuss them nonetheless.
 
Since $\xi$ is a Majorana spinor, the $\xi$-bilinears
$\cB_0(\xi,J\circ\gamma_{a_1...a_k}\xi)$ (which equal
$\cB_1(\xi,\gamma_{a_1...a_k}\xi)$ by virtue of \eqref{B123exp}) give the same
differential forms as $\cB_0(\xi,J\circ D\circ\gamma_{a_1...a_k}\xi)$ (which equals
$-\cB_3(\xi,\gamma_{a_1...a_k}\xi)$), due to the `selection rule':
\ben
\label{Jgammasel}
\cB_0(\xi,J\circ\gamma_{a_1...a_k}\xi)=-\cB_0(\xi,J\circ D\circ\gamma_{a_1...a_k}\xi)= 0~~\mathrm{unless}~~k=\odd~~.
\een
The first of identities \eqref{JDtranspose} takes the form $J^t=J$ since $\frac{d(d+1)}{2}=28$ is even in our case. 
Together with \eqref{gammaAtranspose}, this gives:
\ben
\label{Jgammatranspose}
(J\circ\gamma_{a_1...a_k})^t=(-1)^{\frac{k(k+1)}{2}}J\circ\gamma_{a_1...a_k}~~,
\een
where we used $\epsilon_0=+1$ and the fact that $J$ commutes with $\gamma_a$. 
Relation \eqref{Jgammatranspose} 
implies that the differential form 
$\frac{1}{k!}\cB(\xi,J\circ\gamma_{a_1...a_k}\xi)e^{a_1\ldots a_k}$ vanishes 
unless $k(k+1)\equiv_4 0\Longleftrightarrow k\equiv_4 0,3$. Together with 
\eqref{Jgammasel}, this shows that the nontrivial
homogeneous form-valued bilinears constructed using $J$ and $\gamma$ have ranks 
$k =3$ or $k=7$ and can thus be written as follows, up to multiplication 
by elements of $\cinf$:
\ben
\addtolength{\fboxsep}{3pt}
\label{psietadef}
\boxed{
\begin{split}
&\psi\eqdef \frac{1}{3!}\cB_0(\xi,J\circ\gamma_{a_1\ldots a_3}\xi)e^{a_1\ldots a_3} ~~,\nn\\
&\eta\eqdef \frac{1}{7!}\cB_0(\xi,J\circ\gamma_{a_1...a_7}\xi)e^{a_1...a_7}~~.
\end{split}
}
\een
Using the case $d=7$ of the well-known formula:
\ben
\label{gmatid}
\gamma_{a_1...a_k}=\frac{(-1)^\frac{k(k-1)}{2}}{(d-k)!}\epsilon_{a_1...a_k}{}^{a_{k+1}...a_d}\gamma_{a_{k+1}...a_d}\gamma^{(d+1)}
\een
and the fact that $J=\gamma^{(8)}$,  we find:
\beqa
&&\gamma_{a_1\ldots a_3}=-\frac{1}{4!}\epsilon_{a_1\ldots a_3}{}^{b_1 \ldots b_4}\gamma_{b_1 \ldots b_4}J~~,\\
&&\gamma_{a_1...a_7}=-\epsilon_{a_1...a_7}\gamma^{(8)}=-\epsilon_{a_1...a_7}J~~.
\eeqa 
Since $J$ commutes with $\gamma^a$ and $J^2=-\id_S$, substitution of these relations into \eqref{psietadef} gives:
\beqa
\psi &=& \frac{1}{4!}\cB_0(\xi,\gamma_{b_1\ldots b_4}\xi)\epsilon_{a_1\ldots a_3}{}^{b_1 \ldots b_4}e^{a_1\ldots a_3} 
~\Longrightarrow~ \boxed{\psi={\tilde \ast}\phi=\ast \phi}~~,\\
\eta &=&\epsilon_{a_1 \ldots a_7}\cB_0(\xi,\xi)e^{a_1 \ldots a_7}~\Longrightarrow~ \boxed{\eta=\nu}~~,
\eeqa
where we noticed that ${\tilde \ast}\phi=\ast \phi$ since $\rk\phi=4$ implies $\tau(\phi)=\phi$. 
This shows that the $\xi$-bilinears constructed using $\gamma$ and $J$ or $J\circ D$ contain no information 
beyond that already contained in $\phi$. 

\paragraph{Some identities for $\phi$ and $\psi=\ast \phi$.}
The identity $||\ast\omega||=||\omega||$ (which holds for any $\omega\in \Omega(M)$),
implies that the norm of $\psi$ equals that of $\phi$: 
\be
||\psi||=||\phi||~~.
\ee
Since $d=7$ and $q=0$ in our example, we have: 
\be
\ast\circ\ast=(-1)^q\pi^{d-1}=+\id_{\Omega(M)}~~{\rm and}~~\nu\diamond\nu=(-1)^{q+\left[\frac{d}{2}\right]}=-1_M~~. 
\ee
This implies that relation $\psi=\ast\phi=\phi\diamond\nu$ is
equivalent with $\phi=\ast\psi=-\psi\diamond\nu$.  Since $\rk\phi=4$,
we have $\pi(\phi)=\phi$, which implies
$\nu\diamond\phi=\phi\diamond\nu$. Using these observations, we
compute:
$\psi\diamond\psi=(\ast\phi)\diamond(\ast\phi)=\phi\diamond\nu\diamond\phi\diamond\nu
=\phi\diamond\phi\diamond\nu\diamond\nu=-\phi\diamond\phi$, which
gives $\psi\diamond\psi=-\phi\diamond\phi$.  Combining this with
\eqref{phidiamondphi} and the generalized product expansion:
\ben
\label{psidiamondpsi}
\psi\diamond\psi=-||\psi||^2+\psi\bigtriangleup_1\psi
\een
gives:
\ben
\label{ident1}
\boxed{\psi\bigtriangleup_1\psi=\phi\bigtriangleup_2\phi}~~.
\een
Similarly, we have $\psi\diamond\phi=-(\psi\diamond\psi)\diamond\nu=||\psi||^2\nu 
-\ast (\psi\bigtriangleup_1\psi)$, where we noticed that 
${\tilde \ast} (\psi\bigtriangleup_1\psi)=\ast (\psi\bigtriangleup_1\psi)$ since $(\psi\bigtriangleup_1\psi)$ has rank $4$. 
Comparing this with the generalized product expansion:
\ben
\label{psidiamondphi}
\psi\diamond\phi=\psi\wedge\phi+\psi\bigtriangleup_1\phi-\psi\bigtriangleup_2\phi-
\psi\bigtriangleup_3\phi~~
\een
and separating ranks gives:
\ben
\label{ident2}
\boxed{\psi\wedge\phi=||\psi||^2\nu ~~,~~ \psi\bigtriangleup_1\phi=0 ~~,~~ \psi\bigtriangleup_2\phi=\ast (\psi\bigtriangleup_1\psi) ~~,~~
\psi\bigtriangleup_3\phi=0}~~.
\een
Combining \eqref{ident1} and \eqref{ident2} with the Fierz identities \eqref{F7d} gives the following system of relations for the four-form $\phi$ and its Hodge dual $\psi=\ast \phi$:
\beqan
||\phi||^2=||\psi||^2=7~~&,&~~\psi\bigtriangleup_1\psi=\phi\bigtriangleup_2\phi=-6\phi~~,\label{Fierz7d1}\\
\psi\wedge \phi=7 \nu~~~~~&,&~~~\psi\bigtriangleup_1\phi=\psi\bigtriangleup_3\phi=0~~~,~~~~\psi\bigtriangleup_2\phi=-6\psi~~.\label{Fierz7d2}
\eeqan

\paragraph{Comparison with the literature.} 
As discussed at the end of Subsection \ref{sec:pairingscasesac}, the spinor
bilinears can also be written using the fiberwise $\C$-bilinear paring $\beta$
on $S$ obtained by complexification of the restriction $\cB_0|_{S_+\otimes S_+}$ (cf. \eqref{BetaDef}):
\be
\beta(\xi,\xi')=\cB_0(\xi,\xi')-i\cB_0(\xi,J\xi')~~,~~\forall\xi,\xi'\in\Gamma(M,S)~~.
\ee
For $J\xi=i\xi$, this gives\footnote{The expressions in the
  right hand side hold in a local frame
$(\epsilon_\alpha,J(\epsilon_\alpha))$ of $S$ defined above 
an open subset $U$ of $M$, where $(\epsilon_\alpha)_{\alpha=1,\ldots,\Delta}$ 
is a local frame of $S_+$ -- see Subsection \ref{sec:pairingscasesac} for details.}:
\beqa
 \beta(\xi,\xi)=\cB_0(\xi,\xi) ~~~&\Longleftrightarrow&~~~\hat{\xi}^T\hat{\xi}=1~~,\\
 \beta(\xi,\gamma_{a_1\ldots a_3}\circ J\xi)=\cB_0(\xi,\gamma_{a_1\ldots a_3}\circ J\xi) 
~~~&\Longleftrightarrow&~~~i\hat{\xi}^T\hat{\gamma}_{a_1\ldots a_3}\hat{\xi}=\psi_{a_1\ldots a_3}~~,\\
 \beta(\xi,\gamma_{a_1\ldots a_4}\xi)=\cB_0(\xi,\gamma_{a_1\ldots a_4}\xi) 
~~~&\Longleftrightarrow&~~~\hat{\xi}^T\hat{\gamma}_{a_1\ldots a_4}\hat{\xi}=\phi_{a_1\ldots a_4}~~,\\
\beta(\xi,\gamma_{a_1\ldots a_7}\circ J\xi)=\cB_0(\xi,\gamma_{a_1\ldots a_7}\circ J\xi) 
~~~&\Longleftrightarrow&~~~i\hat{\xi}^T\hat{\gamma}_{a_1\ldots
  a_7}\hat{\xi}=\eta_{a_1\ldots a_7}~~.
\eeqa
The authors of \cite{BehrndtJeschek} use the bilinears $\varpi\eqdef-\psi, \varphi\eqdef-\phi$ and $\epsilon\eqdef-\eta$, i.e.:
\beqa
\varpi_{a_1...a_3}&=&-i{\hat \xi}^T{\hat \gamma}_{a_1...a_3}{\hat \xi}=-i\beta(\xi, \gamma_{a_1...a_3}\xi)~~, \\
\varphi_{a_1\ldots a_4}&=&-{\hat \xi}^T{\hat \gamma}_{a_1\ldots a_4}{\hat \xi}=-\beta(\xi, \gamma_{a_1\ldots a_4}\xi)~~,\\
\epsilon_{a_1...a_7}&=&-i{\hat \xi}^T{\hat \gamma}^{a_1...a_7}{\hat \xi}=-i\beta(\xi, \gamma_{a_1\ldots a_7}\xi)~~.
\eeqa 
The minus sign in the correspondence
$\phi\leftrightarrow-\varphi$, $\psi\leftrightarrow-\varpi$,
$\eta\leftrightarrow-\epsilon$ arises from the fact that
\cite{BehrndtJeschek} use the complex structure $J'=-\gamma(\nu)=-J$ on
the bundle $S$, which is the conjugate of the complex structure
$J=+\gamma(\nu)$ on $S$ used in the present paper. This translates
into the relation $J'\gamma(\nu)=+\id_S$ and implies $J'\xi=-i\xi$ if multiplication by $i$ is defined using $J$.
Notice the Hodge duality relation:
\be
\varpi_{a_1...a_3}=\beta(\xi,\frac{1}{4!}\epsilon_{a_1...a_3}{}^{a_4...a_7} \gamma_{a_4...a_7}\gamma(\nu)\circ J\xi) =
-\frac{1}{4!}\beta(\xi,\epsilon_{a_1...a_3}{}^{a_4...a_7} \gamma_{a_4...a_7}\xi)\eqdef \ast(\varphi_{a_4...a_7})~~.
\ee
The Fierz identities listed in \cite[page 4]{MinasianTomasiello}:
\beqan
\label{Fierz7ind}
\varpi_{abe}\varpi^{ecd}&=&-\varphi_{ab}{}^{cd}+2\delta_{ab}^{cd}~~, \nn\\
\varpi^{abf}\varphi_{fcde}&=&-6\delta^{[a}_{[c}\varpi^{b]}{}_{de]}~~, \nn\\
\varphi_{abcg}\varphi^{defg}&=&6\delta^{abc}_{def}-9\varphi_{[ab}{}^{[de}\delta^{f]}_{c]}-\varpi_{abc}\varpi^{def}~~
\eeqan
imply the relations:
\beqan
\label{indexEq7d}
\varpi_{abe}\varpi^{ecd}=-\varphi_{ab}{}^{cd}+2\delta_{ab}^{cd}~~&\Longrightarrow&~~ \varpi\bigtriangleup_1\varpi=-6\varphi, \nn\\
\varpi^{bfa}\varphi_{fade}=4\varpi_{bde}~~&\Longrightarrow&~~\varpi\bigtriangleup_2\varphi=6\varpi ~~, \nn\\
\varphi_{abcg}\varphi^{decg}=42\delta^{ab}_{de}-2\varphi_{ab}{}^{de}~~&\Longrightarrow&~~\varphi\bigtriangleup_2\varphi=-6\varphi~~,
\eeqan
which indeed agree with \eqref{Fierz7d1}-\eqref{Fierz7d2}. 

\subsection{One real pinor in five dimensions with metric signature
  $(p,q)=(1,4)$ (quaternionic case)}
\label{sec:examplesquaternionic}

Consider one real pinor on a psedo-Riemannian five-manifold $(M,g)$ with 
 the mostly minus signature  $(p,q)=(1,4)$. We have $p-q\equiv_8 5$, which places this
example in the quaternionic non-simple case (see Sections
\ref{sec:pincasesquat}, \ref{sec:pairingscasesquat}). This situation occurs in 
the study of (gauged) ${\cal N}=1$ supergravity in
five-dimensions (see, for example,  \cite{GutowskiReall, BellorinT, BellorinMO, GGHPR}).  
The morphism $\gamma:\wedge T^\ast M\rightarrow \End(S)$ is
neither injective nor surjective, having image equal to $\End_\H(S)$.  It
satisfies $\gamma(\nu)=\epsilon_\gamma\id_S$, where $\epsilon_\gamma\in
\{-1,1\}$ is the signature of $\gamma$. We choose to work with $\epsilon_\gamma=+1$, i.e. 
$\gamma(\nu)=\id_S$, and with the \emph{basic admissible pairing}
$\cB_0$ out of the four independent admissible pairings ($\cB_0,\;\cB_1,\;\cB_2$ and $\cB_3$) 
which exist in this case. Notice that $\cB_0$ is symmetric ($\sigma_0=+1$) and has type
$\epsilon_0=+1$. The restriction $\gamma_\ev$ is fiberwise irreducible, so chiral
projectors cannot be defined in this case. 

Choosing the pin representation $\gamma$ with signature $\epsilon_\gamma=+ 1$,
we realize the \KA algebra $(\Omega^+(M),\diamond)$ through the truncated model
$(\Omega^<(M),\bdiamond_+)$, where $\Omega^<(M)\eqdef\oplus_{k=0}^2\Omega^k(M)$. 
In the case where there is only one globally-defined pinor 
$\xi\in\Gamma(M,S)$, we are interested in the following pinor bilinears: 
\beqa
&&\bcE^{(0,k)}\eqdef
\frac{1}{k!}\cB_0(\xi,\gamma_{a_1...a_k}\xi')e^{a_1\ldots a_k}\in\Omega^k(M)~~,~~\forall k\in\overline{0,2}~~,\\
&&\bcE^{(i,p)}\eqdef
\frac{1}{p!}\cB_0(\xi,J_i\circ\gamma_{a_1...a_p}\xi')e^{a_1\ldots a_p}\in\Omega^p(M)~~,~~\forall p\in\overline{0,2}~,
~\forall i\in\overline{1,3}~.
\eeqa
\paragraph{Form-valued bilinears constructed using only $\gamma$.}
The symmetry $\sigma_0=1$ of $\cB_0 $ implies:
\be
\cB_0(\xi,\gamma_A\xi)=\cB_0(\gamma_A\xi,\xi)~~.
\ee
Since $\cB_0$ has type $\epsilon_0=1$, we also have:
\be
\cB_0(\gamma_A\xi,\xi)=(-1)^{\frac{|A|(|A|-1)}{2}}\cB_0(\xi,\gamma_A\xi)~~,
\ee
which further implies that $\bcE^{(0,k)}=\frac{1}{k!}\cB_0(\xi,\gamma_{a_1...a_k}\xi)e^{a_1...a_k}$ vanishes unless $k$ satisfies
$k(k-1)\equiv_4 0$. The latter holds when $k=0,1,4,5$, which gives the non-trivial pinor bilinears:
\be
f=\cB_0(\xi,\xi)~,~~V=\cB_0(\xi,\gamma_a\xi)e^a~,
~~W=\frac{1}{4!}\cB_0(\xi,\gamma_{a_1...a_4}\xi)e^{a_1...a_4}~,~~\nu=\frac{1}{5!}\cB_0(\xi,\gamma_{a_1...a_5}\xi)e^{a_1...a_5}~.
\ee
\paragraph{Form-valued bilinears constructed using $\gamma$ and $J_i$.} Since $\sigma_0=+1$ while
$J_i$ commutes with $\gamma_a$, we have: 
\be
\cB_0(\xi,J_i\circ\gamma_A\xi)=\cB_0(J_i\circ\gamma_A\xi,\xi)=\cB_0(\gamma_A\circ J_i\xi,\xi)~~.
\ee
Using the fact that the $\cB_0$-transpose of $J_i$ is given by $J_i^t=-J_i$, we find:
\be
\cB_0(\gamma_A\circ J_i\xi,\xi)=(-1)^{\frac{|A|(|A|-1)}{2}+1}\cB_0(\xi,J_i\circ\gamma_A\xi)~~.
\ee
The last two identities imply that the bilinears $\bcE^{(i,p)}$ vanish unless $p(p-1)+2\equiv_4 0$, i.e unless $p=2,3$.
This gives the nontrivial bilinears:
\be
\Phi^i=\frac{1}{2!}\cB_0(\xi,J^i\circ\gamma_{a_1a_2}\xi)e^{a_1a_2}~~,~~
\Psi^i=\frac{1}{3!}\cB_0(\xi,J^i\circ\gamma_{a_1...a_3}\xi)e^{a_1...a_3}~~,~~\forall i=1,...,3~~.
\ee
Identity \eqref{gmatid} and the relation $\tilde\ast=\ast\circ\tau$ imply that the form-valued bilinears of rank greater than two can be written as: 
\be
\nu=\ast f=\tilde\ast f~~,~~W=\ast V=\tilde\ast V~~,~~ \Psi^i=-\ast\Phi^i=\tilde\ast \Phi^i
\ee
and are thus dual to the scalar, vector and two-forms respectively.
As expected, this means that the non-truncated inhomogeneous forms: 
\be
\check{E}^{(0)}=\frac{1}{2^4}(f+V+W+\nu)~~~ \mathrm{and} ~~~\check{E}^{(i)}=\frac{1}{2^4}(\Phi^i+\Psi^i)~~
\ee
are {\em twisted selfdual}. The factor of $\frac{1}{2^4}$ comes from the prefactor $\frac{\Delta}{2^d}$ in 
the general expressions for the generators of the Fierz algebra, where we have 
substituted $\Delta=2$ and $d=5$. 

Since for the truncated model we only need the form bilinears of rank less than or equal to 2:
\ben
\label{Bilinears7d}
\boxed{f=\cB_0(\xi,\xi)~,~~V=\cB_0(\xi,\gamma_m\xi)e^m~,~~\Phi^i=\cB_0(\xi,J^i\circ\gamma_{mn}\xi)e^{mn}~,
~\forall i=1,...,3~,~\forall m,n=1,...,5},
\een
we build the truncated generators of the Fierz algebra as:
\ben
\label{gens}
\boxed{\check{E}^{(0)}_<=\frac{1}{2^4}(f+V)~~~\mathrm{and}~~~\check{E}^{(i)}_<=\frac{1}{2^4}\Phi^i}~.
\een

\paragraph{Analysis of Fierz relations for the truncated algebra $(\Omega^<(M),\bdiamond_+)$.} 
In what follows we shall omit the subscript `$+$' of $\bdiamond_+$. Using the expression \cite{ga1}:
\be
\check{E}^{(0)}=2P_+(\check{E}^{(0)}_<)~~~\mathrm{and}~~~
\check{E}^{(i)}=2P_+(\check{E}^{(i)}_<)~~ \nn
\ee
and the relation $P_+(\omega\bdiamond\eta)=P_+(\omega)\diamond P_+(\eta)$ (where $P_+=\frac{1}{2}(1+\tilde\ast)$), 
the Fierz identities for the truncated generators become:
\beqa
&& \check{E}^{(0)}_<\bdiamond \check{E}^{(0)}_<- \sum_i \check{E}^{(i)}_< \bdiamond  \check{E}^{(i)}_< 
= \frac{1}{2} \cB_0(\xi,\xi) \check{E}^{(0)}_<~~,~~\forall i=1\ldots 3~~,\nn\\
&& \check{E}^{(0)}_< \bdiamond  \check{E}^{(i)}_< +  \check{E}^{(i)}_< \bdiamond \check{E}^{(0)}_<
+\sum_{j,k=1}^3 \epsilon_{ijk} \check{E}^{(j)}_< \bdiamond  \check{E}^{(k)}_< = \frac{1}{2}\cB_0(\xi,\xi) \check{E}^{(i)}_<~~, \nn
\eeqa
i.e.:
\beqan
\label{quatFierz5d}
&& (f+V)\bdiamond(f+V)-\sum_i\Phi^i\bdiamond\Phi^i=8f(f+V) ~~,\nn\\
&& (f+V)\bdiamond\Phi^i+\Phi^i\bdiamond(f+V)+\sum_{j,k}{\epsilon_{ijk}\Phi^j\bdiamond\Phi^k}=8f\Phi^i ~~.
\eeqan
The latter simplify to:
\beqan
\label{quatFierz1}
&& V\bdiamond V -\sum_i \Phi^i\bdiamond\Phi^i=7f^2+6fV~~,\\
\label{quatFierz2}
&& V\bdiamond\Phi^i+\Phi^i\bdiamond V+\sum_{j,k}\epsilon_{ijk}\Phi^j\bdiamond\Phi^k=6f\Phi^i ~~.
\eeqan
The various terms on the left hand side expand as:
\beqan
\label{quatexp1}
&& V\bdiamond V=||V||^2~~,\nn\\
&& \sum_i\Phi^i\bdiamond\Phi^i=\sum_i\left( \tilde\ast(\Phi^i\wedge\Phi^i)-||\Phi^i||^2\right)~~,\nn\\
&& V\bdiamond\Phi^i+\Phi^i\bdiamond V=2\tilde\ast(V\wedge\Phi^i)~~,\\
&&\sum_{j,k}\epsilon_{ijk}\Phi^j\bdiamond\Phi^k=\sum_{j,k}\epsilon_{ijk}
\left( \tilde\ast(\Phi^j\wedge\Phi^k)-\Phi^j\bigtriangleup_1\Phi^k-\Phi^j\bigtriangleup_2\Phi^k\right)~~.\nn
\eeqan
We have ~$\iota_V\Phi^i=0$ ~since $V$ is the only non-trivial pinor bilinear which can be constructed at rank one and $\iota_V\Phi^i$ is a rank one bilinear, which must therefore be proportional to $V$:
\be 
\iota_V\Phi^i= c_1 V ~\Rightarrow~ \iota_V\iota_V\Phi^i= c_1 ||V||^2
\ee 
for some constant $c_1$. However, we have:
\be
\iota_V\iota_V\Phi^i=\iota_{V\wedge V}\Phi^i=0
\ee
since $V\wedge V=0$, which implies that $c_1||V||^2=0$ and thus $c_1=0$, which means that $\iota_V\Phi^i$ vanishes. This further leads to:
\ben
\label{VPhi}
 V\bdiamond\Phi^i=\Phi^i\bdiamond V=\tilde\ast(V\wedge\Phi^i)~~.
\een
Let us introduce the following notations:
\beqan
\label{quatexp2}
||V||^2=a~~,~~\sum_i||\Phi^i||^2=c~~,~~\sum_i \tilde\ast(\Phi^i\wedge\Phi^i)=eV~&,&~ \tilde\ast(V\wedge\Phi^i)=b\Phi^i~~, \nn\\
\sum_{j,k}\epsilon_{ijk}\tilde\ast(\Phi^j\wedge\Phi^k)=w^iV~~,~~\sum_{j,k}\epsilon_{ijk}\Phi^j\bigtriangleup_1\Phi^k=u\Phi^i~&,
&~\sum_{j,k}\epsilon_{ijk}\Phi^j\bigtriangleup_2\Phi^k=l^i~~,
\eeqan
where $a,b,c,e,u,l^i,w^i$ are non-negative smooth real-valued functions on $M$.
Using relations \eqref{quatexp1} in \eqref{quatFierz1}-\eqref{quatFierz2}, we find:
\ben
\label{const1}
a+c=7f^2~~,~~e=-6f~~,~~2b-u=6f~~,~~w^i=l^i=0~~.
\een
With \eqref{VPhi}-\eqref{const1}, the system \eqref{quatexp1} simplifies to:
\beqan
\label{exp1}
&& V\bdiamond V=||V||^2=a~~,\\
\label{exp2}
&& \sum_i\Phi^i\bdiamond\Phi^i=-6fV-c~~,\\
\label{exp3}
&& V\bdiamond\Phi^i=\Phi^i\bdiamond V=b\Phi^i~~,\\
\label{exp4}
&&\sum_{j,k}\epsilon_{ijk}\Phi^j\bdiamond\Phi^k=-(6f-2b)\Phi^i ~~.
\eeqan
In what follows we use associativity of the reduced geometric product $\bdiamond$.
First notice that $\bdiamond$-multipling \eqref{exp3} with $\Phi^i$ from the right and summing the result over $i=1\ldots3$ gives:
\be
\sum_i (V\bdiamond\Phi^i)\bdiamond\Phi^i=b\sum_i\Phi^i\bdiamond\Phi^i~~,
\ee
which upon using \eqref{exp2} leads to:
\ben
\label{re1}
\sum_i V\bdiamond\Phi^i\bdiamond\Phi^i=-6bfV-6bc~~.
\een
Furthermore, $\bdiamond$-multiplying \eqref{exp2} with $V$ from the left gives:
\be
V\bdiamond(\sum_i\Phi^i\bdiamond\Phi^i)=-6fV\bdiamond V-6cV
\ee
which takes the following form upon using \eqref{exp1}: 
\ben
\label{re2}
V\bdiamond\sum_i\Phi^i\bdiamond\Phi^i=-6fa-6cV~~.
\een
Subtracting \eqref{re2} from \eqref{re1} gives:
\ben
\label{const2}
c=6bf~~~,~~~a=b^2~~.
\een
From \eqref{const2} and the first relation in \eqref{const1}, we find a second order equation for $b$:
\be
  b^2 + 6 b f - 7 f^2 = 0~~ ,
\ee
which has the solutions $b=f$ and $b= -7f$. The solution $b=f$ further
gives $a=f^2$, $c=6f^2$ and $u=-4f$, while the second solution $b=-7f$
gives $c=-42f^2$, which cannot hold since $c=\sum_i||\Phi^i||^2$ must be non-negative. We conclude that the
Fierz relations \eqref{quatFierz1}-\eqref{quatFierz2} are equivalent with the following system of
conditions on the forms $V$ and $\Phi^i$:
\beqan
\label{FierzSol5d}
||V||^2=f^2 &,&\sum_{i=1}^3||\Phi^i||^2=6f^2~~,~~\iota_V\Phi^i=0~~,\nn\\
\tilde\ast(V\wedge\Phi^i)= f\Phi^i ~&\Longrightarrow& ~\iota_V\tilde\ast\Phi^i=f\Phi^i~\Longleftrightarrow ~\iota_V\ast\Phi^i=-f\Phi^i~~, \nn\\
\sum_{i=1}^3\tilde\ast(\Phi^i\wedge\Phi^i)=-6fV~~&\Longrightarrow& ~ \sum_{i=1}^3(\Phi^i\wedge\Phi^i)=-6f\tilde\ast V=-6f\ast V ~~, \\
\sum_{j,k}\epsilon_{ijk}\tilde\ast(\Phi^j\wedge\Phi^k)= 0~~&\Longrightarrow&~~\Phi^j\wedge\Phi^k=0~~,~~\forall j,k=1\ldots 3~~,~~j\neq k~~,\nn\\
\sum_{j,k}\epsilon_{ijk}\Phi^j\bigtriangleup_2\Phi^k= 0~~&\Longrightarrow&~~\iota_{\Phi^j}\Phi^k=0~~,~~\forall j,k=1\ldots 3~~,~~j\neq k~~,\nn\\
\sum_{j,k}\epsilon_{ijk}\Phi^j\bigtriangleup_1\Phi^k= -4f\Phi^i~~&\Longrightarrow&~~\Phi^j\bigtriangleup_1\Phi^k=-2f\epsilon^{ijk}\Phi^i~~,~~\forall i,j,k=1\ldots 3~~.\nn
\eeqan
To arrive at the equations above, we made use  of the definitions:
\be
\tilde\ast\omega=\omega\diamond\nu~~,~~ \ast\omega=\tau(\omega)\diamond\nu=\iota_\omega\nu~~
\ee
and of the properties $\tilde\ast\circ\tilde\ast=\ast\circ\ast=\id_{\Omega(M)}$ as well as of the following identities (see \cite{Rodr-Oliv}) which hold for
homogeneous forms $\omega\in \Omega^r(M)$ and $\eta\in\Omega^s(M)$:
\beqan
\label{Contractions}
&& \omega\wedge\ast\eta=(-1)^{r(s-1)}\ast \iota_{\tau(\omega)}\eta ~~,~~\mathrm{when}~~r\leq s~~,\nn\\
&& \iota_{\omega}(\ast\eta)=(-1)^{r s}\ast(\tau(\omega)\wedge\eta)~~,~~\mathrm{when}~~r+s\leq d~~.
\eeqan

\paragraph{Comparison with the literature.}
Using one `generalized symplectic-Majorana spinor' (see below), which appears in the literature as a pair of
spinors of real dimension $N=2^{\left[\frac{d}{2}\right]}=4$, the authors of 
\cite{GutowskiReall, BellorinT, BellorinMO, GGHPR} construct a scalar, a vector and three two-forms as
spinor bilinears. Specifically, writing the symplectic Majorana spinor as the 
spinor-pair $(\xi_1,\xi_2)$, these bilinears are written in \cite{GGHPR} as:
\be
s=i\bar{\xi}_j\xi^j~~,~~v_m=i\bar{\xi}_j\Gamma_m\xi^j~~,~~
\phi^r_{mn}=\bar{\xi}_j(\boldsymbol{\sigma}^r)_k{}^j\Gamma_{mn}\xi^k~~, 
\ee
for all $j,k\in{1,2}$~,~ $m,n\in{0,\ldots,4}$~, ~$ r\in{1,\ldots,3}$ ~and 
where $\boldsymbol{\sigma}^r$ are the Pauli matrices. 

Using the \emph{biquaternion formalism} of Subsection \ref{sec:pincasesquat} for the case $p=1, ~q=4$, we have
$(\Gamma^4)^2=-\id_{S_\C^+}$ and $\Gamma^4$ is a complex structure on $S_\C^+$. 
Furthermore, one can choose $\Gamma^4$ such that $[\fZ_0, \Gamma^4]_{-,\circ}=0$,
where the endomorphism $\fZ_0\in\Gamma(M,\End_\R(S_\C^+))$ was defined in 
Subsection \ref{sec:pincasesquat}, and hence the endomorphism 
$\rho\in \Gamma(M,\End_\C(S_\C))$, which has the matrix form:
\be
{\hat \rho}\eqdef i\boldsymbol{\sigma}_2 \Gamma^4=\left[\begin{array}{cc} 0 & \Gamma^4 \\ -\Gamma^4 &  0 
 \end{array}\right]
\ee
satisfies $[\fZ,\rho]_{-,\circ}=0$. Recall the notation of Subsection \ref{sec:pincasesquat}: 
$\fZ$ is a natural real structure on $S_\C$, $\fR$ is a $\C$-linear product 
structure anticommuting with $\fZ$ and $\fJ_2$ is the complexification of 
the endomorphism $J_2\in\Gamma(U,\End_\C(S_\C))$. It is easy to check that $\rho$ satisfies: 
\be
\rho^2=\id_{S_\C}~~,~~[\rho,\fR]_{+,\circ}=0~~,~~[\rho, \fJ_2]_{-,\circ}=0~~.
\ee
We can thus define another real structure $~^\ast$ on $S$ through:
\be
\xi^\ast\eqdef \rho\circ \fZ(\xi)~~,
\ee
since the properties listed above insure that $~^\ast$ is antilinear and that
it squares to $+\id_{S_\C}$. We have: 
\be
\xi^\ast=\Gamma^4\circ \fZ_0(\xi)~\Longleftrightarrow~
\fZ_0(\xi)=-\Gamma^4(\xi^\ast)~~,~~\forall \xi\in \Gamma(M,S^+_\C)~~. 
\ee
Condition $[\fZ,\rho]_{-,\circ}=0$ is equivalent with $(\Gamma^4)^\ast=\Gamma^4$, i.e. $(\Gamma^4\circ T^\ast)^\ast=\Gamma^4 \circ T$ for any
$T\in \Gamma(M,\End(S_\C^+))$. Since $\fZ$ anticommutes with $\fR$ and commutes with $\fJ_2$, 
it follows that the real structure $~^\ast$ commutes with
$\fR$ and with $\fJ_2$ and hence it satisfies:
\be
(\xi^\ast)^1=(\xi^1)^\ast~~,~~(\xi^\ast)^2=(\xi^2)^\ast~\Longrightarrow~
\widehat{\xi^\ast}=\left[\begin{array}{c} (\xi^1)^\ast\\ (\xi^2)^\ast\end{array}\right]~~,~~\forall \xi\in \Gamma(M,S)~~.
\ee
The restriction of $~^\ast$ to $S_+$ is a real structure on $S_\C^+$. It follows that $\fZ$ can be expressed as:
\be
\fZ(\xi)=\rho(\xi^\ast)~\Longrightarrow~ \widehat{\fZ(\xi)}= 
\left[\begin{array}{cc} 0 & \Gamma_4 \\ -\Gamma_4 & 0  \end{array}\right]\left[\begin{array}{c} (\xi^1)^\ast \\ (\xi^2)^\ast \end{array} \right]~~.
\ee
The reality condition $\xi\in \Gamma(M,S)\Longleftrightarrow
\fZ(\xi)=\xi$ can be written as: 
\ben
\label{SymplecticMajoranaCondition} 
\rho(\xi^\ast)=\xi~\Longleftrightarrow~(\xi^1)^\ast=\Gamma^4\xi^2~\Longleftrightarrow~ (\xi^2)^\ast=-\Gamma^4 \xi^1~~.
\een
Condition \eqref{SymplecticMajoranaCondition} is sometimes called the `(generalized) symplectic Majorana condition' in the literature on
supergravities in five dimensions (see, for example,  \cite{GutowskiReall, BellorinT, BellorinMO, GGHPR}). In those references, one also chooses the
morphism $\Gamma:\wedge T^\ast M\rightarrow \End_\C(S_\C^+)$ such that $\Gamma^0, \ldots, \Gamma^3$ are purely
imaginary with respect to the complex conjugation $~^\ast$ on $S_\C^+$, i.e. :
\be
(\Gamma^0)^\ast=-\Gamma^0~~,~~(\Gamma^1)^\ast=-\Gamma^1~~,~~(\Gamma^2)^\ast=-\Gamma^2~~,~~(\Gamma^3)^\ast=-\Gamma^3~~,~~(\Gamma^4)^\ast=+\Gamma^4~~.
\ee
The matrices of $\Gamma^m$ with respect to a local frame of $S_\C^+$ are the
(complex) `gamma matrices' used in  \cite{GutowskiReall, BellorinT, BellorinMO, GGHPR}.

The `generalized symplectic Majorana' pinors $(\xi_1,\xi_2)$ therefore correspond to those denoted by $(\xi_+,\xi_-)$
in \cite{GutowskiReall, BellorinT, BellorinMO, GGHPR}, where $\xi_{\pm}\in\Gamma(M,S^{\pm}_{\mathbb{C}})$. Using the correspondence between the endomorphisms 
$J_k\in\Gamma(U,\mathrm{End}_{\mathbb{C}}(S_{\mathbb{C}}))$ and the Pauli matrices:
\be
J_1\rightarrow i\boldsymbol{\sigma}^3~~,~~J_2\rightarrow i\boldsymbol{\sigma}^2~~,~~J_3\rightarrow i\boldsymbol{\sigma}^1~~,
\ee
the fact that \cite{GutowskiReall, BellorinT, BellorinMO, GGHPR} defines the Majorana 
conjugate (using the charge conjugation matrix $\mathcal{C}=i\Gamma^0\Gamma^4$) through the relation  $\bar{\xi}^i=\xi^{iT}\mathcal{C}$ 
and expressing $\cB_0(\xi,\xi')$ as $\xi^{T}\mathcal{C}\xi'=\bar{\xi}\xi'$, we can re-write the definitions of $s, v$ and $\phi^r$ as:
\beqan
&&\quad\quad\quad\quad\quad   s=i\cB(\hat{\xi},\hat{\xi})=is'~~,~~v=i\cB_0(\hat{\xi},\Gamma_m\hat{\xi})e^m=iv'~~,\nn\\
\label{redefQuatF}
&&\phi^1=-i\cB(\hat{\xi},\hat{J}_3\otimes\Gamma_{mn}\hat{\xi})e^{mn}=-i\phi'^3  ~~,~~ 
\phi^2=i\cB(\hat{\xi},\hat{J}_2\otimes\Gamma_{mn}\hat{\xi})e^{mn}=i\phi'^2~~,\\
&&\quad\quad\quad\quad\quad \quad   \phi^3=-i\cB(\hat{\xi},\hat{J}_1\otimes\Gamma_{mn}\hat{\xi})e^{mn}=-i\phi'^1~~\nn
\eeqan
for all $m,n\in\overline{1,5}$. For future reference, we identify the form-valued bilinears 
corresponding to our notations by decorating them with primes.

One can compare \eqref{FierzSol5d} with the Fierz identities given in \cite{GutowskiReall, BellorinT, GGHPR}, 
in particular with those of \cite[page 35]{BellorinMO}, which read:
\beqa
&&v_a v^a=s^2~~~,~~~v^a\phi^r{}_{ab}=0~~~,~~~ v^a(\ast\phi^r)_{abc}=-s\phi^r{}_{bc}~~, \\
&&\phi^r{}_a{}^c\phi^s{}_{cb}=-\delta^{rs}(\eta_{ab}s^2-v_a v_b)-\epsilon^{rst}s\phi^t{}_{ab} ~~,\\
&&\phi^r{}_{[ab}\phi^s{}_{cd]}=-\frac{1}{4}s\delta^{rs}\epsilon_{abcde}v^e~~.    
\eeqa 
The latter can be written as follows in terms of the redefined forms $s', v'$ and $\phi'^r$:
\beqan
\label{FierzLit}
 v'_a v'^a=-s'^2~~&\Longrightarrow&~~||v'||^2=-s'^2~~,\nn\\
 v'^a\phi'^r{}_{ab}=0~~&\Longrightarrow&~~\iota_{v'}\Phi'^r=0~~,\\
 v'^a(\ast\phi'^r)_{abc}=-s'\phi'^r{}_{bc}~~&\Longrightarrow&~~\iota_{v'}\ast\Phi'^r=-s'\phi'^r~~,\nn\\
\phi'^r{}_a{}^c\phi'^s{}_{cb}=-\delta^{rs}(\eta_{ab}s'^2-v'_a v'_b)+\epsilon^{rst}s'\phi'^t{}_{ab} ~~
&\Longrightarrow&~~ \phi'^r\bigtriangleup_1\phi'^s=-2\epsilon^{rst}s'\phi'^t ~~,~~||\Phi'^r||^2=6s'^2~~,\nn\\
 \phi'^r{}_{[ab}\phi'^s{}_{cd]}=-\frac{1}{4}s'\delta^{rs}\epsilon_{abcde}v'^e 
~~&\Longrightarrow&~~ \phi'^r\wedge\phi'^s=-2s'\delta^{rs}(\ast v')~~.    \nn
\eeqan
These agree with \eqref{FierzSol5d} (which was derived through geometric algebra techniques) upon identifying $f\leftrightarrow s'$, 
$V\leftrightarrow v'$ and $\Phi^i\leftrightarrow\phi'^i$.

\section{Conclusions and further directions}
\label{sec:conclusions}

We gave a geometric algebra reformulation of Fierz identities for
form-valued pinor bilinears in arbitrary dimensions and signatures,
which displays the underlying Schur algebra structure in a
conceptually transparent manner. This approach to Fierz identities
uncovers the underlying reason for phenomena which occur in various
applications and allows for a unified and efficient treatment of 
Fierz identities in various problems of interest in 
supergravity and string theory. In particular, this formulation opens the way for unified
studies of flux vacua and of other questions arising in string and M-theory compactifications. 
Our approach is highly-amenable to implementation in various symbolic computation
systems such as {\tt Ricci}\cite{Ricci}  (and we have carried out such
implementations as illustrated in our previous papers \cite{ga1, ga2} and briefly
touched upon in Section 5).  Using this implementation, we
illustrated our approach by recovering certain well-known results
within our formalism.  Further applications of our
methods will be discussed in forthcoming publications.

\acknowledgments{This work was supported by the CNCS projects
PN-II-RU-TE (contract number 77/2010), PN-II-ID-PCE (contract
numbers 50/2011 and 121/2011) and PN 09 37 01 02 / 2009. 
The work of C.I.L. is also supported by the 
Research Center Program of IBS (Institute for Basic Science) in Korea
(grant CA1205-01). E.M.B. thanks the Center for
Geometry and Physics, Institute for Basic Science and Pohang University of Science and
Technology (POSTECH), Korea and especially Jae-Suk Park for providing 
 excellent working conditions during her visit in the final stages of preparation for this work.
I.A.C. acknowledges the student scholarship from the Dinu
Patriciu Foundation ``Open Horizons'', which supported part of her
studies.}

\appendix

\section{Systematics of pin bundles for Riemannian and Lorentzian manifolds of dimension up to eleven}
\label{sec:appsystematics}

In Table \ref{table:Euclidean} and Table \ref{table:Minkowskian}, we
systematize the properties of real pinors for pseudo-Riemannian
manifolds $(M,g)$ of dimension up to $11$ and such that $g$ has
Euclidean or Minkowskian signature.

\begin{table}[H]
\centering
{\tiny
\begin{tabular}{|c|c|c|c|c|c|c|c|c|c|c|c|}
\hline
$d$& $d~{\rm mod}~8$ & $\S$ & $\Delta$ & N & $\Cl(p,q)$& $\begin{array}{c}\mathrm{Irrep.}\\\mathrm{image}\end{array}$& $\begin{array}{c} \mathrm{No.~of}\\ \mathbb{R}\mathrm{-irreps.}\end{array}$& Injective ? & 
$\begin{array}{c}\mathrm{Chirality}\\ \mathrm{operator}\\ \cR \end{array}$ &  $\gamma^{(d+1)}$ & 
$\begin{array}{c} \mathrm{Name~of}\\\mathrm{pinors}\\\mathrm{(spinors)}\end{array}$ \\
\hline\hline
\rowcolor{cyan}$1$ & $1$ & $\R$ & $1$ & $1$ & $\Mat(1,\mathbb{R})^{\oplus 2}$& $\Mat(1,\mathbb{R})$ & $2$ & no & N/A  & $\pm \hbox{1\kern-.27em l}$ & M \\
\hline
$2$ & $2$ & $\R$ & $2$ & $2$ & $\Mat(2,\mathbb{R})$ & $\Mat(2,\mathbb{R})$& $1$ & yes & N/A & $\gamma^{(3)}$ & M \\
\hline
$3$ & $3$ & $\C$ & $2$ & $4$ & $\Mat(2,\mathbb{C})$ & $\Mat(2,\mathbb{C})$ & $1$ & yes & N/A & $\pm J$ & M \\
\hline
$4$ & $\mathbin{\textcolor{ForestGreen}{\mathbf{4}}}$ & $\H$ & $2$ & $8$ & $\Mat(2,\mathbb{H})$ & $\Mat(2,\mathbb{H})$ & $1$ & yes & $\gamma(\nu)$ & $\gamma^{(5)}$   & \cellcolor{green} SM (SMW)\\
\hline
\rowcolor{cyan}$5$ & $5$ & $\H$ & $2$ & $8$ & $\Mat(2,\mathbb{H})^{\oplus 2}$ & $\Mat(2,\mathbb{H})$ & $2$ & no & N/A  & $\pm \hbox{1\kern-.27em l}$ & SM \\
\hline
$6$ & $\mathbin{\textcolor{ForestGreen}{\mathbf{6}}}$& $\H$ & $4$ & $16$ & $\Mat(4,\mathbb{H})$ & $\Mat(4,\mathbb{H})$ & $1$ & yes & $\gamma(\nu)\circ J$  & $\gamma^{(7)}$  & \cellcolor{green} DM (M) \\
\hline
$7$ & $\mathbin{\textcolor{ForestGreen}{\mathbf{7}}}$ & $\C$ & $8$ & $16$ & $\Mat(8,\mathbb{C})$ & $\Mat(8,\mathbb{C})$& $1$ & yes & $D$  & $\pm J$ & \cellcolor{green} DM (M) \\
\hline
$8$ & $\mathbin{\textcolor{ForestGreen}{\mathbf{0}}}$ & $\R$ & $16$ & $16$ & $\Mat(16,\mathbb{R})$ & $\Mat(16,\mathbb{R})$ & $1$ & yes & $\gamma(\nu)$  & $\gamma^{(9)}$ & \cellcolor{green} M (MW) \\
\hline
\rowcolor{cyan}$9$ & $1$ & $\R$ & $16$ & $16$ & $\Mat(16,\mathbb{R})^{\oplus 2}$ & $\Mat(16,\mathbb{R})$& $2$ & no & N/A  & $\pm \hbox{1\kern-.27em l}$ & M \\
\hline
$10$ & $2$ & $\R$ & $32$ & $32$ & $\Mat(32,\mathbb{R})$ & $\Mat(32,\mathbb{R})$ & $1$ & yes & N/A & $\gamma^{(11)}$  & M \\
\hline
\end{tabular}}
\caption{{\scriptsize Clifford algebras, representations and character of (s)pinors for Riemannian manifolds. In this case, one has $q=0$ and $d=p$.}}
\label{table:Euclidean}
\end{table}

\begin{table}[H]
\centering
{\tiny
\begin{tabular}{|c|c|c|c|c|c|c|c|c|c|c|c|}
\hline
$d$ & $\begin{array}{c}d-2\\\mathrm{mod}~8\end{array}$ & $\S$ & $\Delta$ & N & $\Cl(p,q)$& $\begin{array}{c}\mathrm{Irrep.}\\\mathrm{image}\end{array}$& $\begin{array}{c} \mathrm{No.~of}\\ \mathbb{R}\mathrm{-irreps.}\end{array}$ & Injectve ? & $\begin{array}{c}\mathrm{Chirality}\\ \mathrm{operator}\\ \cR \end{array}$ 
&  $\gamma^{(d+1)}$ & $\begin{array}{c}\mathrm{Name~of}\\\mathrm{pinors}\\\mathrm{(spinors)}\end{array}$\\
\hline\hline
$1$ & $\mathbin{\textcolor{ForestGreen}{\mathbf{7}}}$ & $\C$ & $1$ & $2$ & $\Mat(1,\mathbb{C})$& $\Mat(1,\mathbb{C})$ & $1$ & yes & $D$  & $\pm J$ & \cellcolor{green} DM (M) \\
\hline
$2$ & $\mathbin{\textcolor{ForestGreen}{\mathbf{0}}}$ & $\R$ & $2$ & $2$ & $\Mat(2,\mathbb{R})$ & $\Mat(2,\mathbb{R})$& $1$ & yes & $\gamma(\nu)$  & $\gamma^{(3)}$  & \cellcolor{green}M (MW) \\
\hline
\rowcolor{cyan}$3$ & $1$ & $\R$ & $2$ & $2$ & $\Mat(2,\mathbb{R})^{\oplus 2}$ & $\Mat(2,\mathbb{R})$ & $2$ & no &  N/A & $\pm \hbox{1\kern-.27em l}$ & M \\
\hline
$4$ & $2$ & $\R$ & $4$ & $4$ & $\Mat(4,\mathbb{R})$ & $\Mat(4,\mathbb{R})$ & $1$ & yes & N/A & $\gamma^{(5)}$  & M \\
\hline
$5$ & $3$ & $\C$ & $4$ & $8$ & $\Mat(4,\mathbb{C})$ & $\Mat(4,\mathbb{C})$ & $1$ & yes & N/A & $\pm J$ & SM \\
\hline
$6$ & $\mathbin{\textcolor{ForestGreen}{\mathbf{4}}}$ & $\H$ & $4$ & $16$ & $\Mat(4,\mathbb{H})$ & $\Mat(4,\mathbb{H})$ & $1$ &  yes & $\gamma(\nu)\circ J$  & $\gamma^{(7)}$ & \cellcolor{green}SM (SMW) \\
\hline
\rowcolor{cyan} $\mathbin{\textcolor{ForestGreen}{\mathbf{7}}}$ & $5$ & $\H$ & $4$ & $16$ & $\Mat(4,\mathbb{H})^{\oplus 2}$ & $\Mat(4,\mathbb{H})$& $2$ & no &  N/A  & $\pm  \hbox{1\kern-.27em l}$ & SM \\
\hline
$8$ & $\mathbin{\textcolor{ForestGreen}{\mathbf{6}}}$ & $\H$ & $8$ & $32$ & $\Mat(8,\mathbb{H})$ & $\Mat(8,\mathbb{H})$ & $1$ & yes &  $\gamma(\nu)\circ J$  & $\gamma^{(9)}$ & \cellcolor{green}DM (M) \\
\hline
$9$ & $\mathbin{\textcolor{ForestGreen}{\mathbf{7}}}$ & $\C$ & $16$ & $32$ & $\Mat(16,\mathbb{C})$ & $\Mat(16,\mathbb{C})$& $1$ & yes &  $D$  & $\pm J$ & \cellcolor{green} DM (M) \\
\hline
$10$ & $\mathbin{\textcolor{ForestGreen}{\mathbf{0}}}$ & $\R$ & $32$ & $32$ & $\Mat(32,\mathbb{R})$ & $\Mat(32,\mathbb{R})$ & $1$ & yes & $\gamma(\nu)$ & $\gamma^{(11)}$ & \cellcolor{green} M (MW) \\
\hline
\rowcolor{cyan}$11$ & $1$ & $\R$ & $32$ & $32$ & $\Mat(32,\mathbb{R})^{\oplus 2}$ & $\Mat(32,\mathbb{C})$ & $2$ & no &  N/A  & $\pm  \hbox{1\kern-.27em l}$ & M \\
\hline
\end{tabular}
}
\caption{{\scriptsize Clifford algebras, representations and character of (s)pinors for Lorentzian manifolds. 
In this case, one has $p=d-1$, $q=1$ and $p-q=d-2$.}}
\label{table:Minkowskian}
\end{table}
\noindent For the name of (s)pinors we used here the following abbreviations: M=Majorana, MW=Majorana-Weyl,  SM=symplectic Majorana, SMW=symplectic Majorana-Weyl, DM=double Majorana.


\end{document}